\documentclass[longauth]{aa} 

\usepackage{graphicx}
\usepackage[colorlinks=true, citecolor=blue]{hyperref}
\usepackage{txfonts}
\usepackage{longtable}
\begin{document} 

   \title{ESPRESSO high resolution transmission spectroscopy of WASP-76~b\thanks{Based on Guaranteed Time Observations collected at the European Southern Observatory under ESO programme 1102.C-0744 by the ESPRESSO Consortium.}}

   \subtitle{}

   \author{H. M. Tabernero\inst{1,2} \and M. R.  Zapatero Osorio\inst{2} \and R. Allart\inst{3} \and F. Borsa\inst{4} \and N. Casasayas-Barris\inst{5,6} \and O. Demangeon\inst{1,14} \and D. Ehrenreich\inst{3} \and J. Lillo-Box\inst{2} \and C. Lovis\inst{3} \and E. Pall{\'e}\inst{5,6} \and S. G. Sousa\inst{1} \and R. Rebolo\inst{5,6} \and N. C. Santos\inst{1,14} \and F. Pepe\inst{3} \and S. Cristiani\inst{7} \and V. Adibekyan\inst{1,14}  \and C. Allende Prieto\inst{5,6} \and Yann Alibert\inst{8} \and S. C. C. Barros\inst{1} \and F. Bouchy\inst{3} \and V. Bourrier\inst{3}  \and V. D'Odorico\inst{7} \and X. Dumusque\inst{3} \and J. P. Faria\inst{1,14} \and P. Figueira\inst{9,1} \and R. G{\'e}nova Santos\inst{5,6} \and  J. I. Gonz{\'a}lez Hern{\'a}ndez\inst{5,6} \and S. Hojjatpanah\inst{1,14}  \and  G. Lo Curto\inst{9} \and B. Lavie\inst{3} \and C. J. A. P. Martins\inst{1,16} \and J. H. C. Martins\inst{1,14} \and  A. Mehner\inst{9} \and G. Micela\inst{10}
	  \and P. Molaro\inst{7,15} \and  N. J. Nunes\inst{1,11} \and E. Poretti\inst{4,12}
	  \and  J. V. Seidel\inst{3} \and A. Sozzetti\inst{13}   \and A. Su{\'a}rez Mascare{\~n}o\inst{5,6} \and  S. Udry\inst{3}
 \and  M. Aliverti\inst{4} \and M. Affolter\inst{8} \and D. Alves\inst{11} \and M. Amate\inst{5,6} \and  G. Avila\inst{9} \and T. Bandy\inst{8} \and W. Benz\inst{8}  \and A. Bianco\inst{4} \and C. Broeg\inst{8} \and A. Cabral\inst{1,11} \and P. Conconi\inst{4} \and J. Coelho\inst{1,11}  \and C. Cumani \inst{9} \and S. Deiries\inst{9} \and H. Dekker\inst{9} \and B. Delabre\inst{9} \and A. Fragoso\inst{5,6} \and M. Genoni\inst{4} \and L. Genolet\inst{3} \and I. Hughes\inst{3} \and J. Knudstrup\inst{9} \and F. Kerber\inst{9}  \and M. Landoni\inst{4} \and J. L. Lizon\inst{9} \and C. Maire\inst{3}  \and A. Manescau\inst{9} \and P. Di Marcantonio\inst{7} \and D. M{\'e}gevand\inst{3} \and M. Monteiro\inst{1,14} \and M. Monteiro\inst{1,11} \and M. Moschetti\inst{4} \and E. Mueller\inst{9} \and A. Modigliani\inst{9} \and L. Oggioni\inst{4} \and A. Oliveira\inst{1,11} \and G. Pariani\inst{4} \and L. Pasquini\inst{9} \and J.~L. Rasilla\inst{5,6} \and E. Redaelli\inst{4} \and M. Riva\inst{4} \and S. Santana-Tschudi\inst{5,6} \and P. Santin\inst{7} \and P. Santos\inst{1,11} \and A. Segovia\inst{3} \and D. Sosnowska\inst{3} \and P. Spanò\inst{4} \and F. Tenegi\inst{5,6} \and O. Iwert\inst{9} \and A. Zanutta\inst{4} \and F. Zerbi\inst{4}
  } 

   \institute{
	Instituto de Astrof{\'i}sica e Ci{\^e}ncias do Espa\c{c}o, Universidade do Porto, CAUP, Rua das Estrelas, 4150-762 Porto, Portugal                                                                
        \email{hugo.tabernero@astro.up.pt}    
	\and
	Centro de Astrobiolog{\'i}a (CSIC-INTA), Carretera de Ajalvir km 4, Torrej{\'o}n de Ardoz, 28850 Madrid, Spain
	\and
	Université de Genève, Observatoire Astronomique, 51 ch. des Maillettes, 1290 Versoix, Switzerland
	\and
	INAF - Osservatorio Astronomico di Brera, Via Bianchi 46, 23807  Merate, Italy
	\and
	Instituto de Astrof\'{i}sica de Canarias (IAC), 38205 La Laguna, Tenerife, Spain
          \and
           Universidad de La Laguna (ULL), Departamento de Astrof\'{i}sica, 38206 La Laguna, Tenerife, Spain
          \and
	INAF, Osservatorio Astronomico di Trieste, Via Tiepolo 11,  34143  Trieste, Italy
	\and
	Universität Bern, Physikalisches Institut, Siedlerstrasse 5, 3012 Bern, Switzerland
	\and
        European Southern Observatory, Karl-Schwarzschild-Strasse 2, 85748  Garching b. München, Germany 
	\and
	INAF - Osservatorio Astronomico di Palermo, Piazza del Parlamento 1, 90134 Palermo, Italy
	\and
	Faculdade de Ciências da Universidade de Lisboa (Departamento de F{\'i}sica), Edifício C8, 1749-016 Lisboa, 1749-016, Lisboa, Portugal
	\and
	INAF - Fundaci\'on Galileo Galilei, Rambla Jos\'e Ana Fernandez P\'erez 7, 38712 Breña Baja, Tenerife, Spain
	\and
	INAF - Osservatorio Astrofisico di Torino, via Osservatorio 20, 10025 Pino Torine
	\and
          Departamento de Física e Astronomia, Faculdade de Ciências, Universidade do Porto, Rua do Campo Alegre, 4169-007 Porto, Portugal
         \and
         Institute for Fundamental Physics (IFPU), Via Beirut, 2, 34151 Grignano TS, Italy
         \and
         Centro de Astrofísica da Universidade do Porto, Rua das Estrelas, 4150-762 Porto, Portugal
         }

   \date{Received XX XX, 2020; accepted XX XX, 2020}

 
  \abstract
   {}
   {We report on ESPRESSO high-resolution transmission spectroscopic observations of two primary transits of the highly-irradiated, ultra-hot Jupiter-size planet WASP-76b. We investigate the presence of several key atomic and molecular features of interest that may reveal the atmospheric properties of the planet.}
   {
   We extracted two transmission spectra of WASP-76b with $R$~$\approx$~140,000 using a procedure that allowed us to process the full ESPRESSO wavelength range (3800-7880 \AA) simultaneously. We observed that at a high signal-to-noise ratio, the continuum of ESPRESSO spectra shows “wiggles” that are likely caused by an interference pattern outside the spectrograph. To search for the planetary features, we visually analysed the extracted transmission spectra and cross-correlated the observations against theoretical spectra of different atomic and molecular species.}
   {The following atomic features are detected: \ion{Li}{i}, \ion{Na}{i}, \ion{Mg}{i}, \ion{Ca}{ii}, \ion{Mn}{i}, \ion{K}{i}, and \ion{Fe}{i}.  All are detected with a confidence level between 9.2~$\sigma$ (\ion{Na}{i}) and 2.8~$\sigma$ (\ion{Mg}{i}). We did not detect the following species:  \ion{Ti}{i},    \ion{Cr}{i}, \ion{Ni}{i}, TiO, VO, and ZrO. We impose the following  1~$\sigma$ upper limits on their detectability: 60, 77, 122, 6, 8, and 8~ppm, respectively.
  }
   {We report the detection of \ion{Li}{i} on WASP-76b for the first time. In addition, we found the presence of \ion{Na}{i} and \ion{Fe}{i} as previously reported in the literature. We show that the procedure employed in this work can detect features down to the level of $\sim$0.1~$\%$ in the transmission spectrum and $\sim$~10~ppm by means of a cross-correlation method. We discuss the presence of neutral and singly ionised features in the atmosphere of WASP-76b.}
   

\maketitle

\section{Introduction}

Planets orbiting late-type stars are ubiquitous, as demonstrated by the thousands of planets discovered to  date\footnote{\url{https://exoplanetarchive.ipac.caltech.edu}}. Their numbers  have been steadily increasing during the last decades thanks to multiple spectroscopic and photometric surveys carried out by, e.g., the space missions Kepler \citep{kepler} and TESS \citep{ric14} and the ground-based spectrographs  HARPS \citep{may03}, HARPS-N \citep{cos12},  HIRES \citep{vog94}, CARMENES \citep{quir16}, or MARVELS \citep{ala15}. Moreover, their numbers and our knowledge about exoplanets will increase thanks to new and future observing facilities like ESPRESSO \citep{pepe10,pepe20}, CHEOPS \citep{ran18}, HPF \citep{hpf}, JWST \citep{gar06},  PLATO \citep{rau14}, NIRPS \citep{wil17}, and SPIRou \citep{mou15}. Exoplanetary research is now approaching the deep study and characterisation of the atmospheres of the extrasolar planets. \\

Transiting exoplanets are of great interest to investigate their atmospheres and learn about their bulk chemical composition. In particular, highly-irradiated gaseous planets are a key target for atmospheric characterisation due to  their intrinsic properties, such as their proximity to the parent host-star, their transit depth, and transit duration. \citet{char02} reported the first detection of an exoplanet atmosphere by means of HST transmission spectroscopy. However, \citet{cas20} and {\color{blue} Casasayas-Barris et al. (2020b, in prep)}  raised some doubts over that detection showing that it could be explained by the Rossiter-McLaughlin effect. In spite of this, \citet{char02} opened a new era in the study of the atmospheres of transiting exoplanets. The first neutral Sodium detection in HD~209458~b with a ground-based telescope was made by \citet{sne08} as well as for HD~189733~b by \citet{red08}. Several chemical species have already been reported in the atmospheres of tens of the so-called Ultra-Hot Jupiters (UHJs), i.e., giant gaseous planets with typical equilibrium temperatures (T$_{eq}$) above $\approx$~2200~K \citep[see ][]{par18}. In particular, \ion{Fe}{i,ii} and \ion{Ti}{i,ii} have been detected in the atmosphere of the  UHJ Kelt-9~b \citep{hoe18,hoe19}. In addition, \ion{Fe}{i,ii} has also been reported in other UHJs such as MASCARA-2b \citep{cas19}, WASP-121b \citep{gib20,hoe20}, or WASP-76b \citep{ehr20}. Other species such as He, Na, Mg, K, V, Cr, CO, CH$_4$, or water vapor have been reported in the atmospheres of tens of highly irradiated gaseous planets  \citep[e.g.,][]{barm15,wyt15,she17,chen18,nor18,par18,all18,all19,alo19,sei19,hoe20}.\\

Spectroscopy is currently a powerful tool able to detect atomic or molecular features in the transmission spectra of transiting exoplanets, which in turn, are key to reveal their internal chemistry and surface composition. The Echelle Spectrograph for Rocky Exoplanets and Stable Spectroscopic Observations \citep[ESPRESSO, see ][]{pepe10,pepe20} is the new generation high-resolution spectrograph  at the 8-m Very Large Telescope (VLT) at Paranal, Chile. ESPRESSO at the VLT offers an excellent opportunity for atmospheric characterisation, given the large collecting area of the VLT. This is a key  factor for achieving the necessary signal-to-noise ratio (S/N) per resolution element during the critical and limited observational windows of transiting exoplanets. In addition, studying the atmospheres of gas giant planets at optical wavelengths is a complement to future observations at much longer wavelengths by JWST \citep{gar06} and Ariel \citep{ariel}. \\

In this work, we present the detailed analysis of the transmission spectrum of WASP-76b using ESPRESSO data. WASP-76b orbits an F7~V star with m$_V$~$\approx$~9.5~mag with an orbital period of approximately 1.8~d \citep{wes16,ehr20}. WASP-76b is an inflated UHJ with roughly one Jupiter mass and twice its radius with an equilibrium temperature of more than 2200~K. Thus, it is a perfect target for atmospheric characterisation given its high equilibrium temperature and the high-metallicity of the host-star \citep[see ][]{sei19,ehr20,edw20,ess20}. We analysed the same data as in \citet{ehr20} with the objective in mind of exploring which spectral atomic and molecular species can be detected using ESPRESSO. \\

This manuscript is organised into the following sections: the ESPRESSO observations are presented in Sect.~\ref{sec_obs}, the extraction of the transmission spectrum is described in Sect.~\ref{sec_data}, whereas the stellar characterisation of WASP-76 can be found in Sect.~\ref{carac}, the transmission spectrum is analysed in Sect.~\ref{sec_analysis}, finally the summary and conclusions are given in Sect.~\ref{sect_sum}.

\section{Observational data}
\label{sec_obs}

We collected several high-resolution {\it echelle} spectra covering two transits of the ultra-hot Jupiter WASP-76b using the HR21\footnote{HR21 uses a binning of a factor of 2 in the direction perpendicular to wavelength.} mode of ESPRESSO covering the optical wavelengths from 3800 to 7880~\AA{} with $R$~$\approx$~140,000. These two transit observations were carried out as part of the ESPRESSO Guaranteed Time Observation under ESO programme 1102.C-744. The observations were reduced using the ESPRESSO reduction pipeline\footnote{\url{https://www.eso.org/sci/software/pipelines/}}. The pipeline delivers the necessary products to further process the data, e.g.: barycentric corrected radial velocities ($V_r$), the stellar fluxes together with their uncertainties, wavelengths (in vacuum),  and S/N. \\
\begin{table}
\centering
\tiny
\caption{Summary of the  WASP-76b transit observations}             
\label{obs_trans}      
\begin{tabular}{ccccc}   
\hline
\hline 
     &   \multicolumn{3}{c}{Number of Spectra}  & \\ 
Date  &  Total & In-transit & Out-of-transit & $t_{exp}$ (s) \\    
\hline    
\noalign{\smallskip}
2018-09-03 & 36 &   20   & 16  & 600  \\
2018-10-31 & 70 &   38   & 32  & 300 \\
\hline
\end{tabular}
\end{table}
 The two transits of WASP-76b were observed on the following Universal Time (UT) dates: 2018 September 03 (first transit, hereafter T1) and 2018 October 31 (second transit, T2). On the two occasions, we observed the target uninterruptedly during $~\sim$2~hours before, during, and $\sim$2~h after the transit. In T1, we collected a total of 36 ESPRESSO spectra with an individual exposure time of 600 s; in T2, we acquired 70 spectra of 300~s each. The average signal-to-noise (S/N) of each individual observation is about 120 per pixel element at $\sim$~5500~\AA{} \citep[see][]{ehr20}. For T1, we performed longer exposure times than for T2 due to weather constraints (i.e., poor seeing). Unfortunately, the ESPRESSO Atmospheric Dispersion Corrector (ADC) is not built to correct the atmospheric transmission above an airmass of 2.2. Thus, we discarded two spectra for both transits because they were not suitable for any meaningful transmission spectrum retrieval and later analysis. Other details regarding the observations for these two transits of WASP-76b can be found in Tab.~\ref{obs_trans}, whereas information on the dates and radial velocities can be found in \citet{ehr20}. \\

\section{Data analysis}
\label{sec_data}
\begin{table}
\centering
\tiny
\caption{Orbital and physical parameters for the WASP-76 system.}             
\label{parwasp76}      
\begin{tabular}{l c c }   
\hline
\hline                 
Parameter &  Value & Reference \\    
\hline     
 & Stellar parameters & \\
\hline
\noalign{\smallskip}
   $T_{\rm eff}$ & 6316~$\pm$~64~K &  This work \\      
   $\log{g}$ &   4.13~$\pm$~0.14~dex &   This work \\
   $\lbrack$Fe/H$\rbrack$  & 0.34~$\pm$~0.05~dex   &  This work\\ 
   A(Li)  & 2.47~dex    & This work\\
   $\xi$  &  1.38~$\pm$~0.07~km~s$^{-1}$ & This work \\
   $M_{*}$   &  1.45~$\pm$~0.02~$M_{\odot}$  &  This work \\
   $R_{*}$   &  1.77~$\pm$~0.07~$R_{\odot}$  &  This work \\
   $m_{\rm V}$& 9.5~mag & \citet{wes16}  \\
   $\pi$    &    5.12~$\pm$~0.16~mas  & \citet{GDR2} \\
   $d$      &   195~$\pm$~6~pc  & \citet{GDR2} \\
 \hline
 & Planet parameters & \\
 \hline
 \noalign{\smallskip}
   $P$       & 1.809886~$\pm$~0.000001 d & \citet{ehr20}\\ 
   $M_{p}$  &  0.92~$\pm$~0.03~$M_{J}$  & \citet{ehr20} \\
   $R_{p}$  &  1.83~$\pm$~0.05~$R_{J}$  & \citet{ehr20}  \\
   $K_1$    &  0.1193~$\pm$~0.0018~km~s$^{-1}$ & \citet{ehr20} \\
   $\gamma$ &  -1.0733~$\pm$~0.0002~km~s$^{-1}$ & \citet{ehr20} \\
   $e$      & 0  & assumed  \\
   $\omega$  & 90 &  assumed \\
\hline
\end{tabular}
\end{table}

\subsection{Telluric correction}

The telluric lines have to be removed from the data before extracting the transmission spectrum of WASP-76b, as they might contaminate the final result. To that aim, we gathered the processed one-dimensional sky-corrected spectra provided by the ESPRESSO reduction pipeline. We corrected for the telluric absorption lines by means of the {\tt Molecfit\footnote{\url{https://www.eso.org/sci/software/pipelines/skytools/molecfit}}} software suite \citep{sme15,kau15}.  The spectra provided by the ESPRESSO pipeline are already corrected for the Barycentric Earth Radial Velocity (BERV). However, {\tt Molecfit} models the telluric transmission spectrum in the Terrestial reference frame \citep[see, e.g.][]{all17}. In consequence, we shifted the spectra to the Terrestial reference frame before performing any telluric line fitting with {\tt Molecfit}. \\

\begin{figure}
    \includegraphics[width=0.48\textwidth]{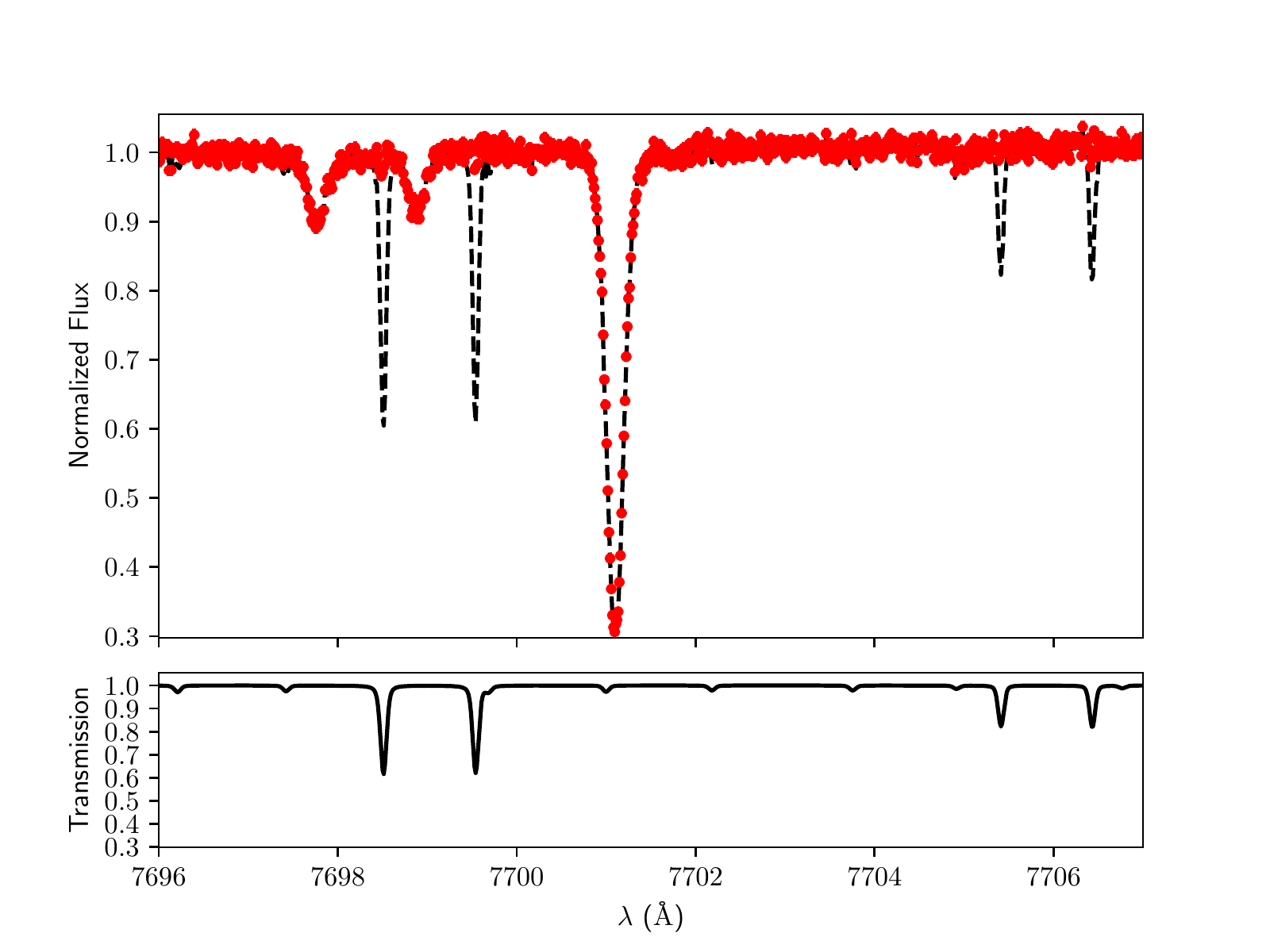}
    \caption{Example of our telluric correction procedure in the \ion{K}{i} region. Top Panel: Uncorrected spectrum (black dashed line), and the corrected spectrum (red dots). Bottom Panel: Best telluric correction solution for a WASP-76 spectrum in this particular wavelength range.}
    \label{telluricex}
\end{figure}

In order to perform the telluric line fitting, we selected three telluric regions in the spectra to fit the H$_2$O and O$_2$ molecular bands (i.e., 6860-6900~\AA{}, 7160-7340~\AA{}, and 7590-7770~\AA{}). Then, we used the  parameters given in Table~\ref{parmolec} and ran {\tt Molecfit} to fit the telluric lines. We refined the {\tt Molecfit} result until the input and output parameters were the same down to the arithmetic precision achievable using the {\tt Molecfit} graphical user interface.  Finally, we gathered the telluric corrected ESPRESSO data and removed the Earth motion using the BERVs already calculated by the ESPRESSO pipeline.

\begin{table*}												
\label{parmolec}
\centering
\caption{{\tt Molecfit} parameters used to correct the telluric lines}
\begin{tabular}{lcc}

\hline
\hline                 
Parameter  &  Value &  Description\\
\hline
\noalign{\smallskip}
{\tt ftol} & 10$^{-5}$ & $\chi^2$ tolerance\\
{\tt xtol} & 10$^{-5}$ & tolerance for the {\tt molectfit} fitted variables\\
{\tt fit\_cont} &  1  & continuum fitting flag\\
{\tt cont\_n} & 3 & degree of polynomial continuum\\
{\tt fit\_res\_gauss} & 1 &  Gaussian kernel\\
{\tt res\_gauss} &  3.5 & kernel size (pixels)\\
{\tt kernfac } & 6.0 & kernel size measured in  units of the kernel $FWHM$\\  
 {\tt list\_molec} & H$_{\rm 2}$O,O$_{\rm 2}$ & molecules to be synthetised \\
 \hline
\end{tabular}
\end{table*}

\subsection{Transmission spectrum extraction}
\label{tran_extract}

The telluric line correction has provided us with a set of clean spectra that we can use to extract the planetary signature. Using these corrected ESPRESSO spectra, we extracted the planet signal by means of a procedure based on the technique described by \citet{wyt15}. First, we shifted the spectra to the stellar reference frame using the RVs calculated with the ephemeris and Keplerian stellar motion given by \cite{ehr20}. Second, each individual spectrum was flux-scaled by means of a second order polynomial \citep[see ][]{tab20} to the lowest air-mass spectrum in its corresponding transit observations. The flux level changes from exposure to exposure, due to variations in airmass and atmospheric transparency, and in consequence each individual spectrum must be corrected for these effects. In other words, thanks to this scaling polynomial the ESPRESSO observations of each transit have been effectively "re-normalized" to the same continuum level.\\

At this point, we have generated a set of aligned observations (in terms of wavelength and flux) that we used to extract the planetary signature. We organised the observed spectra into two categories: in-transit and out-of-transit according to the ephemeris given by \citet{ehr20}. Then, we computed, wavelength by wavelength, the median of the out-of-transit spectra to generate a master stellar spectrum. After this, we divided each individual spectrum by this master spectrum in order to remove the stellar flux from the data. The result of this process is a set of spectra that contain the planetary signal plus the noise left after the removal of the stellar contribution.\\

\begin{figure}
    \includegraphics[width=0.48\textwidth]{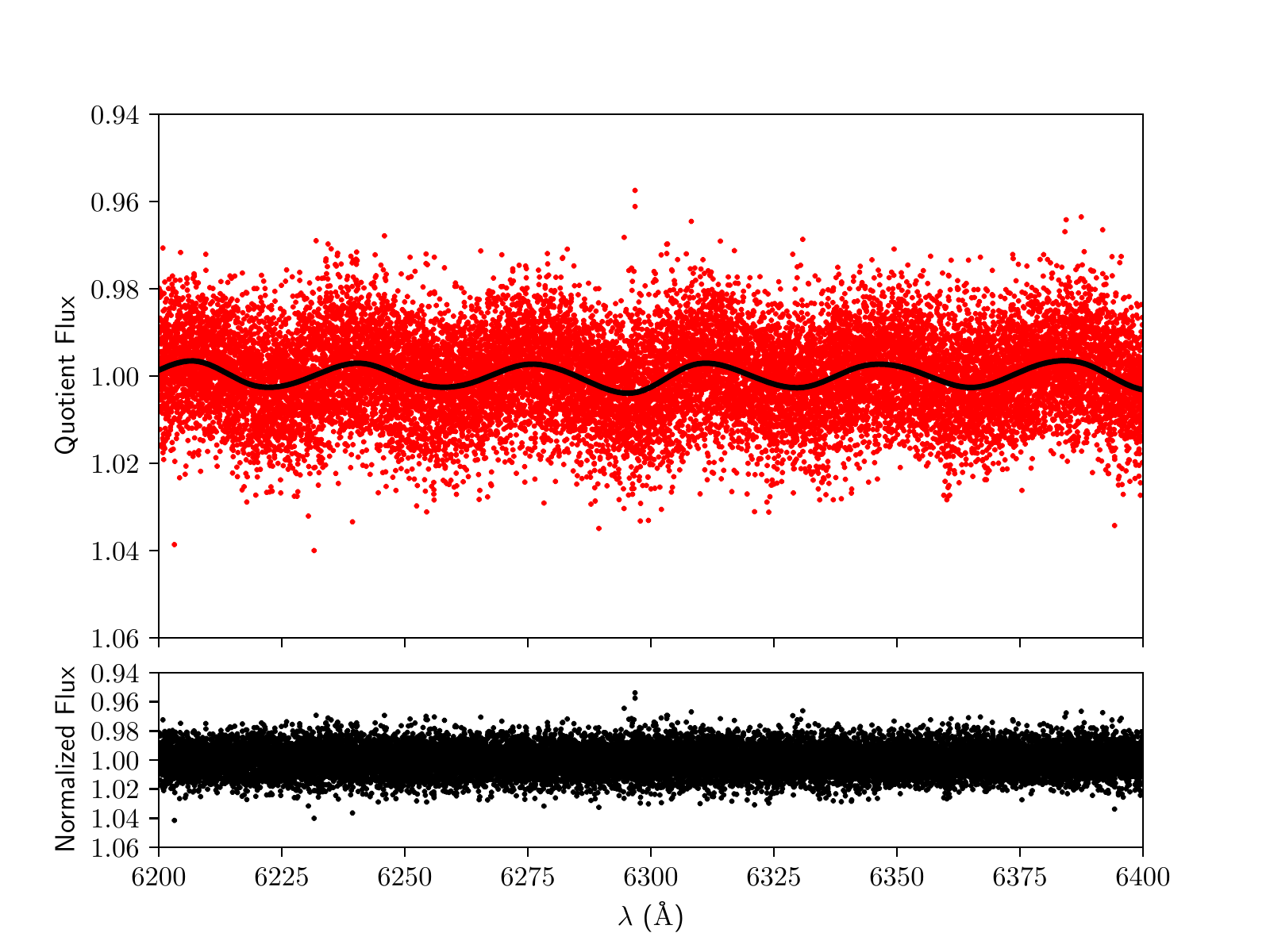}
    \caption{Wiggle correction for a residual spectrum. Upper panel: Original residuals (Red) and the cubic splines used to fit the wiggles (Black). Lower Panel: Final residual spectrum after removing the wiggles. }
    \label{wiggleex}
\end{figure}

After removing the stellar contribution from the data, we found a sinusoidal pattern  (wiggles, thereafter) on each of the resulting residual spectra. These wiggles are likely caused by an interference pattern induced by the coude train optics. Qualitatively speaking, they have an amplitude of ~1\% with a period of $\sim$~30~\AA{} at $\sim$~6300~\AA{} and might affect the final transmission signature, and in consequence they had to be removed from the transit data. To that aim, we removed them by fitting a set of cubic splines to the residual spectra using the {\tt iSpec} code \citep{cua14} (see Fig.~\ref{wiggleex}). Then, the in-transit residual spectra were shifted in velocity to bring them to the planetary-rest frame using the ephemeris given by \citet{ehr20}. These shifted residual spectra were later merged wavelength by wavelength into a single median transmission spectrum. This procedure has been applied independently to T1 and T2 to produce two independent transmission spectra for WASP-76b.\\

\section{Stellar characterisation}
\label{carac}
\subsection{Stellar atmospheric parameters}
\label{stepar}

The stellar atmospheric parameters of WASP-76, namely effective temperature ($T_{\rm eff}$), surface gravity ($\log{g}$),  microturbulence ($\xi$), and metallicity ([Fe/H]), have been calculated using the Equivalent Width ($EW$) method by means of the {\scshape StePar}\footnote{\url{https://github.com/hmtabernero/StePar}} code \citep{tab19}. The atmospheric parameters were computed using the master spectrum from Subsection~\ref{tran_extract}.  The latest state-of-the-art version of {\scshape StePar} relies on the 2017 version of the MOOG code \citep[via the {\tt abfind} driver, see][]{sne73} and a grid of MARCS stellar atmospheric models \citep{gus08}. We employed the selection of $\ion{Fe}{i}$ and $\ion{Fe}{ii}$ following the line list given by \citet{tab19} for metal rich dwarf stars. All the same, the available atomic data of these iron lines were taken from the public version of the Gaia-ESO line list \citep{hei15b}. As damping prescription, we used the Anstee-Barcklem-O'Mara \citep[ABO, see][]{abo98} data (if available), through option 1 of MOOG. In addition, we used ARES\footnote{\url{https://github.com/sousasag/ARES}} \citep{sou15} to measure the Equivalent-widths ($EW$s) of the \ion{Fe}{i,ii} lines used.\\

We only considered measured lines with 10 m\AA~$<$~$EW$~$<$~120~m\AA~to avoid problems with line profiles of very intense lines and tentatively incorrect $EW$ measurements of extremely weak lines. The atmospheric parameters can then be inferred from the previously measured $\ion{Fe}{i}$-$\ion{Fe}{ii}$ line list. The minimisation procedure of {\scshape StePar} is the Downhill Simplex algorithm \citep{pre02}, which tries to minimise a quadratic form composed of the excitation and ionisation equilibrium conditions to find the best parameters of the target star. {\scshape StePar} iterates until $\log{\epsilon}$(\ion{Fe}{i}), and $\log{\epsilon}$(\ion{Fe}{ii}) stand for the Fe abundance returned by the \ion{Fe}{i} and \ion{Fe}{ii} lines, respectively, and $\log{(EW/\lambda)}$ be their reduced equivalent width, {\scshape StePar} iterates until the slopes of $\chi$ versus $\log{\epsilon(\textrm{\ion{Fe}{i}})}$ and $\log{(EW / \lambda)}$ versus $\log{\epsilon(\textrm{\ion{Fe}{i}})}$ are virtually zero, i.e. excitation equilibrium, and imposing ionisation equilibrium, so that $\log{\epsilon(\textrm{\ion{Fe}{i}})} =\log{\epsilon(\textrm{\ion{Fe}{ii}})}$. The stellar parameters derived for WASP-76 in this work can be found in Table~\ref{parwasp76}.\\

We have also calculated stellar atmospheric parameters using the {\tt ARES+MOOG} code \citep{sou08,san13,sou18} and obtained similar results as those provided by {\scshape  StePar}: $T_{\rm eff}$~$=$~6329~$\pm$~24~K, $\log{g}$~$=$~4.20~$\pm$~0.03~dex, $\xi$~$=$~1.54~$\pm$~0.03~km~s$^{-1}$, and [Fe/H]=0.37~$\pm$~0.02~dex. Both {\tt ARES+MOOG} and {\scshape StePar} are two similar implementations of the $EW$ method. {\tt ARES+MOOG} employs a similar workflow to that of {\scshape StePar}, however it implements KURUCZ stellar atmospheric models \citep{kur93}, a solar calibrated $\log{gf}$  \ion{Fe}{i,ii} linelist \citep{sou08} and the damping option 0 of MOOG. In addition, \citet{wes16} calculated the following stellar parameters of $T_{\rm eff}$~$=$~6250~$\pm$~100~K, $\log{g}$~$=$~4.13~$\pm$~0.02~dex, $\xi$~$=$~1.4~$\pm$~0.1~km~s$^{-1}$, and [Fe/H]~$=$~0.23~$\pm$~0.1~dex. These values are in good agreement with those calculated with {\sc StePar} and {\tt ARES+MOOG} codes.\\

\subsection{Stellar mass and radius}

We used the PARAM web interface\footnote{\url{http://stev.oapd.inaf.it/cgi-bin/param_1.3}} \citep{sil06} to calculate the mass ($M_{*}$) and radius ($R_{*}$) of WASP-76. We used the stellar parameters calculated with {\sc StePar} along with the Gaia DR2 Parallax \citep{GDR2}, the visual magnitude ($m_V$) given by \citep{wes16}, and the PARSEC stellar evolutionary tracks and isochrones \citep{bre12} to obtain a mass of 1.45~$\pm$~0.02~$M_{\odot}$ and a radius of 1.77~$\pm$~0.07~$R_{\odot}$. Our values are consistent with those provided by \citet{ehr20} who reported $M_{*}$~$=$~1.46~$\pm$~0.02~$M_{\odot}$ and $R_{*}$~$=$~1.76~$\pm$~0.07~$R_{\odot}$. Whereas \citet{wes16} obtained 1.46~$\pm$~0.07~$M_{\odot}$ and  1.73$\pm$0.04~$R_{\odot}$, being both consistent with the values derived in this work. 

\subsection{Chemical abundances}

We calculated the chemical abundances of WASP-76 by means of the $EW$ method for the following atomic species: \ion{Li}{i} \ion{C}{i}, \ion{O}{i}, \ion{Na}{i}, \ion{Mg}{i}, \ion{Si}{i}, \ion{Ca}{i}, \ion{Ti}{i}, \ion{Cr}{i}, \ion{Mn}{i}, and \ion{Ni}{i}. Moreover, the atomic data for each atomic species under analysis were collected from the Gaia-ESO (GES) line list \citep{hei15b}. We measured an $EW$~$=$29.2~m\AA{} for the \ion{Li}{i} line at 6709.61~\AA{} by means of a Gaussian fit performed by means of the Levenberg–Marquardt algorithm (LMA) implemented in the python library {\tt SciPy} \citep{scipy}. The $EW$s of the other elements were measured by means of the ARES code. Then, we interpolated a model atmosphere from the  MARCS stellar atmospheric grid \citep{gus08} and we used the MOOG code \cite{sne73} to derive the abundances for WASP-76. The \ion{O}{i} abundance was later corrected for NLTE effects thanks to the corrections given by \citet{sit13} by means of the web interface at \url{http://nlte.mpia.de/}. Then, we used the \ion{C}{i} and \ion{O}{i} to derive a value a carbon-to-oxygen ratio (C/O) of 0.51~$\pm$~0.03, which in turn is consistent with the solar value. Finally, the atomic parameters employed  to calculate the abundances can be found in Table~\ref{linelists}. 

\begin{table}
  \caption{Chemical abundances relative to the solar value ([X/H]) for WASP-76.}             
  \label{Abundances}
  \begin{tabular}{cccc}
  \hline
  \hline
Species &   [X/H]  & $\sigma_{lines}$& $N_{lines}$ \\
        &   (dex)  &  (dex)   &   \\
  \hline
  \ion{C}{i}    &  0.19 &  0.03  & 3\\
  \ion{O}{i}     & 0.21 & 0.01   &  3\\
   \ion{Na}{i}  &  0.48  & 0.05  &  4 \\ 
   \ion{Mg}{i}  &  0.32  & 0.04  &  2\\
   \ion{Si}{i}  &  0.36  & 0.09  &  22\\
   \ion{K}{i}   & 0.39  &  --  & 1 \\ 
   \ion{Ca}{i}  &  0.41  & 0.09  & 18\\
    \ion{Ti}{i}  &  0.35  & 0.06   & 18 \\
     \ion{Cr}{i}  &  0.38  & 0.07  & 10\\
     \ion{Mn}{i}  &  0.32  & 0.01 & 3 \\
    \ion{Ni}{i}  & 0.35   & 0.05  & 33 \\
   \noalign{\smallskip}  
\hline
\end{tabular}
\end{table}

\section{Transmission spectrum analysis}
\label{sec_analysis}

\begin{figure}
    \centering
    \includegraphics[width=0.48\textwidth]{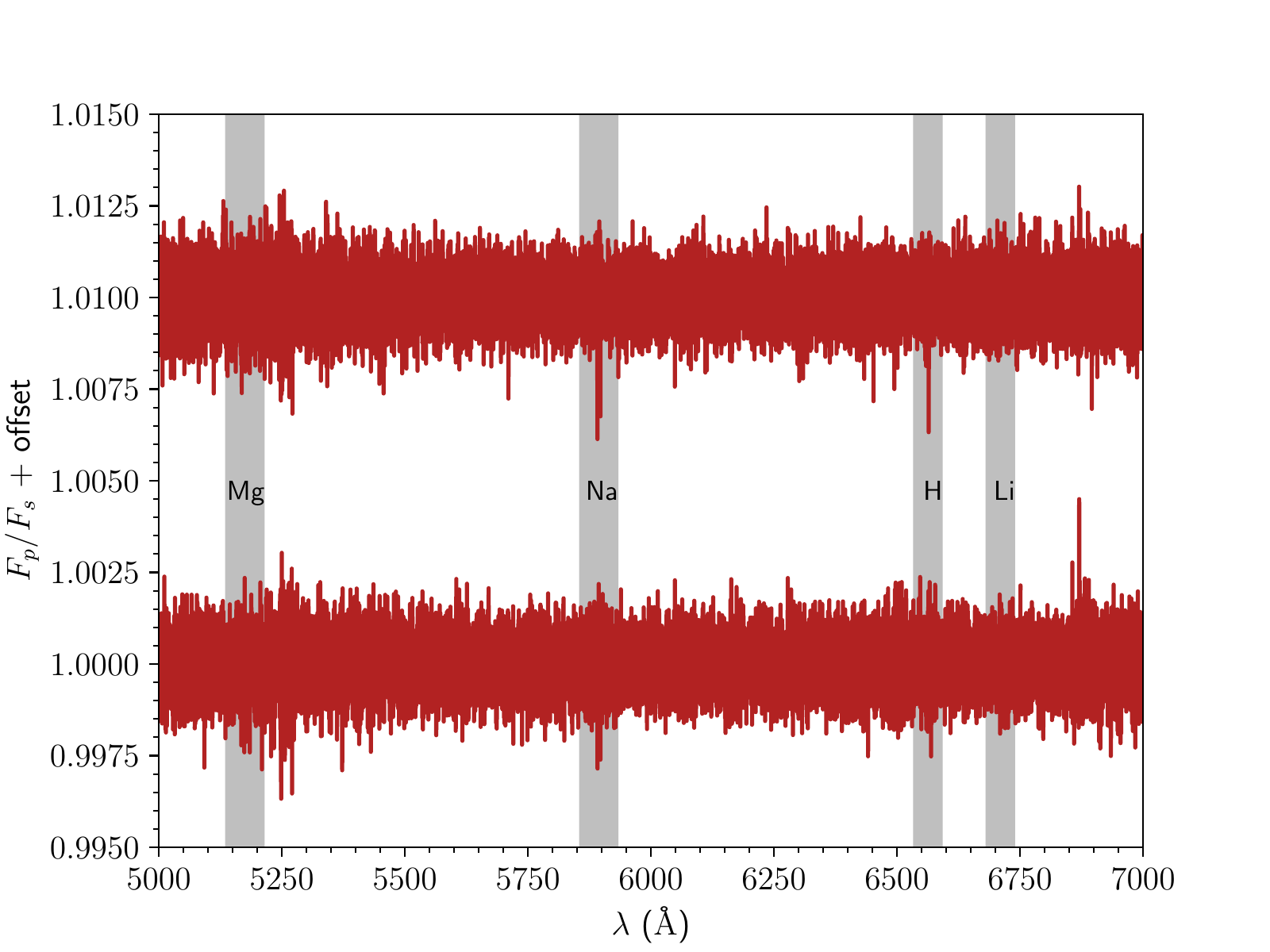}
    \caption{Transmission spectra for WASP-76b in the 5000~$\AA$-7000$\AA$, a few key features are shaded in gray. The T1 is the top spectrum, whereas T2 is the bottom one.  A binning of 0.1~\AA{} has been applied to the data for clarity.
    All wavelengths are given in vacuum.}
    \label{full_spec_planet}
\end{figure}

The procedure employed in this work delivers the entire ESPRESSO spectrum of WASP-76b simultaneously.  This is possible via the stable and consistent wavelength calibration of the ESPRESSO data. The extracted transmission spectra for T1 and T2 are depicted in Fig.~\ref{full_spec_planet}. Using these two spectra we can identify tentative spectral features by visual inspection and perform a cross-correlation by means of a binary mask built for a given atomic or molecular species.  We explored the presence of the following atomic species by direct inspection of the transmission spectrum: \ion{H}{i}, \ion{Li}{i}, \ion{Na}{i}, \ion{Mg}{i}, \ion{Ca}{ii},  \ion{Mn}{i}, \ion{Fe}{i}, \ion{K}{i}. In addition, we explored the presence of other species by means of a cross correlation function (CCF) analysis: \ion{Ti}{i}, \ion{Cr}{i}, \ion{Fe}{i}, and \ion{Ni}{i} in addition to the diatomic molecules TiO, VO, and ZrO. \\ 

Each tentative atomic feature directly seen in the transmission spectrum of WASP-76b has to be scrutinised for its significance. Thus, we modelled each individual feature with a Gaussian profile plus the continuum level. The model fitting is then performed by means of the Levenberg–Marquardt algorithm (LMA) implemented in the python library {\tt SciPy} \citep{scipy}. The LMA  explores the parameter space that  provides us with valuable information about each spectroscopy feature under scrutiny.  In summary, our modelling provides the Gaussian parameters of the line under analysis (i.e., centre, amplitude, depth, and width) alongside their uncertainties. We also calculated the Doppler shift ($V_{wind}$) of each line to explore any tentative planetary winds \citep[see, e.g.,][]{hoe18,cas19}. Moreover, we derived effective planetary radii at center of each feature ($R_{\lambda}$) in units of the radius of the planet ($R_{p}$) at the centre of each individual atomic feature. Thus, we used the following expression: $R_{\lambda}$~$=$~$\sqrt{1+h/\delta}$~$R_p$, where $h$ is the line depth of a given absorption feature in the transmission spectrum and $\delta$ is the transit depth of the planet. All these results can be found in Table~\ref{line_intent}. Regarding these lines we produced a series of tomography plots (see Figs.~\ref{map_na_planet}, and \ref{map_HK} to \ref{map_Ha}) in order to explore the passage of the planet with time  for each transit, which allows us to confirm if the signal. However, some lines are perhaps too  weak to be directly  seen in the tomography plots. \\

\begin{figure}
      \centering
      \includegraphics[width=0.48\textwidth]{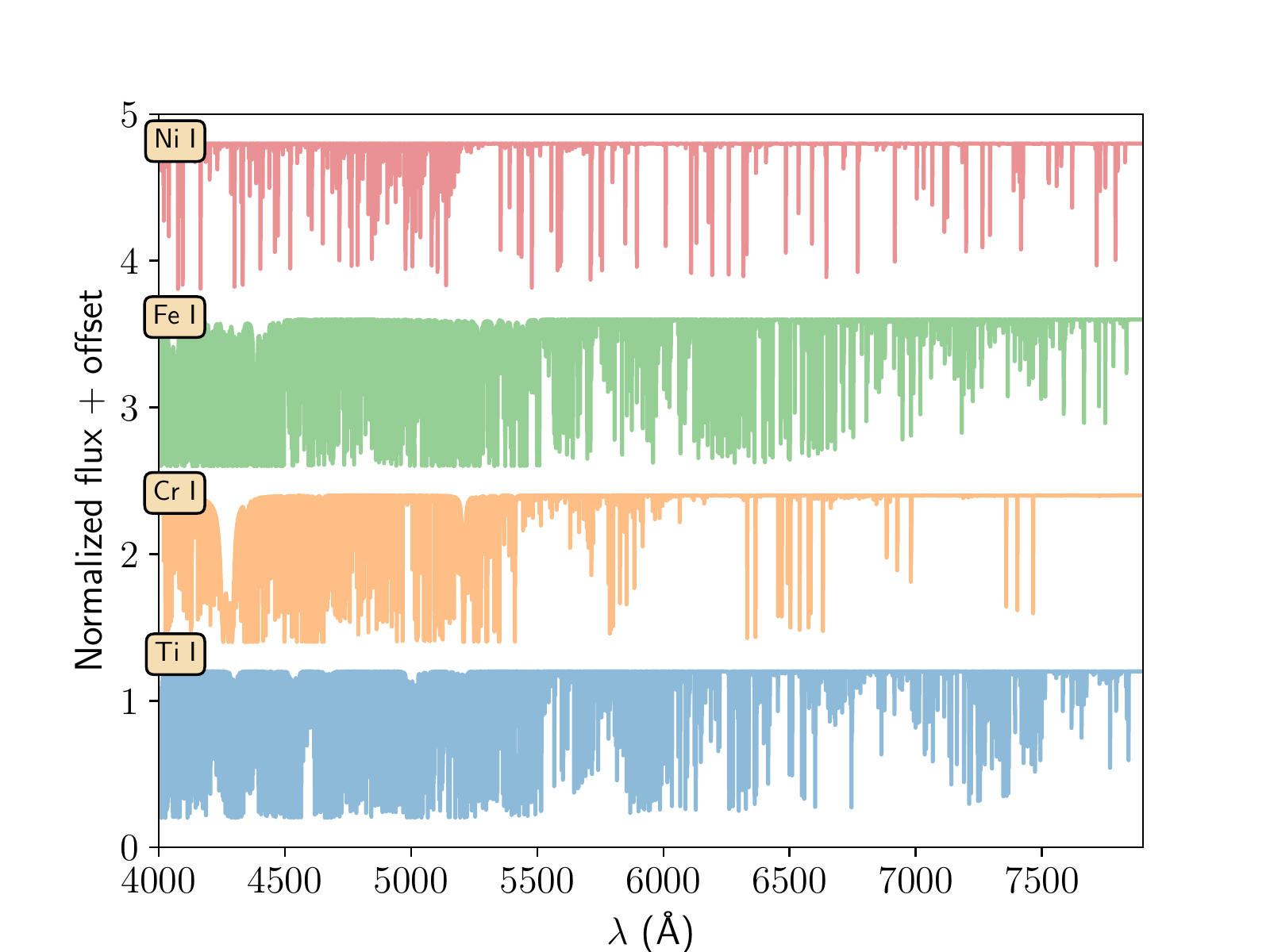}
      \includegraphics[width=0.48\textwidth]{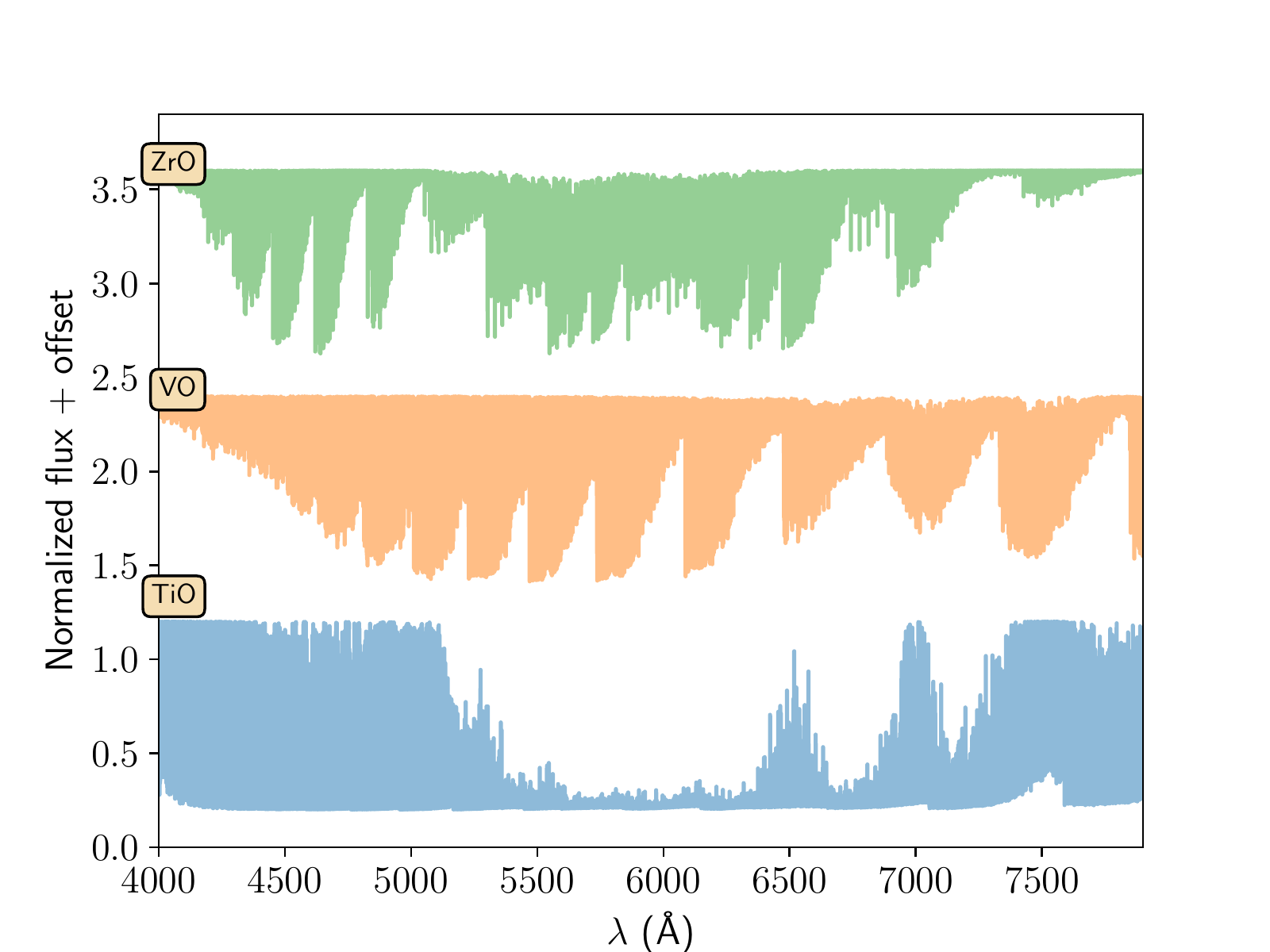}
\caption{Synthetic models generated with {\tt turbospectrum} for the different atomic and molecular species. All of them have been degraded to $R$~$=$~140,000.}
      \label{atom_molec_planet}
  \end{figure}

In addition, the detection of some atomic or molecular species in the atmosphere of the planet can be achieved by cross-correlating the planetary signal with a synthetic spectrum.  To that particular aim, we calculated an atmospheric structure of WASP-76b by means of the HELIOS code\footnote{\url{https://github.com/exoclime/HELIOS}} \citep{mal17,mal19}. In order to generate a model spectrum, we need a radiative transfer code to input the planet atmospheric structure calculated by HELIOS. The aforementioned atmospheric structure consists of a set of atmospheric layers containing  the gaseous pressure (P$_g$), temperature (T), and geometrical depth. These quantities are necessary to solve the radiative transfer equation in order to produce a synthetic model spectrum. We employed {\tt turbospectrum}\footnote{\url{https://github.com/bertrandplez/Turbospectrum2019}} to solve the radiative transfer problem \citep{ple12} alongside seven different line lists comprised of the following atomic and molecular species: \ion{Cr}{i}, \ion{Ti}{i}, \ion{Fe}{i}, \ion{Ni}{i}, TiO, VO, and ZrO. The atomic data were downloaded using the VALD3 interface\footnote{\url{http://vald.astro.uu.se/}}  \citep{rad15} whereas the molecular data for TiO, VO, and ZrO were gathered from \citet{tioplez}, \citet{voexomol}, and \citet{zroplez}, respectively. Furthermore, {\tt turbospectrum} is a general purpose radiative transfer tool that can generate high resolution synthetic spectra by solving the problem of radiative transfer in spherical geometry. The resulting synthetic spectra (see Fig.~\ref{atom_molec_planet}) were later converted into binary line-masks that we used to perform a cross-correlation against our two transmission spectra. To perform the cross-correlation we employed  the {\tt iSpec} \citet{cua14} code, which in turn implements the algorithm described in \citet{pepe02} to calculate the cross correlation function (CCF). The goal of the cross-correlation technique is to combine thousands of spectral features to produce an imprint that we will be able detect. 

\subsection{\ion{Na}{i}}

\begin{figure*}
\centering
    \includegraphics[width=0.48\textwidth]{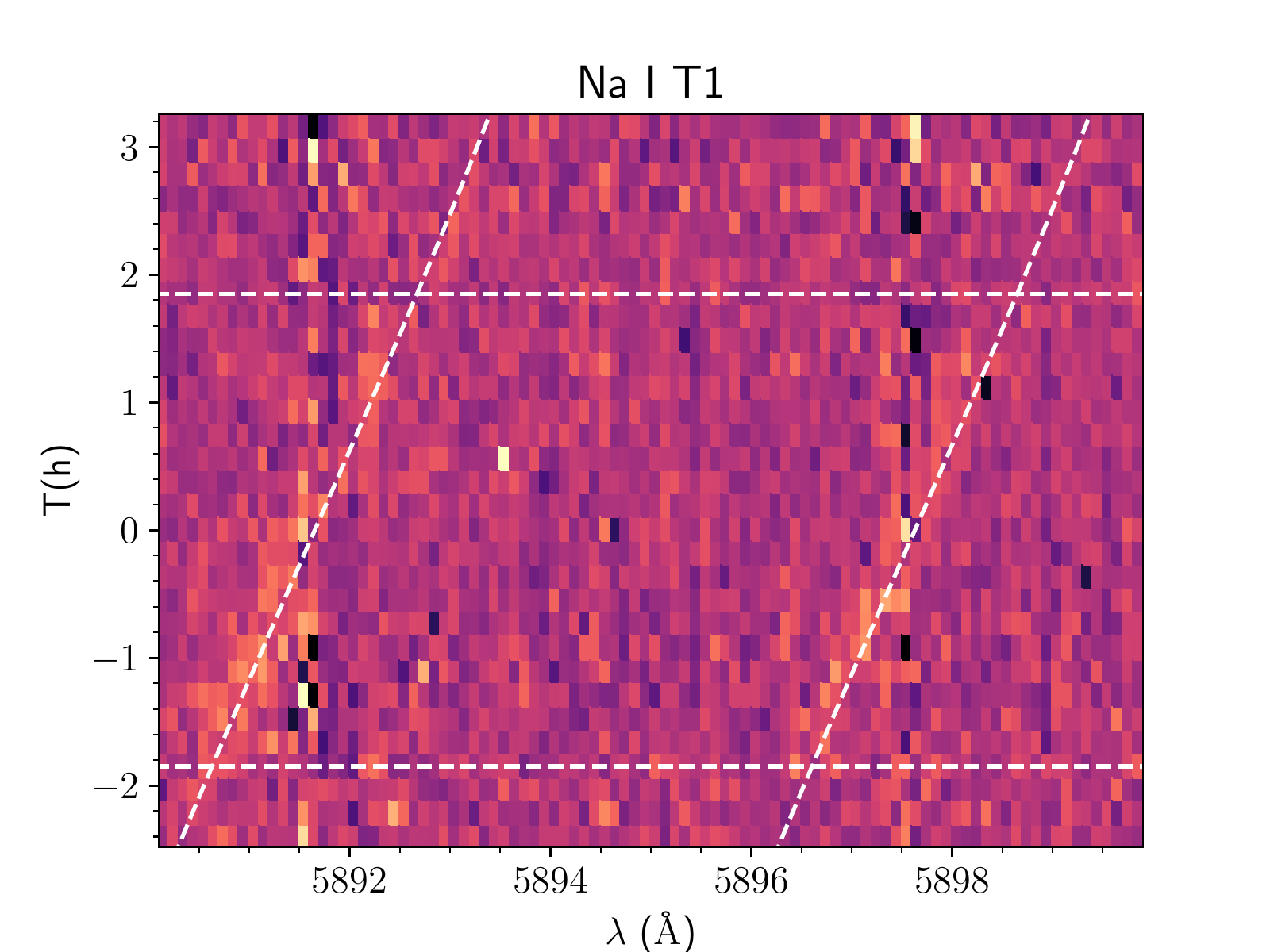}
    \includegraphics[width=0.48\textwidth]{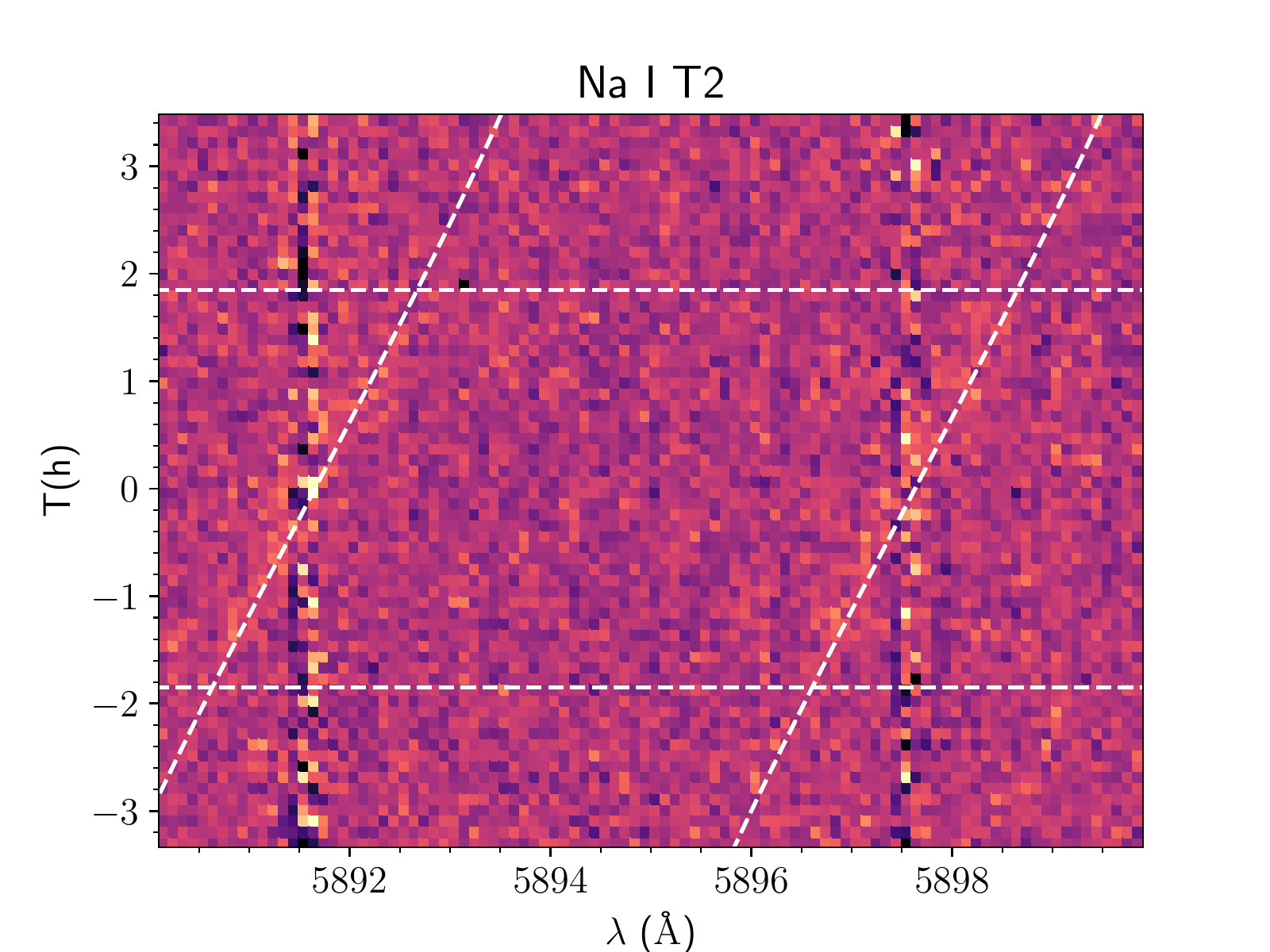}
    \caption{\ion{Na} Double tomography for transit 1 (left) and transit 2 (right).  Time 0.0 h corresponds to the mid-transit according to the planetary ephemeris given in Table~\ref{parwasp76}. Each tile represents a  0.1~\AA{} bin. These plots are qualitative and they have been generated only for visualization purposes.}
    \label{map_na_planet}
\end{figure*}

\begin{figure}
    \centering
    \includegraphics[width=0.48\textwidth]{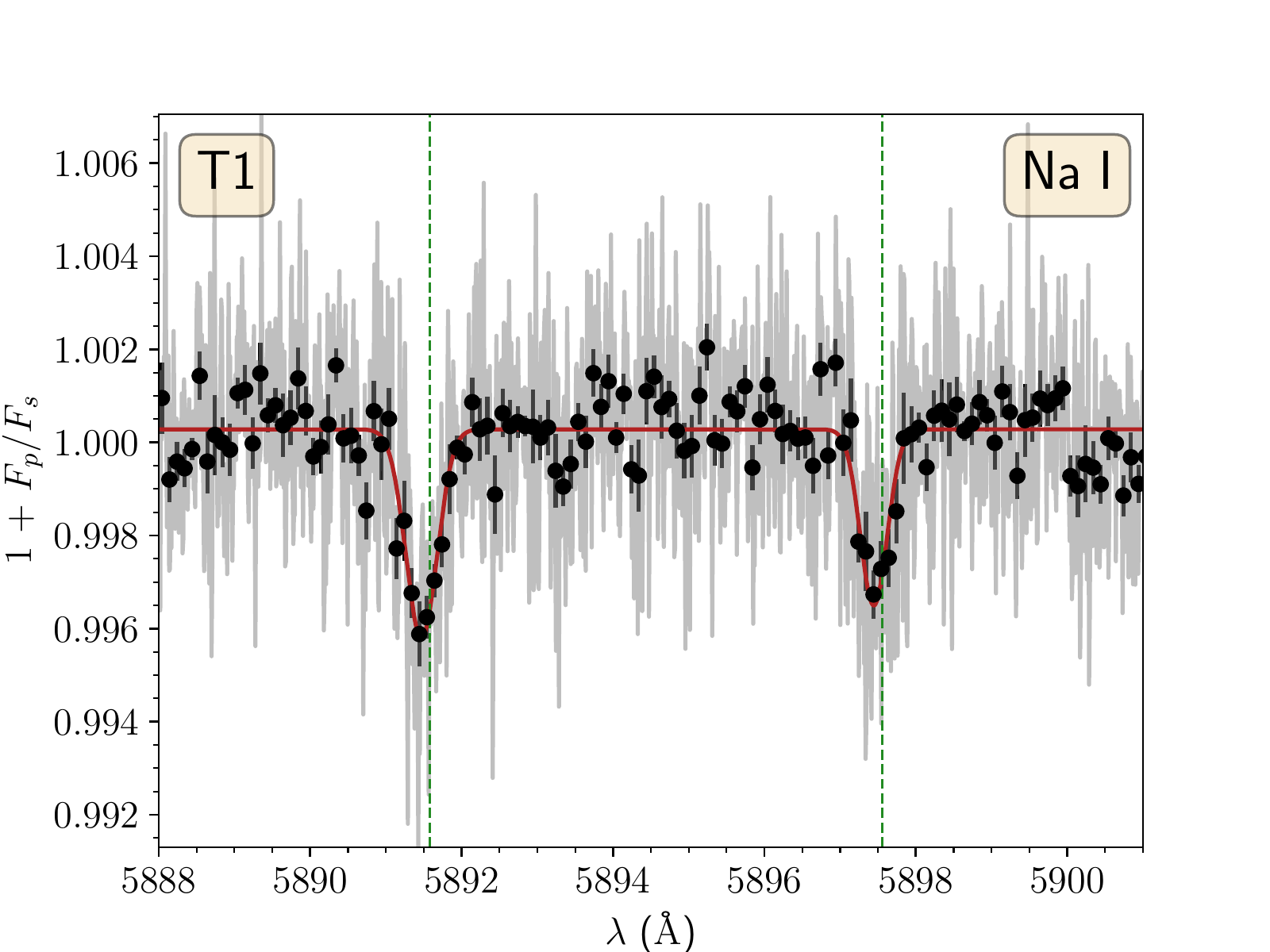}
    \includegraphics[width=0.48\textwidth]{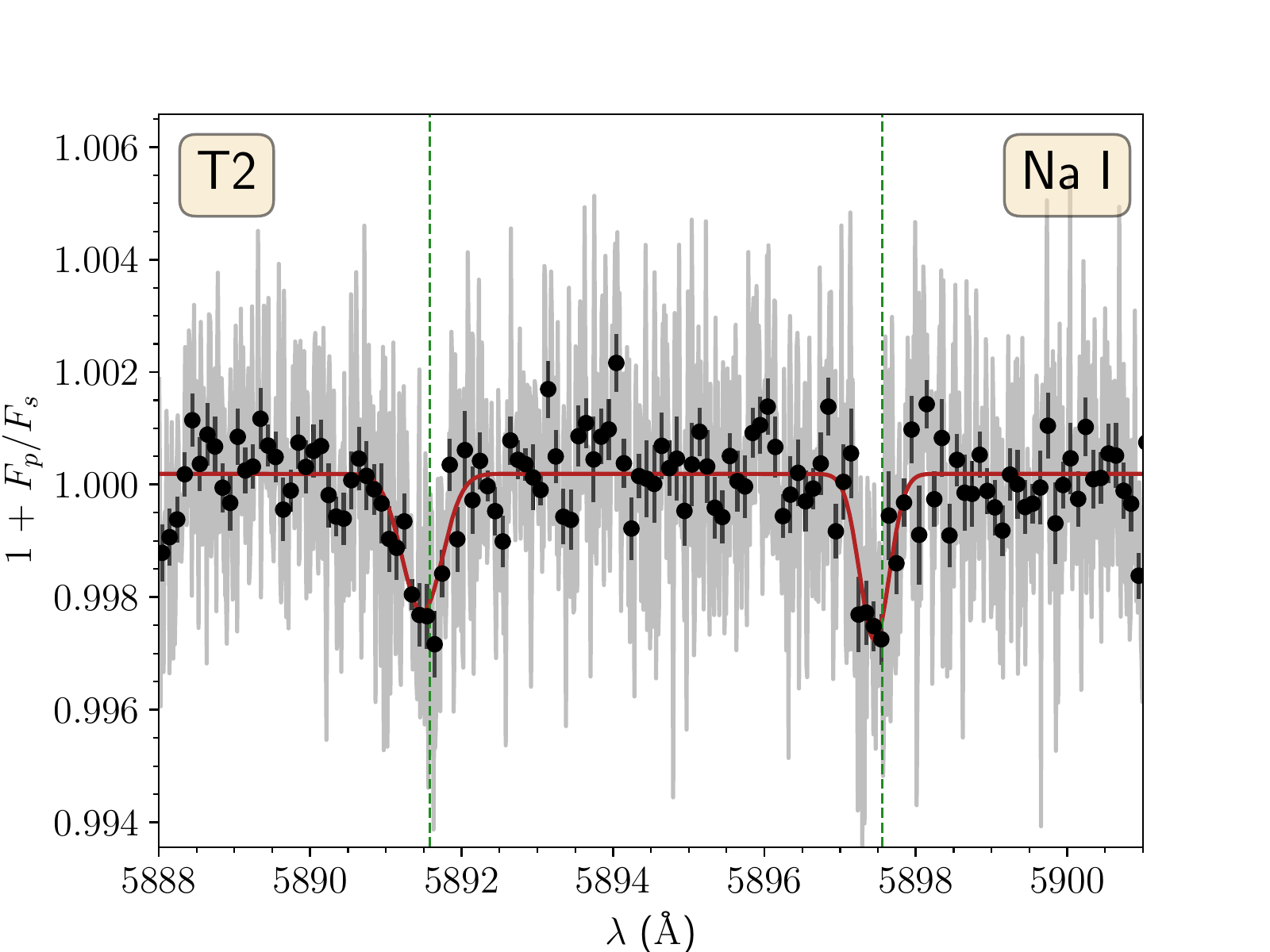}
    \caption{Transmission spectra around the \ion{Na}{i} doublet: top (T1), bottom (T2). The grey line represents the original transmission spectrum of WASP-76b, whereas the black dots represent a binning of 0.1~\AA{}. The red line represents the best fit to the data. The rest-frame wavelength of each individual line is represented by a dashed green vertical line.  }
    \label{Na_doublet_lines_planet}

\end{figure}

\begin{figure}
    \centering
    \includegraphics[width=0.48\textwidth]{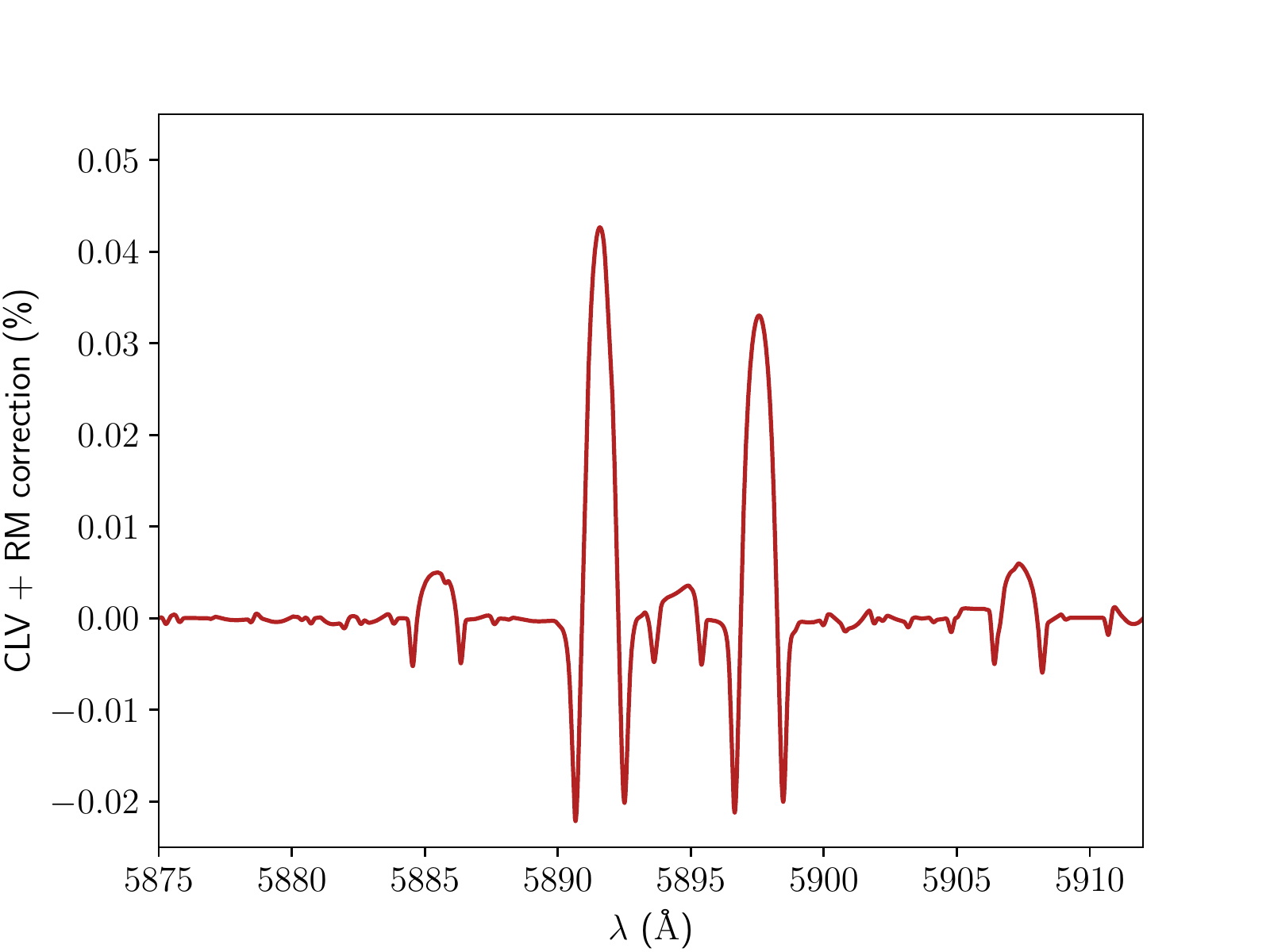}
    \caption{CLV~$+$~RM correction around the \ion{Na}{i} lines}
    \label{Na_doublet_corr}
\end{figure}

First, we tested that our procedure is extracting the signals for lines already reported in the literature, such as the \ion{Na}{i} doublet. The detection of the \ion{Na}{i} Doublet in WASP-76b has been already reported by  \citet{sei19} and \citep{edw20}. We clearly detected both lines in the 2D-map moving with the planet velocity (see Fig.~\ref{map_na_planet}). Our final spectrum of the \ion{Na}{i} doublet is shown in Fig.~\ref{Na_doublet_lines_planet}. Our detection is  9.2 and 7.5~$\sigma$ levels for T1, and 6.5 and 7~$\sigma$ for T2. These detections are more significant than those obtained by \citet{sei19} using HARPS data.  In addition, we verified the effect of the Rossiter-McLaughlin on our extraction by means of the  modelling described in \citet{cas19} using the ephemeris given by \citet{ehr20}. We found that the effect is not significant (0.04\%) given our error bars for the two transits of WASP-76~b (see Fig.~\ref{Na_doublet_corr}). 

\subsection{\ion{Ca}{ii}}

\begin{figure*}                                                                                                                                                                                           
      \centering
      \includegraphics[width=0.48\textwidth]{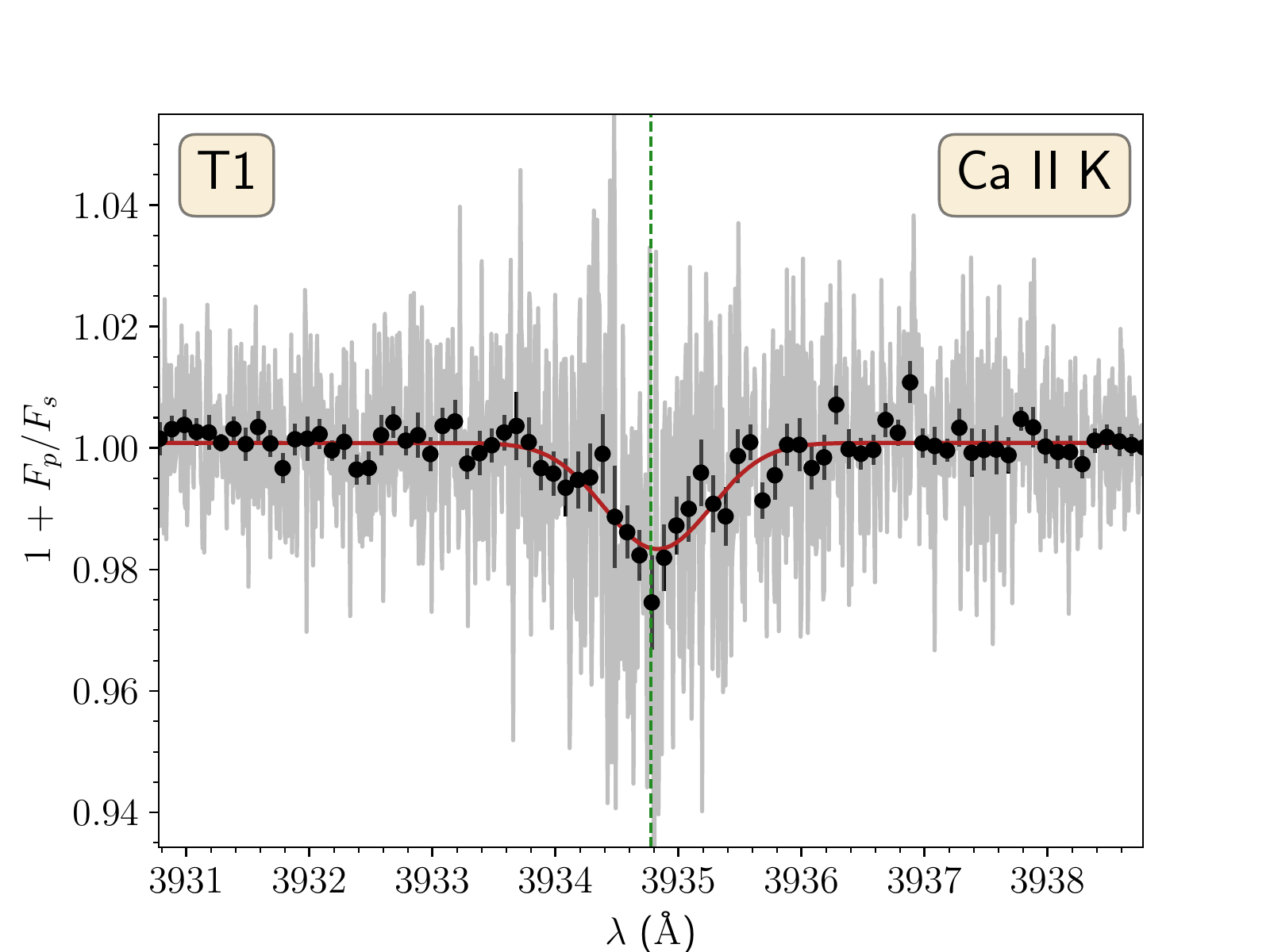}
      \includegraphics[width=0.48\textwidth]{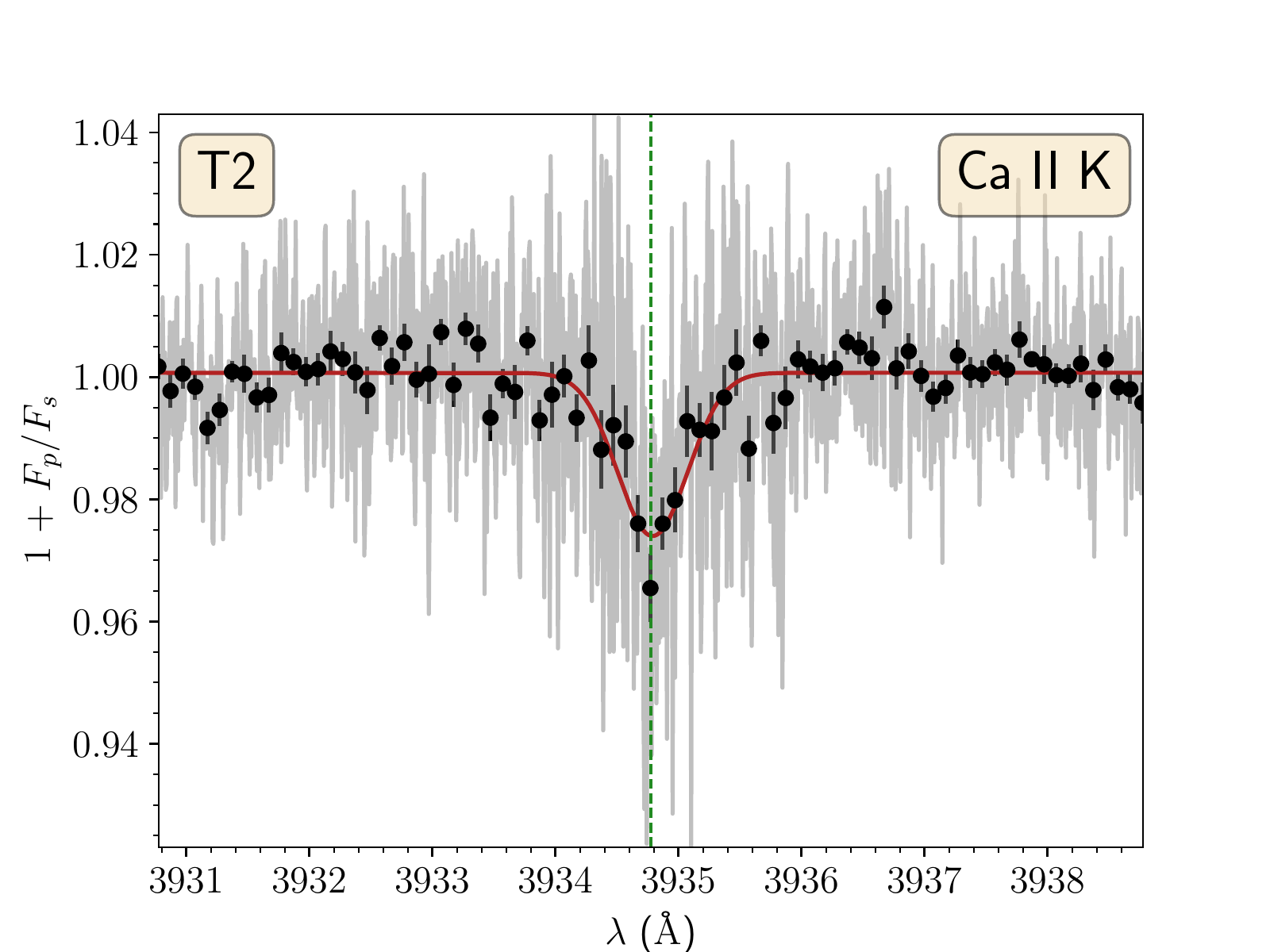}
      \includegraphics[width=0.48\textwidth]{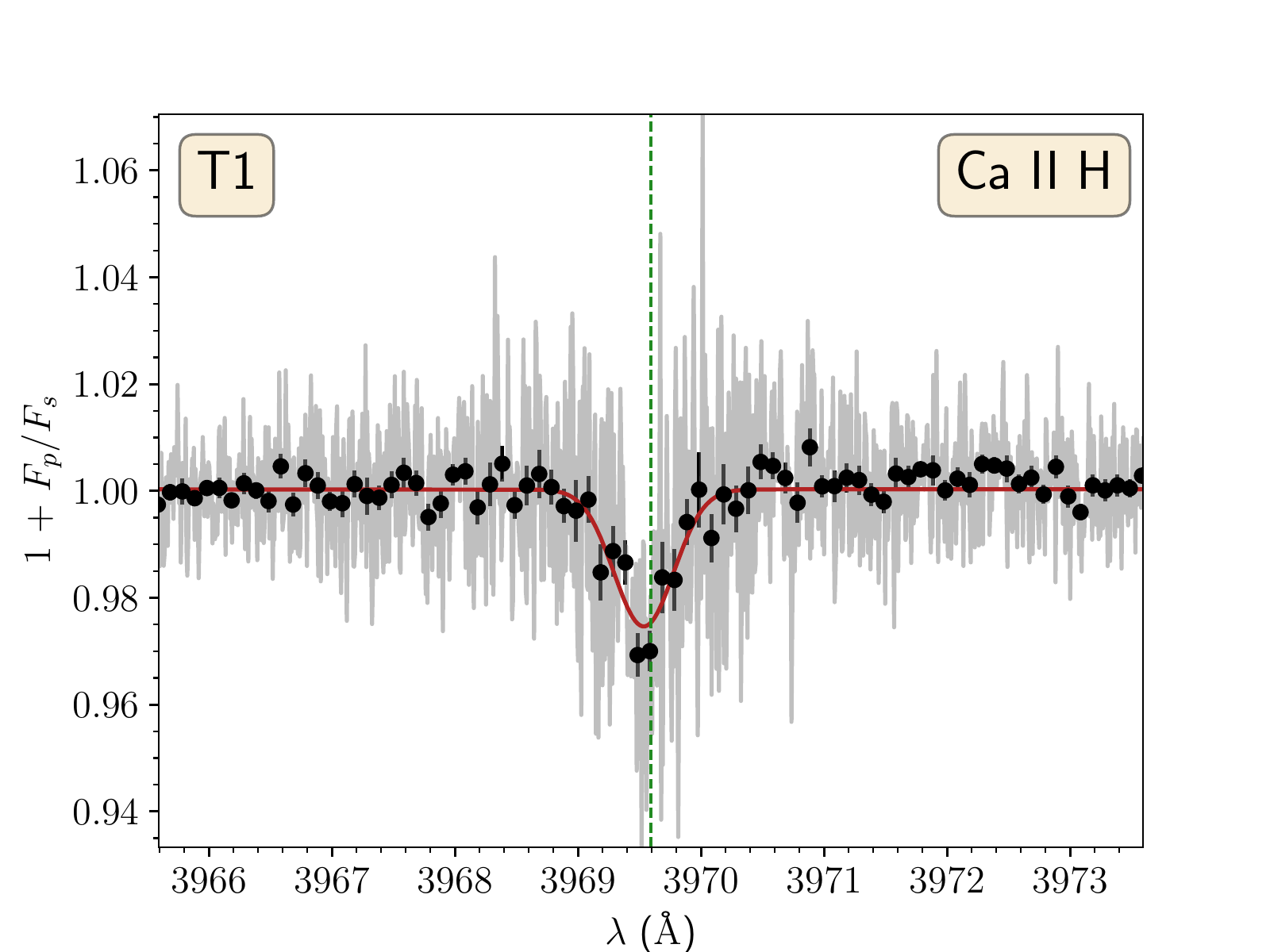}
      \includegraphics[width=0.48\textwidth]{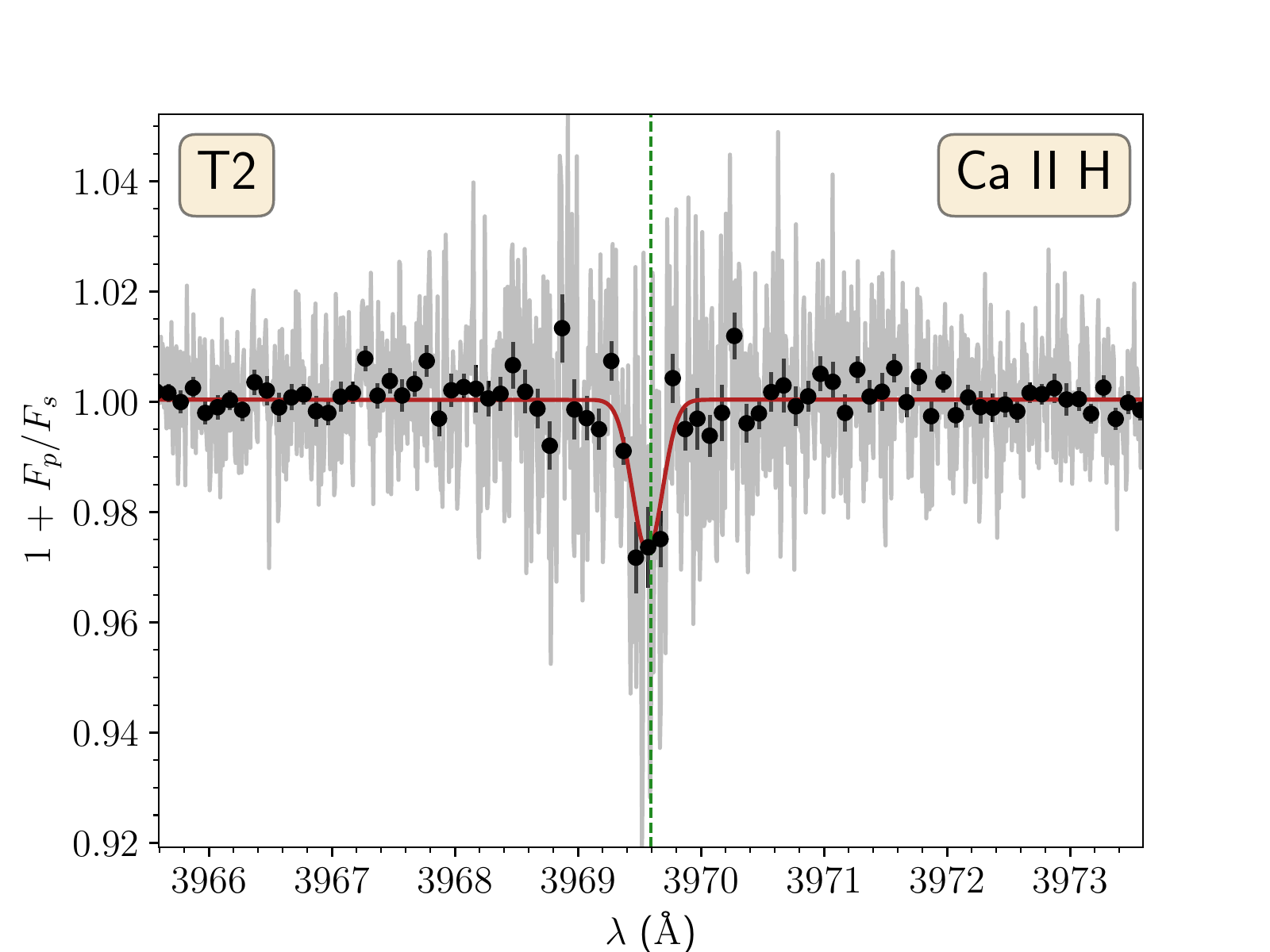}
      \caption{Same as Fig.~\ref{Na_doublet_lines_planet}  but for the \ion{Ca}{ii} H and K lines.}
      \label{Ca_HK_lines}                                                                                                                                                                       
  \end{figure*}

The \ion{Ca}{ii} H (3969.59~\AA{}) and \ion{Ca}{ii} K (3934.78~\AA{}) lines are visible in the transmission spectrum (see Fig.~\ref{Ca_HK_lines}), our Gaussian fits shows that they are significant at the level of 8.5-7.1 $\sigma$ for T1 and 5.6-8.4 $\sigma$ for T2. In all, these two lines are the most prominent features of the transmission spectrum of WASP-76, being one order of magnitude deeper than the other atomic features ($\sim$2-3\%). In fact, they might be originated by photo-ionisation in the upper part of the highly irradiated planetary exosphere \citet[see ][]{yan19}. Moreover, their intrinsic depth points towards a high formation altitude (see Table~\ref{line_intent}) as our calculated R$_{\lambda}$ is 1.78-1.57 $R_p$ (T1) and 1.82-1.80 $R_p$ (T2). 

\subsection{Balmer lines}

\begin{figure}                                                                                                                                                                                           
        \centering                                                                                                                                                                                          
        \includegraphics[width=0.48\textwidth]{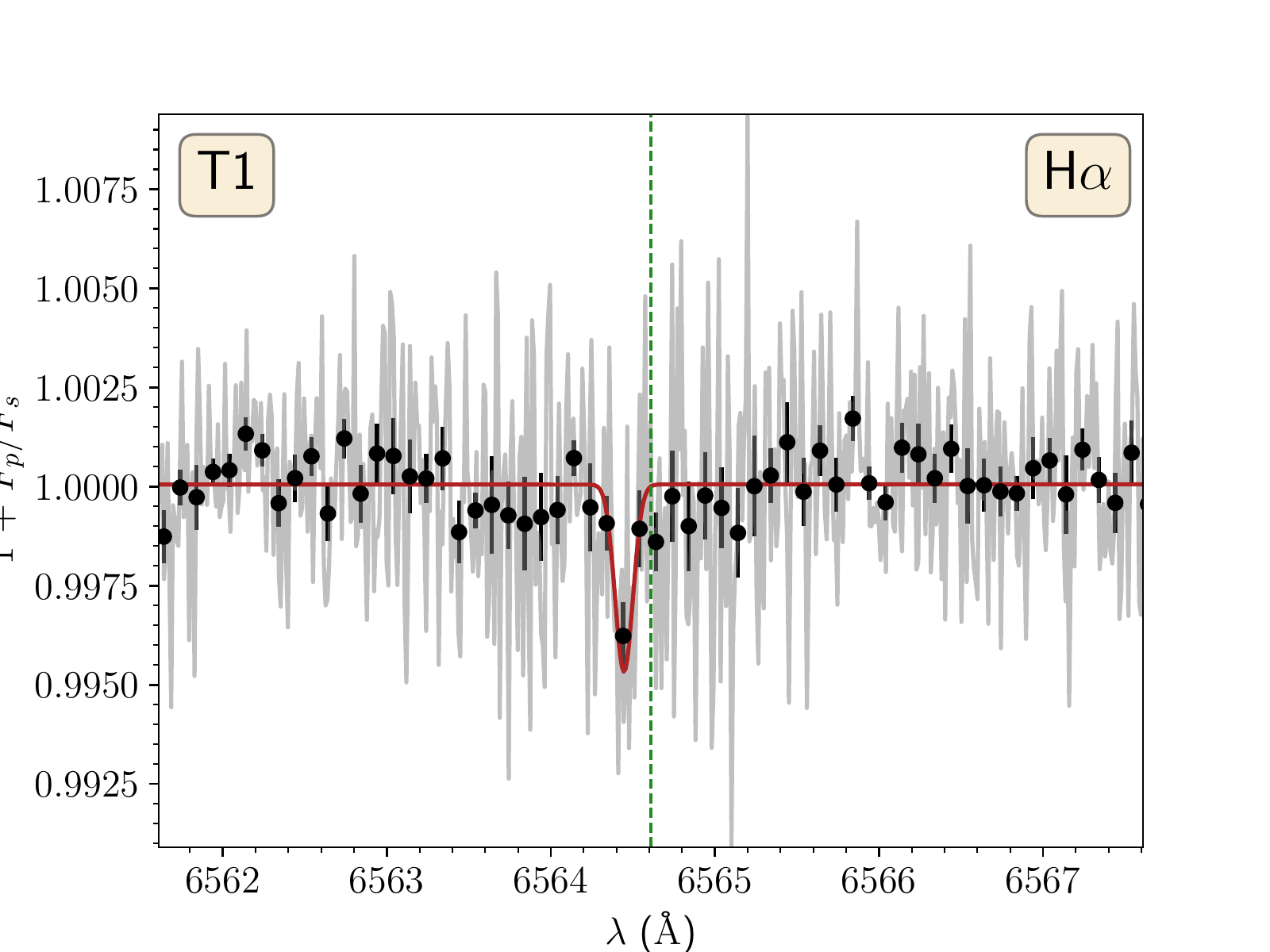}                                                                                                                                         
        \includegraphics[width=0.48\textwidth]{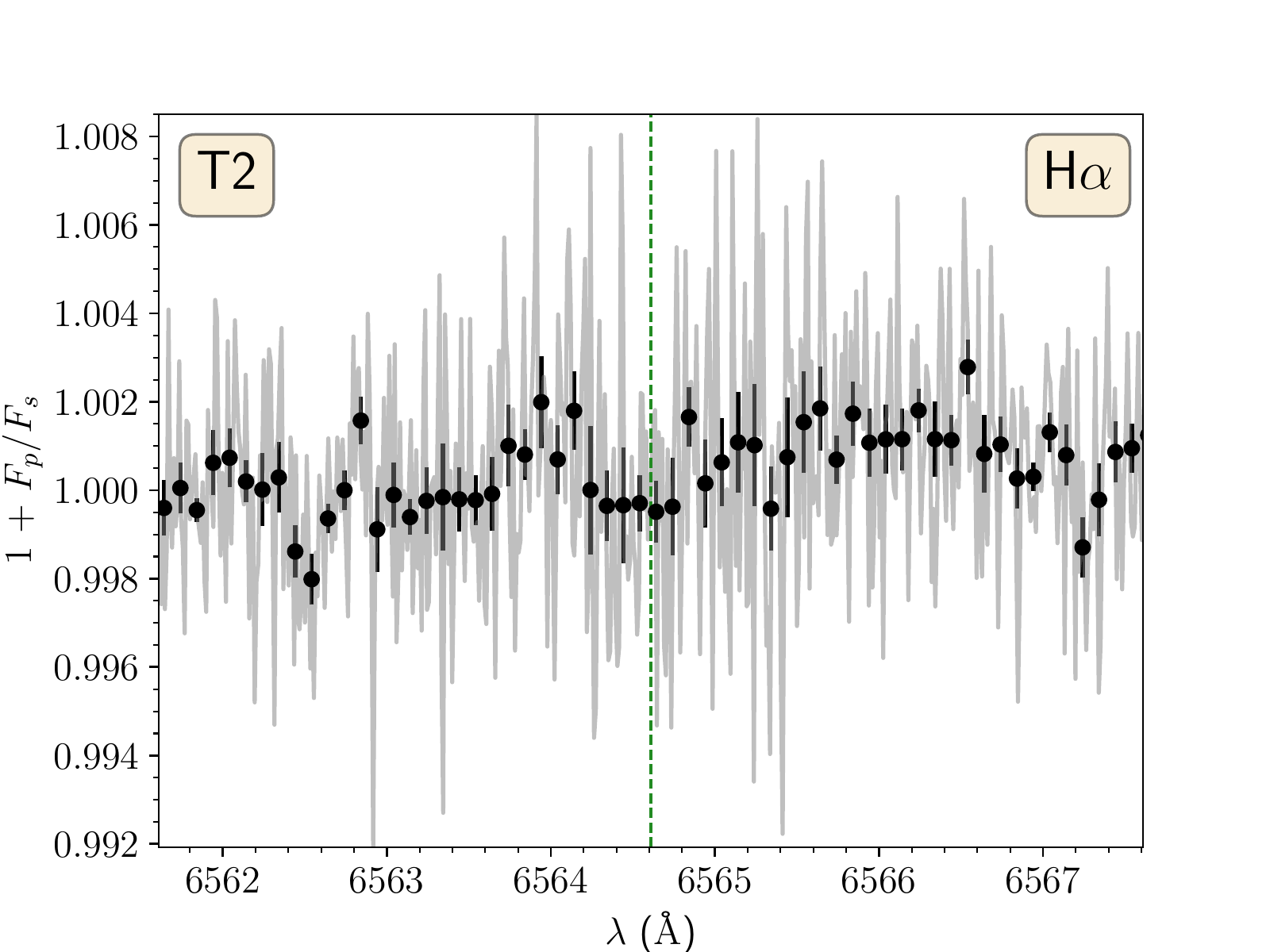}                                                                                                                                         
        \caption{Same as Fig.~\ref{Na_doublet_lines_planet}  but for the H$\alpha$ line.}                                                                                                       
        \label{ha_line_planet}                                                                                                                                                                          
    \end{figure}

The H$_{\alpha}$ line at 6564.61~\AA{} is present in T1, whereas T2 shows a little bump, rather than a proper line (see Fig.~\ref{ha_line_planet}). Thus, it is not possible to confirm the presence of atomic hydrogen in the atmosphere using only T2 and for T1 we found a depth at  0.48~$\pm$~0.12\%. We have also explored other Balmer lines (H${\beta}$, H${\gamma}$), but none of them are visible in the planetary transmission spectrum. The T2 data have a higher quality than T1, which in turn implies that if H$\alpha$ absorption were present in T2 with the same intensity as in T1, we should have detected it.  Interestingly enough, no study to the present date has provided an upper limit to the presence of H$\alpha$ in the spectrum of WASP-76b. \citet{ess20} explored the transmission of spectrum of WASP-76b using HST optical spectra and did not find any evidence of strong absorption due to H$\alpha$. In contrast, our findings at higher resolution might be evidence for H$\alpha$ variability from the planetary atmosphere given that the line is not seen in T2. However, the available data are not sufficient to confirm this scenario.

\subsection{\ion{Li}{i}}

WASP-76b is expected to have Li in its atmosphere because substellar objects with masses below $\approx$55 M$_{\rm Jup}$ do not deplete this element at any moment during their evolution \citep{cha00,bar15}. In consequence, WASP-76b should maintain in its atmosphere the amount of lithium that was present in the protoplanetary disc from which it was supposedly formed. Actually, the presence of Li was modelled by \citet{chen18} for another giant planet. The \ion{Li}{i} absorption feature at 6709.61~\AA{}~is detected in the ESPRESSO spectra of the planet WASP-76b. We found that \ion{Li}{i} is moving with the planet velocity in T2, whereas we find a small trace during T1 (see Fig.~\ref{map_Li}). The \ion{Li}{i} line is more significant for T2 than T1 (5.7$\sigma$ vs 4.4$\sigma$) and it can be easily explained by the cadence of the observations of T2 in contrast with T1. To the best of our knowledge, this is the first reported detection of Lithium using high-resolution spectroscopy \citep[in][ there is also a report on the detection of lithium in another giant planet using ESPRESSO data]{bor20}.

 \begin{figure}
        \centering
        \includegraphics[width=0.48\textwidth]{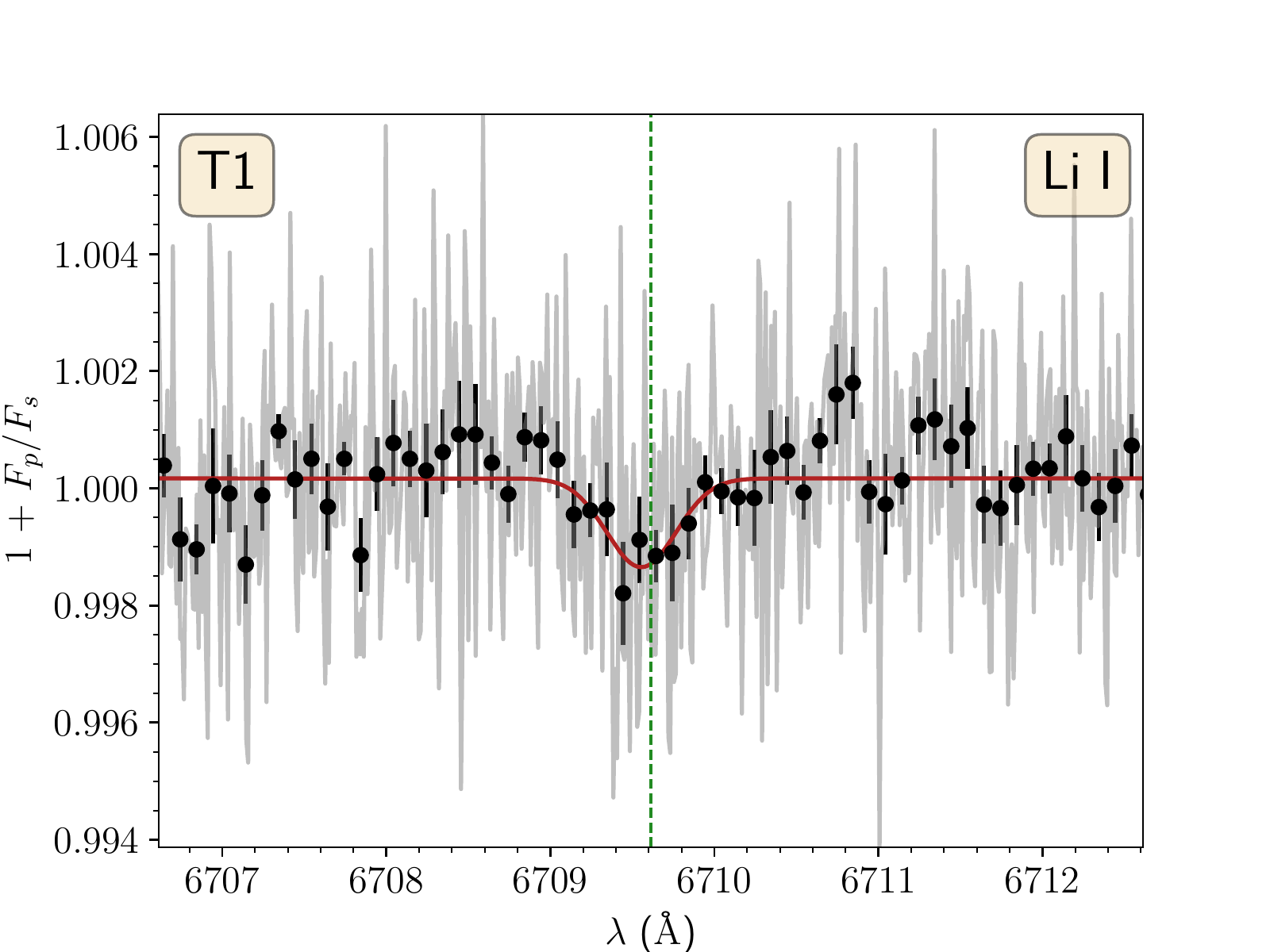}
        \includegraphics[width=0.48\textwidth]{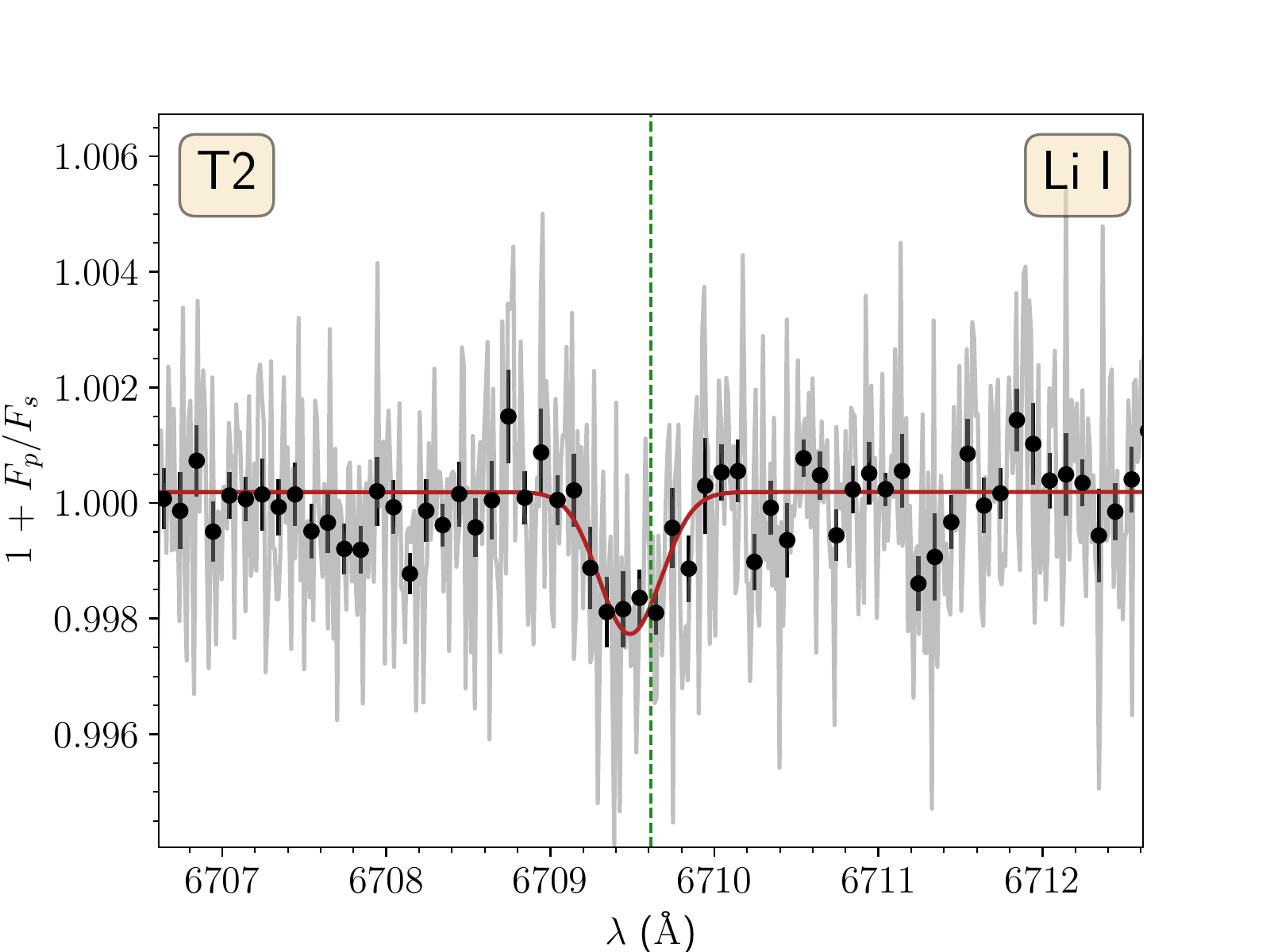}
        \caption{Same as Fig.~\ref{Na_doublet_lines_planet}  but for the \ion{Li}{i} line.}
        \label{li_line_planet}
\end{figure}

\subsection{\ion{Fe}{i}}

Interestingly enough, the accuracy of the ESPRESSO data has  allowed us to detect actual \ion{Fe}{i} lines in the transmission spectra of WASP-76~b. In particular, our two transit spectra show a few tentative neutral Iron lines in the range of $\approx$~4377-4430~\AA{}. The following lines are seen in the spectrum of WASP-76b: 4377.16~\AA{}, 4384.78~\AA{}, 4406.26~\AA{}, 4416.17~\AA{}, and 4428.55~\AA{}. To increase the S/N of the profile of the Fe lines, we merged all of them into a single line in velocity space (see Fig.~\ref{weak_lines_planet_1}). The combination of these lines is stronger for T2  (6.7$\sigma$) rather than T1 (4.7~$\sigma$). The detection of these lines is in agreement with the results of the CCF against a \ion{Fe}{i} binary mask (see Fig.~\ref{ccf_lines_planet}). In this work, we found a depth of 255~$\pm$~20~ppm (T1) and 182~$\pm$~12~ppm (T2). These two values correspond to a central velocity of -8.27~$\pm$~0.25~km~s$^{-1}$ and -8.76~$\pm$~0.56~km~s$^{-1}$ respectively and to a detection at 12.8~$\sigma$  and the 18.2~$\sigma$ levels for T1 and T2, respectively (see Table~\ref{ccf_intent}). In all, our results reinforce those presented  by \citet{ehr20} for WASP-76b.

\subsection{\ion{Mg}{i}, \ion{K}{i}, and \ion{Mn}{i}}

\begin{figure*}
      \centering
      \includegraphics[width=0.48\textwidth]{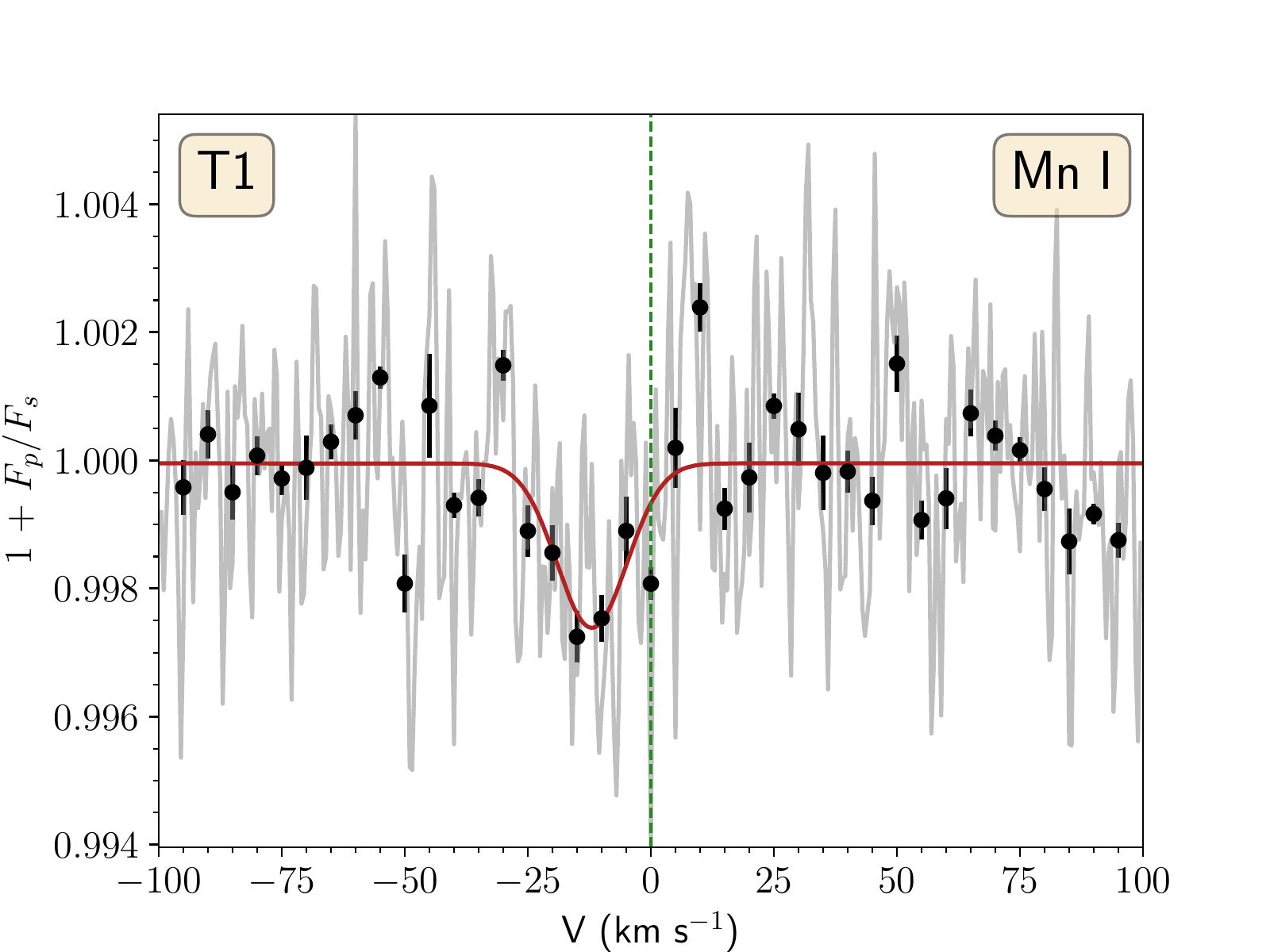}
      \includegraphics[width=0.48\textwidth]{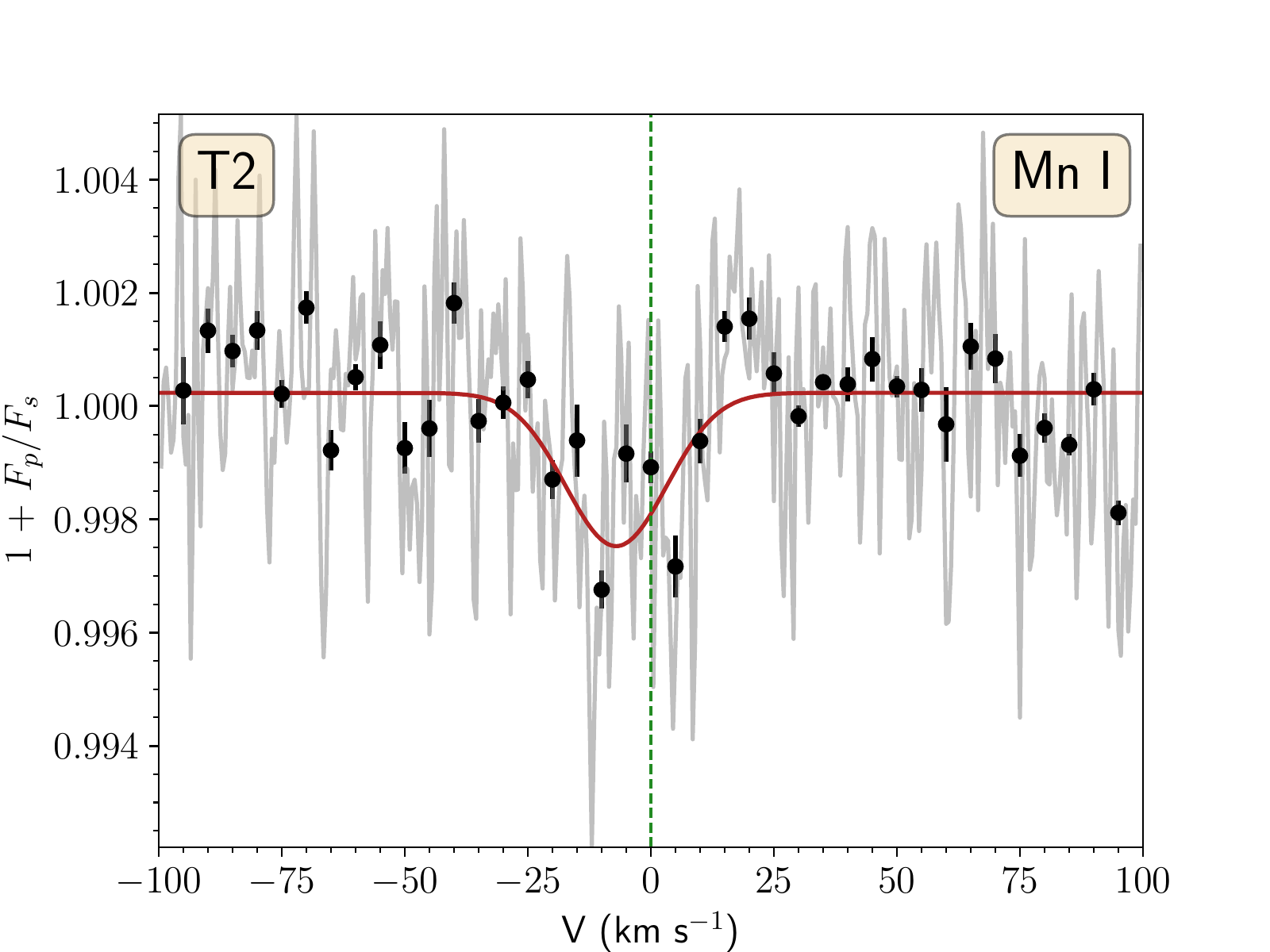}
      \includegraphics[width=0.48\textwidth]{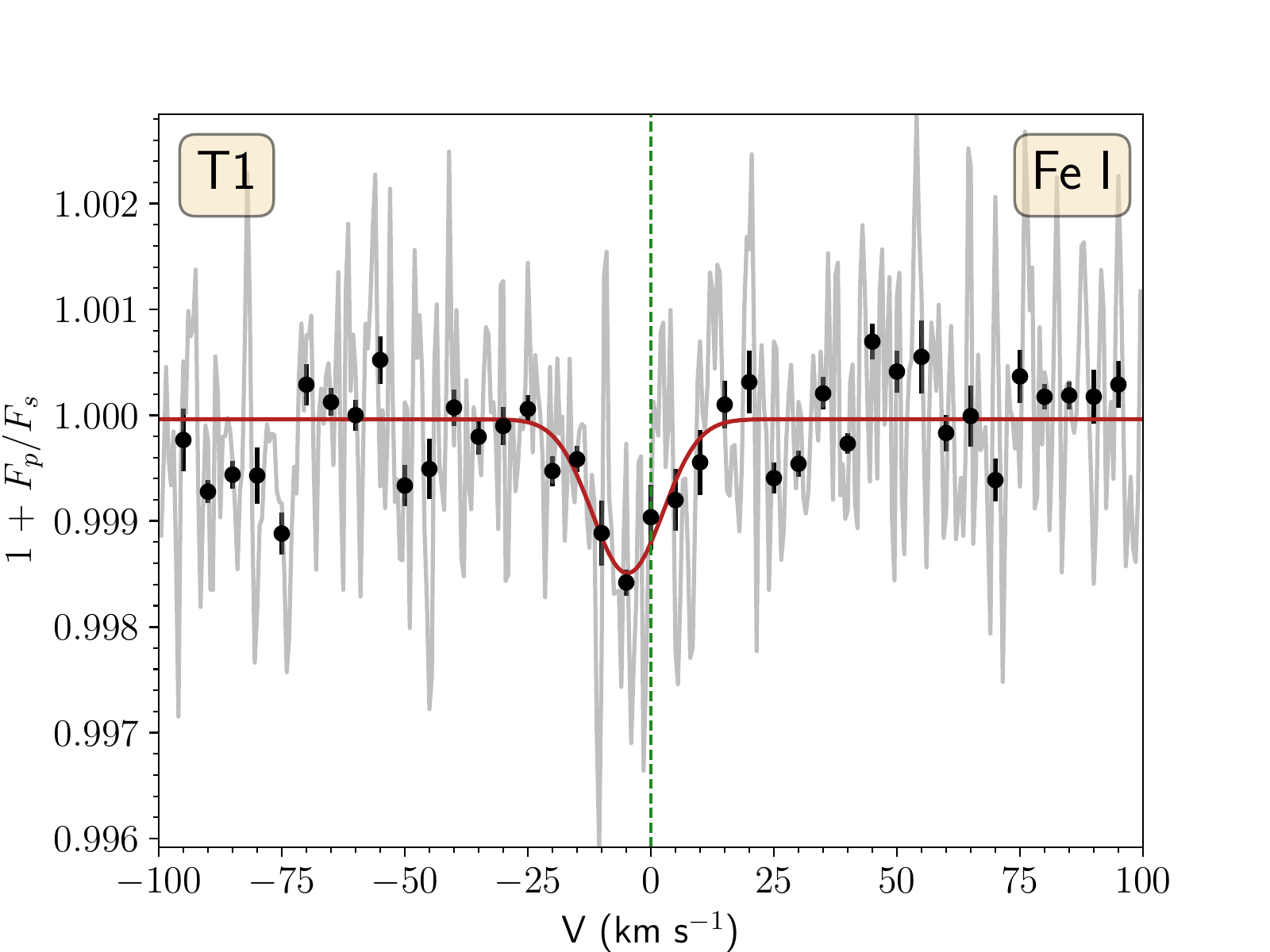}                                                                                                       
      \includegraphics[width=0.48\textwidth]{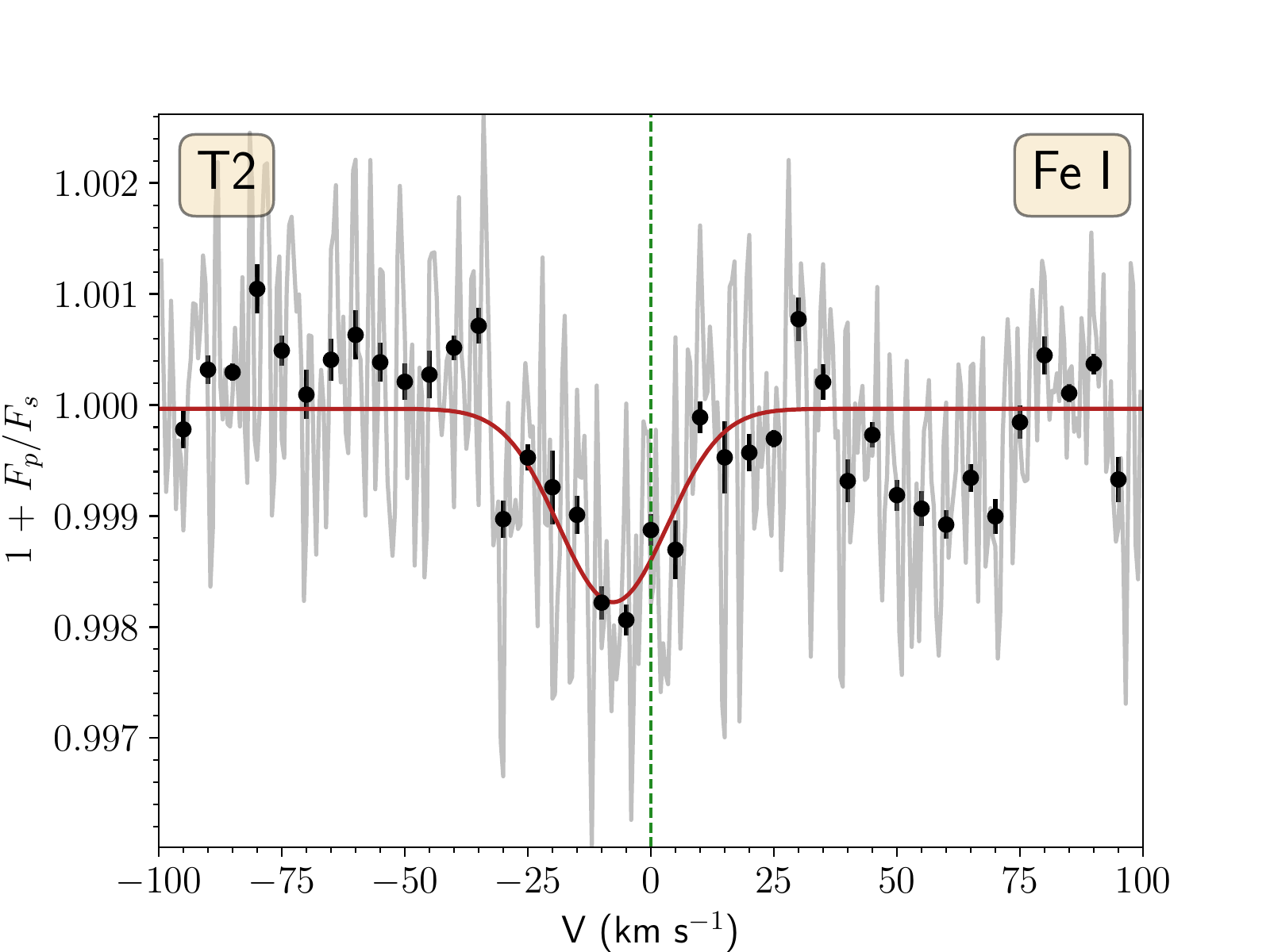}                                                                                                       
      \includegraphics[width=0.48\textwidth]{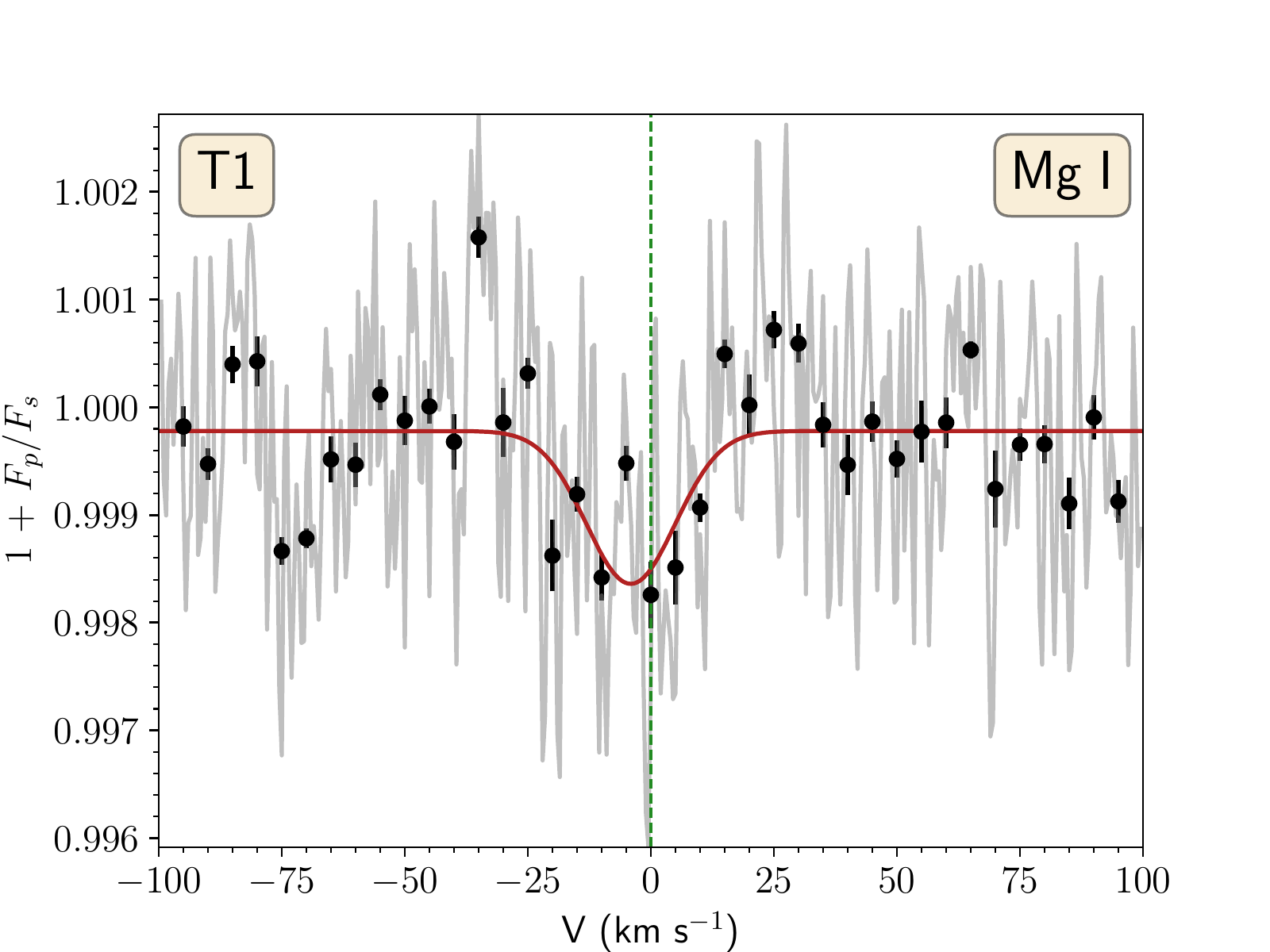}
      \includegraphics[width=0.48\textwidth]{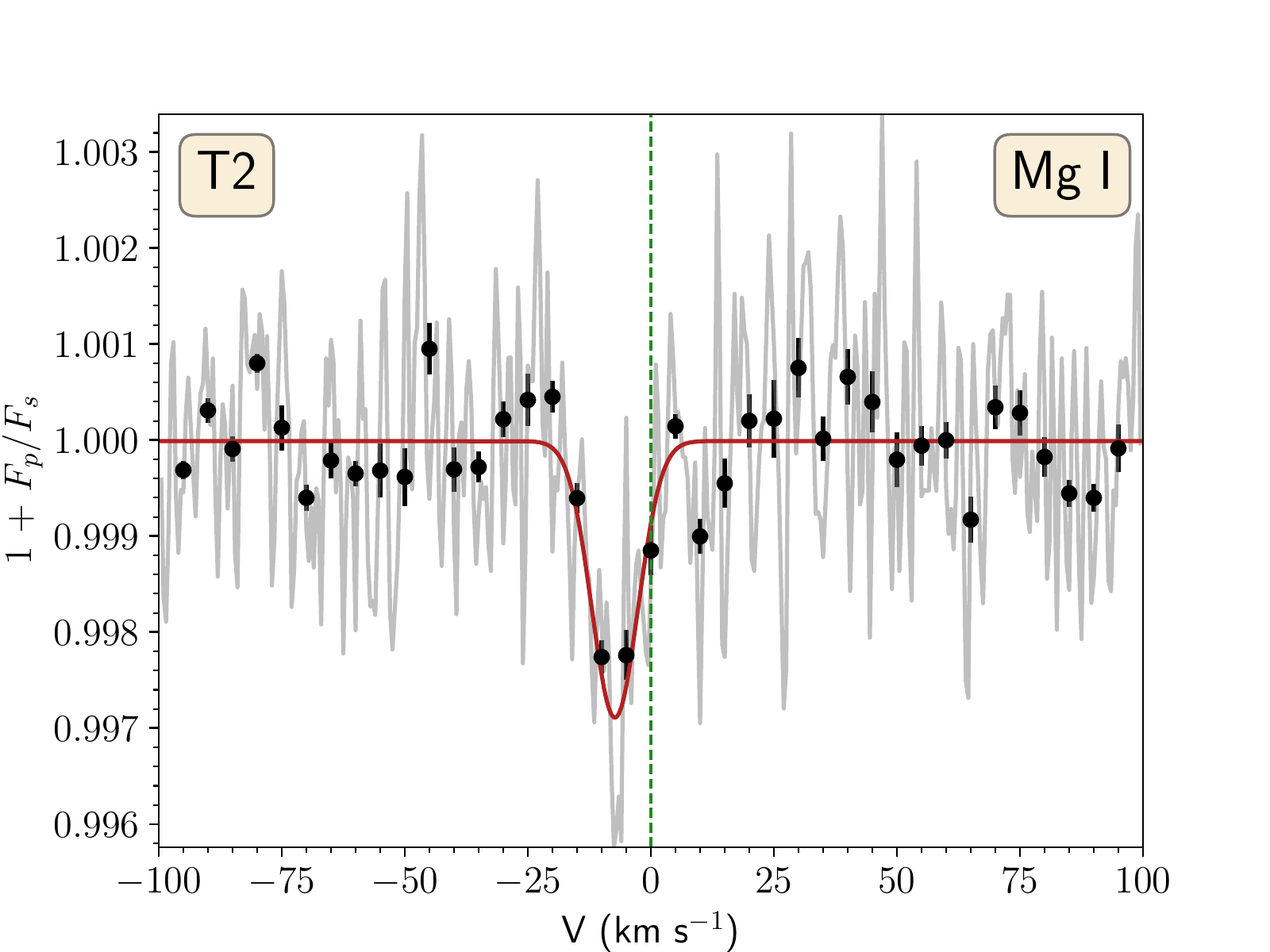}
\caption{ Same as Fig.~\ref{Na_doublet_lines_planet} but for other elements. Here we represent the combination of several lines in velocity space to strength the signal. Here black points represent a binning of 10~km~s$^{-1}$ and the green dashed line represents the 0~km~s$^{-1}$ mark.}
      \label{weak_lines_planet_1}
  \end{figure*}

The \ion{Mg}{i} line at 4572.38~\AA~seems to be in the transmission spectrum with a significance of $\sim$~2.8~$\sigma$ for T1 and 3.7~$\sigma$ for T2. In addition to this \ion{Mg}{i}, we found that the space-velocity combined Magnesium triplet lines (5168.76,5174.13, and 5185.05~\AA{}) are clearly seen at the 7.5~$\sigma$ level for T2 (see Fig.~\ref{weak_lines_planet_1}). Regarding \ion{K}{i} the transmission maps show a small imprint leaving a partial trace that we can barely see in our maps (see Fig.~\ref{map_K}). However, the line is present at the 5.7~$\sigma$ level in T2. We also report the detection of the \ion{Mn}{i} triplet lines at 4031.89,4034.20, and 4035.62~\AA{}. Again, we combined these three lines in velocity space and we found that their significance is in the range of 4.7-5.0~$\sigma$.

\subsection{\ion{Ti}{i}, \ion{Cr}{i}, and \ion{Ni}{i}}

\begin{figure*}
    \centering
     \includegraphics[width=0.38\textwidth]{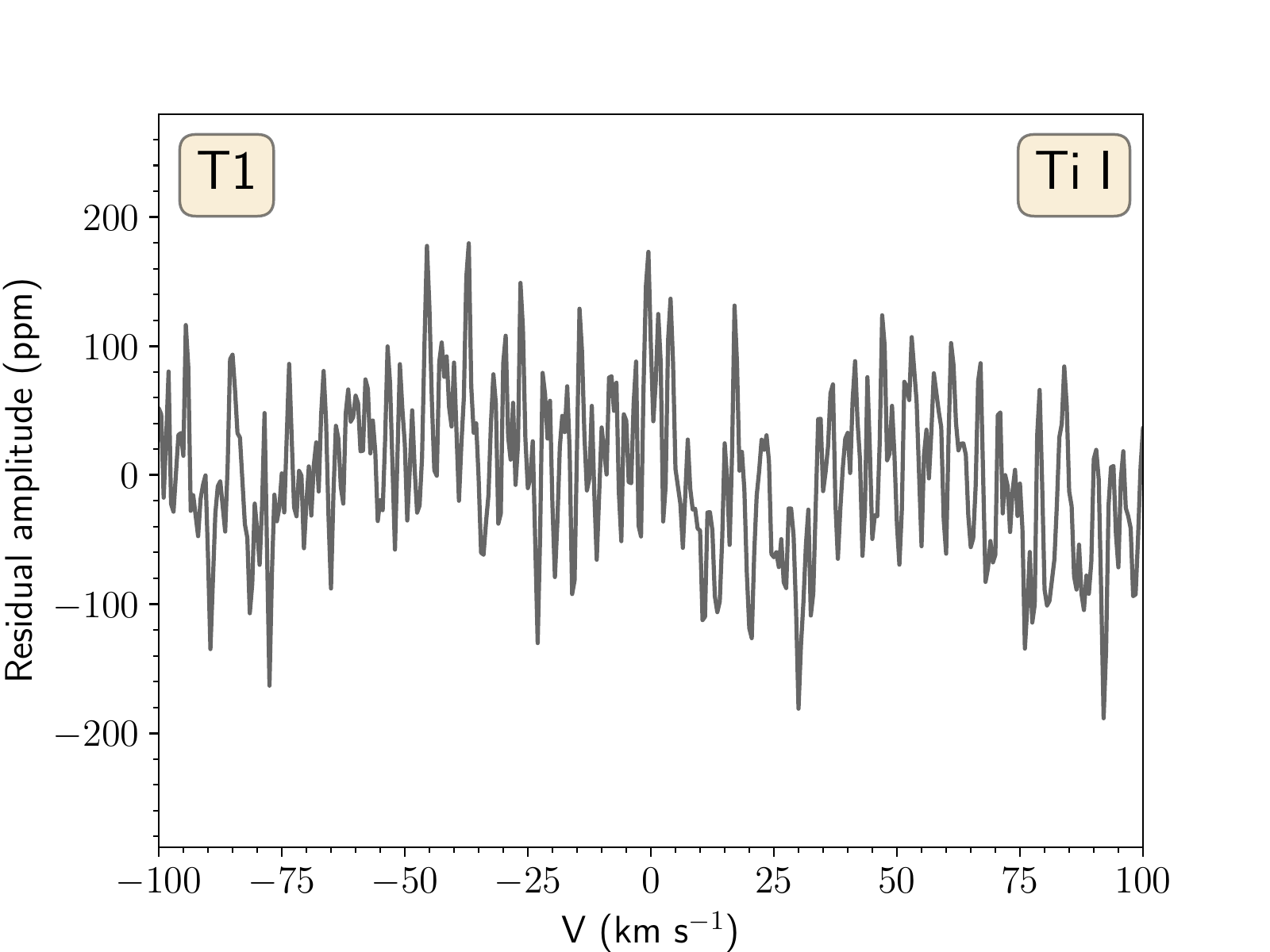}                                                                                                                                                 
      \includegraphics[width=0.38\textwidth]{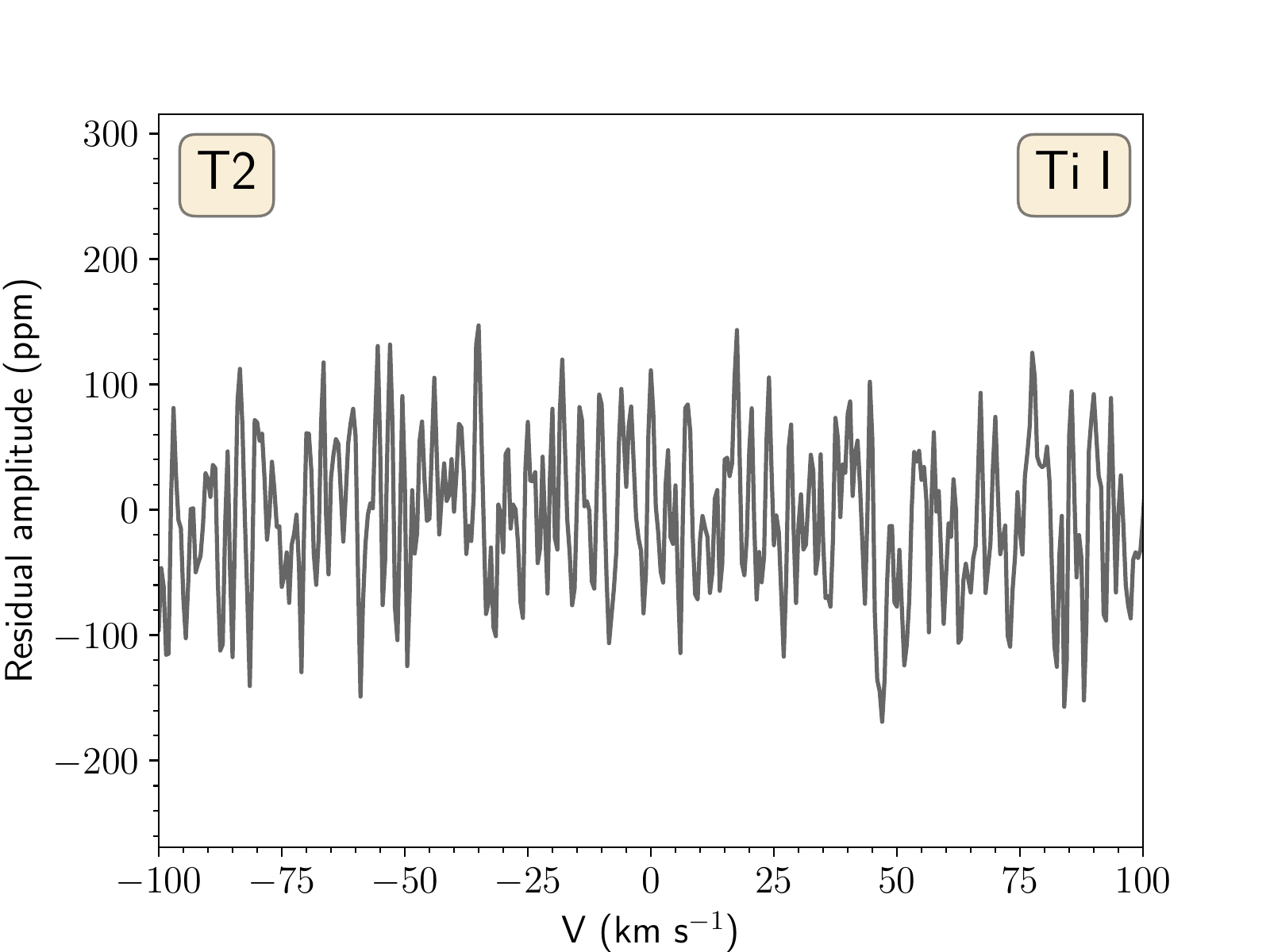}
           \includegraphics[width=0.38\textwidth]{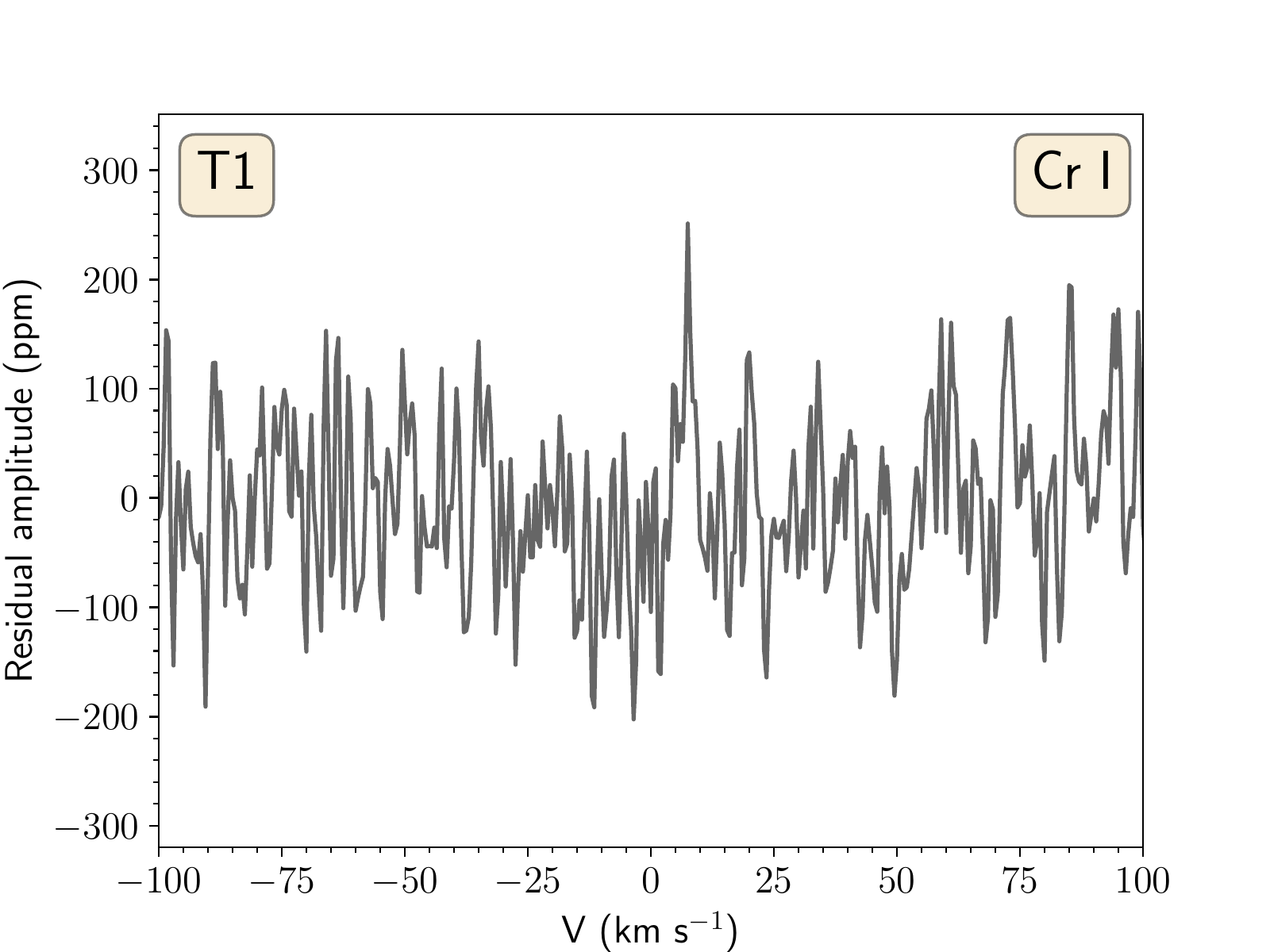}                                                                                                                                                 
        \includegraphics[width=0.38\textwidth]{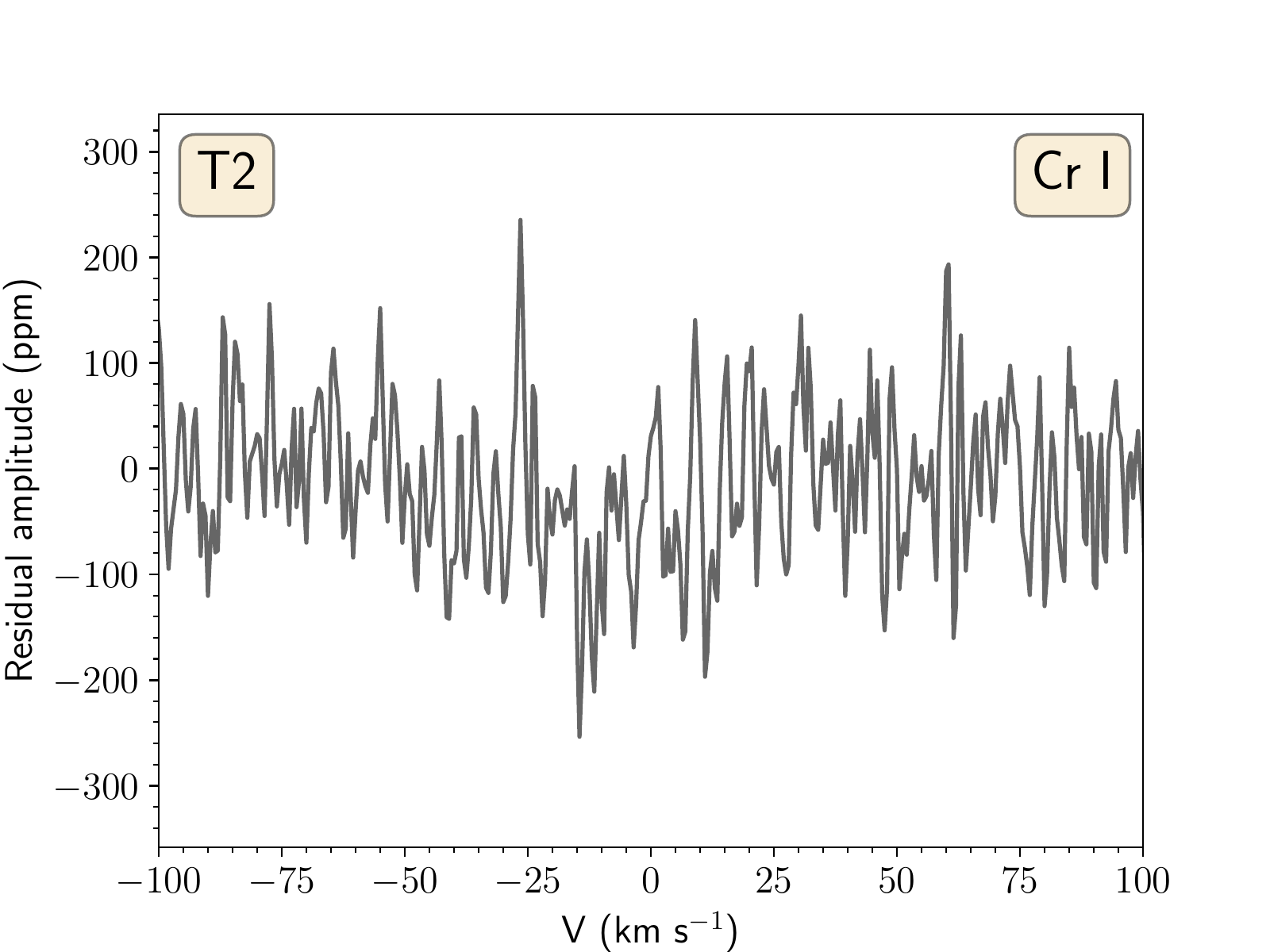}  
     \includegraphics[width=0.38\textwidth]{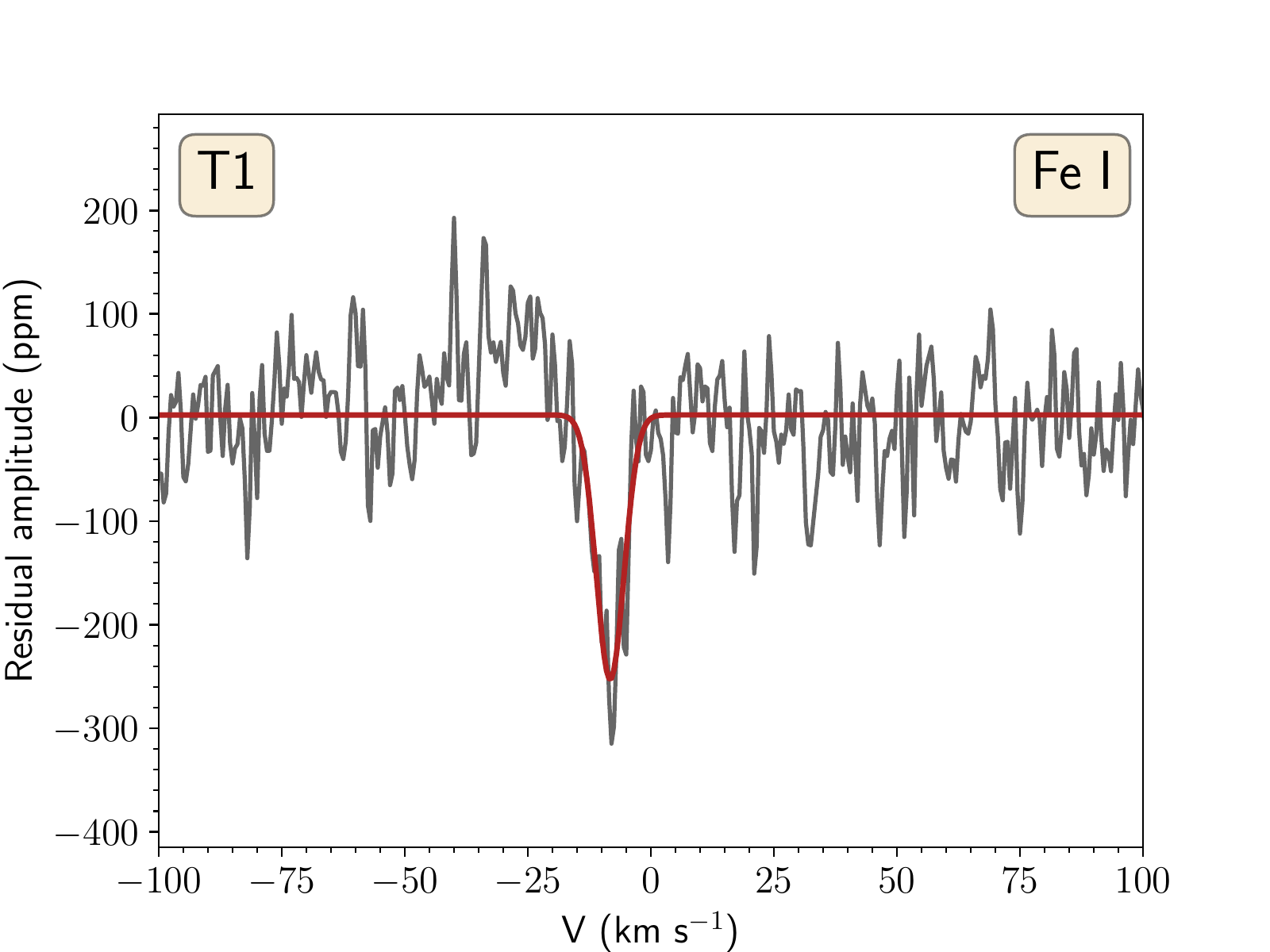}
      \includegraphics[width=0.38\textwidth]{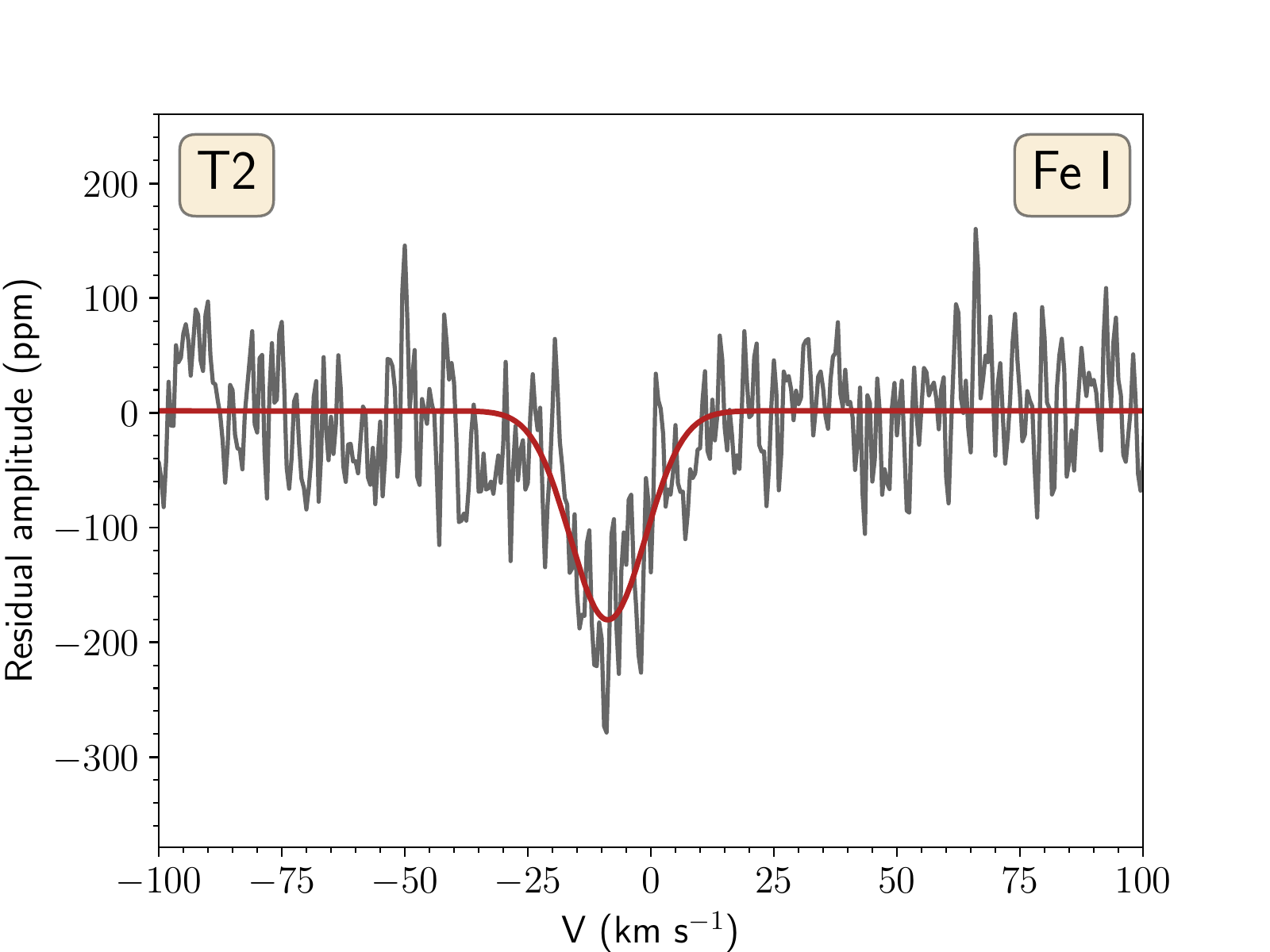}
       \includegraphics[width=0.38\textwidth]{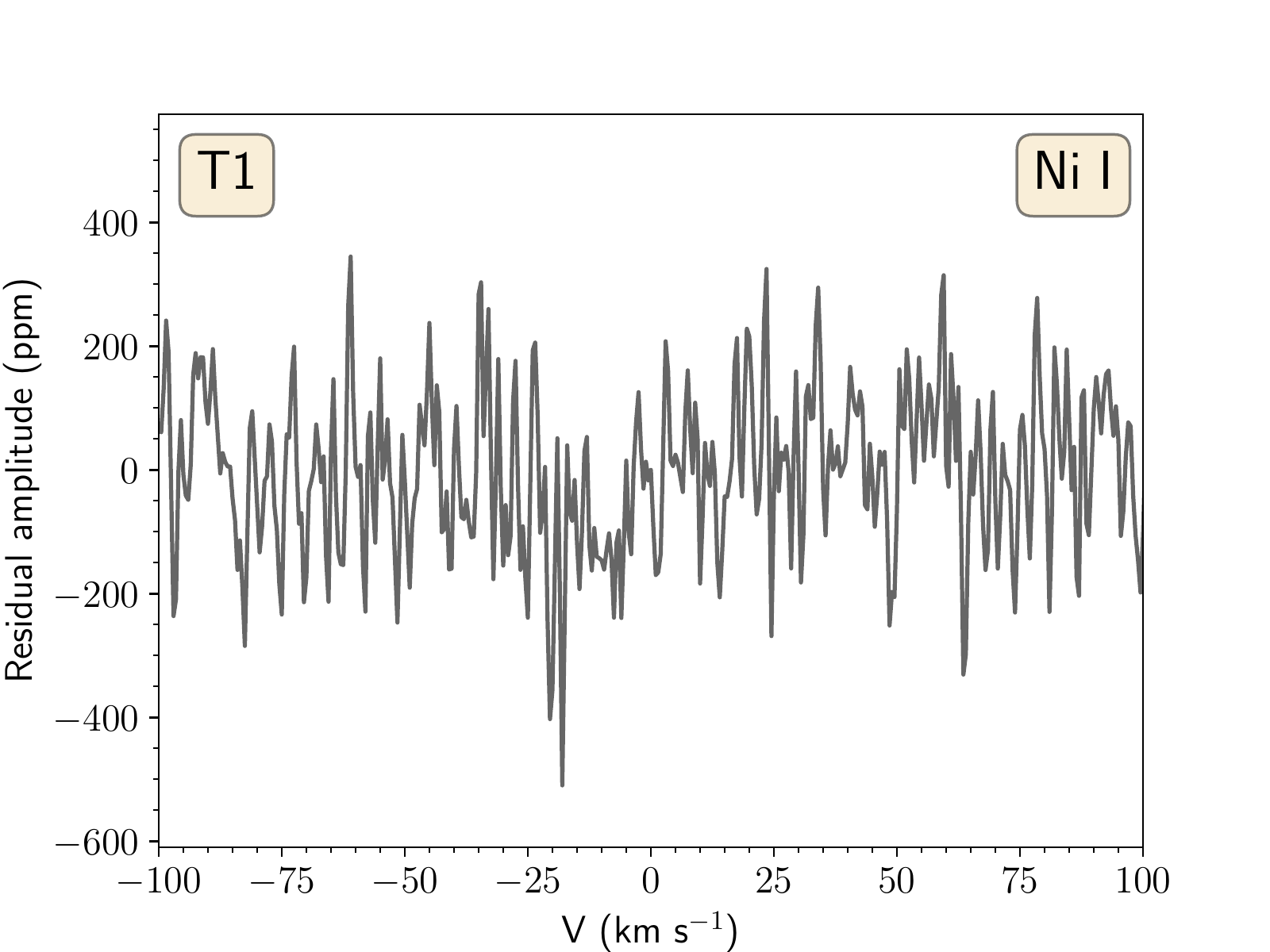}                                                                                                                                            
       \includegraphics[width=0.38\textwidth]{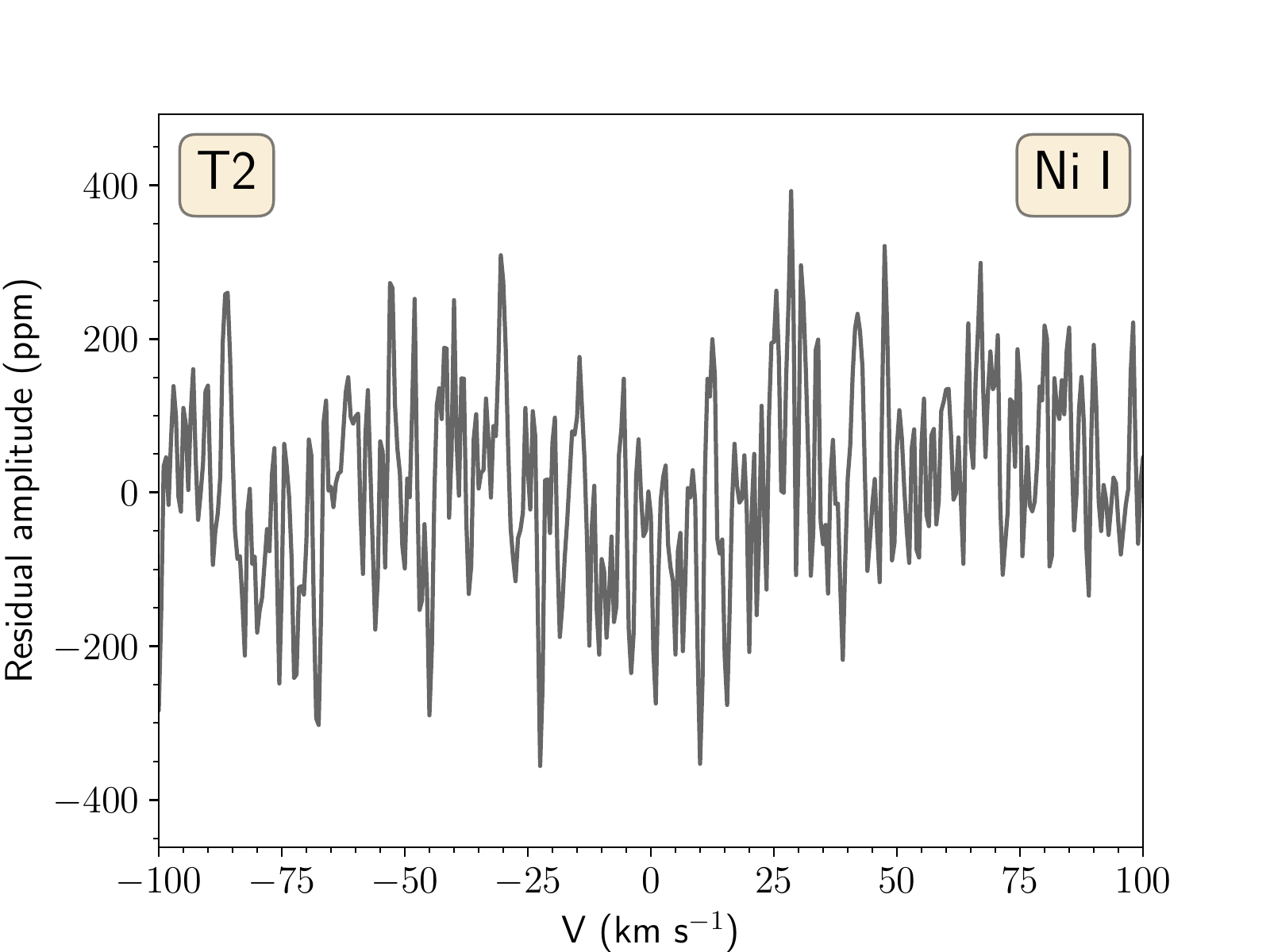} 
    \caption{ Cross-correlation function against our line masks for \ion{Ti}{i}, \ion{Cr}{i}, \ion{Fe}{i}, and \ion{Ni}{i}. The black line represents CCF, whereas the red line represents the best fit to the \ion{Fe}{i} CCF.}
    \label{ccf_lines_planet}
\end{figure*}

We also explored the presence of  \ion{Ti}{i}, \ion{Cr}{i}, and \ion{Ni}{i} in our data by means of a cross-correlation method. For the most part, our search was unsuccessful and we could only calculate upper limits to their presence of $\sim$~60~ppm (\ion{Ti}{i}),  $\sim$77~ppm (\ion{Cr}{i}), and $\sim$130~ppm (\ion{Ni}{i}). Finally, the data regarding these elements and their respective CCFs can be found in Table~\ref{ccf_intent} and in Fig.~\ref{ccf_lines_planet}

\subsection{Diatomic molecules: TiO, VO, and ZrO}

These three diatomic molecules  are important absorbers in cool stellar atmospheres with a non-negligible opacity source as already shown by \citet{vaneckgrid}. The CCFs calculated in this work do not show any trace of them in our data as shown in Fig.~\ref{ccf_lines_planet_mol}. However, we can place a conservative upper limit to their presence of $\leq$~10~ppm (see Table~\ref{ccf_intent}).  Interestingly, the TiO atomic data has been recently improved by the exomol team \citep[see, ][]{tioexomol}. In consequence, we generated a new TiO model spectrum using the exomol line list \citep{tioexomol} and performed a CCF against it. We found an upper limit to presence of TiO of 6~ppm which in turn is in agreement to what we obtained with the \citet{tioplez} line list (see Table~\ref{ccf_intent}). In all, these state-of-the-art line lists might be not accurate enough to confidently retrieve the presence of TiO in the atmosphere of WASP-76b. Despite this, we can think of two scenarios in this particular  planet: These features are really weak  or they are hidden due to other effects. However, clouds might be responsible for weakening of the molecular bands an in the planetary spectrum \citep{char15}. In addition, another explanation might stem from the transport of molecules to the night side of the planet \citep{nug17}. In all, TiO and VO should dominate the spectrum in the optical wavelength range \citep{vaneckgrid}, given the equilibrium temperature of the planet. Other studies have explored the atmosphere of WASP-76b using HST data \citep{edw20,ess20} and modelled the atmosphere of the planet. In fact, the transmission spectrum and the modelling of \citet{ess20} did not present any evidence for TiO absorption in the optical wavelength range. This is consistent with our non-detections using ESPRESSO data. Interestingly, the nearly solar C/O calculated here for WASP-76 does not suggest that the atmosphere of the WASP-76b is driven by a carbon-rich chemistry. However, clouds might be responsible for weakening of the molecular bands an in the planetary spectrum \citep{char15}. In addition, another explanation might stem from the transport of molecules to the night side of the planet \citep{nug17}. \\

\begin{figure*}
    \centering
    \includegraphics[width=0.48\textwidth]{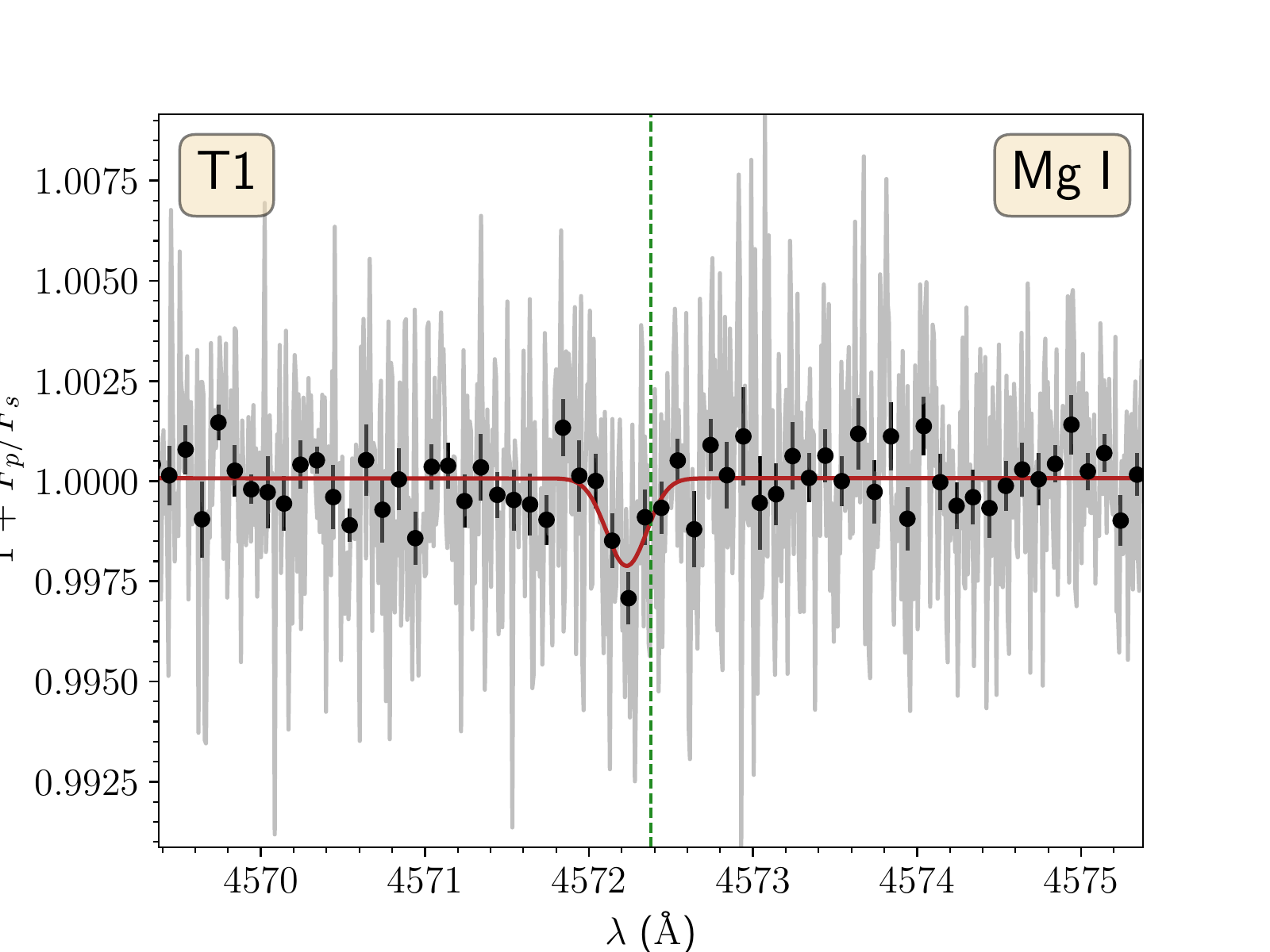}
    \includegraphics[width=0.48\textwidth]{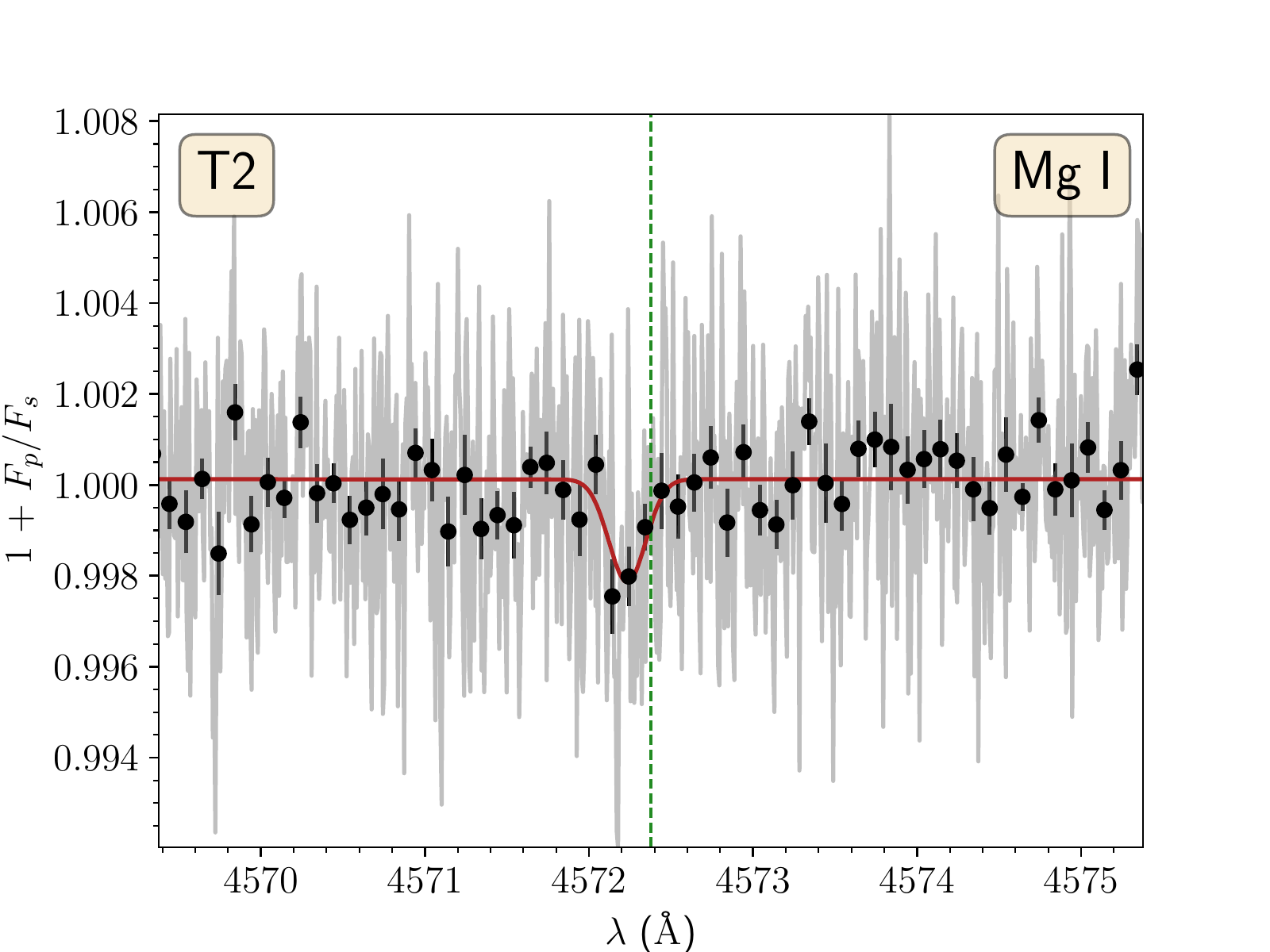}
    \includegraphics[width=0.48\textwidth]{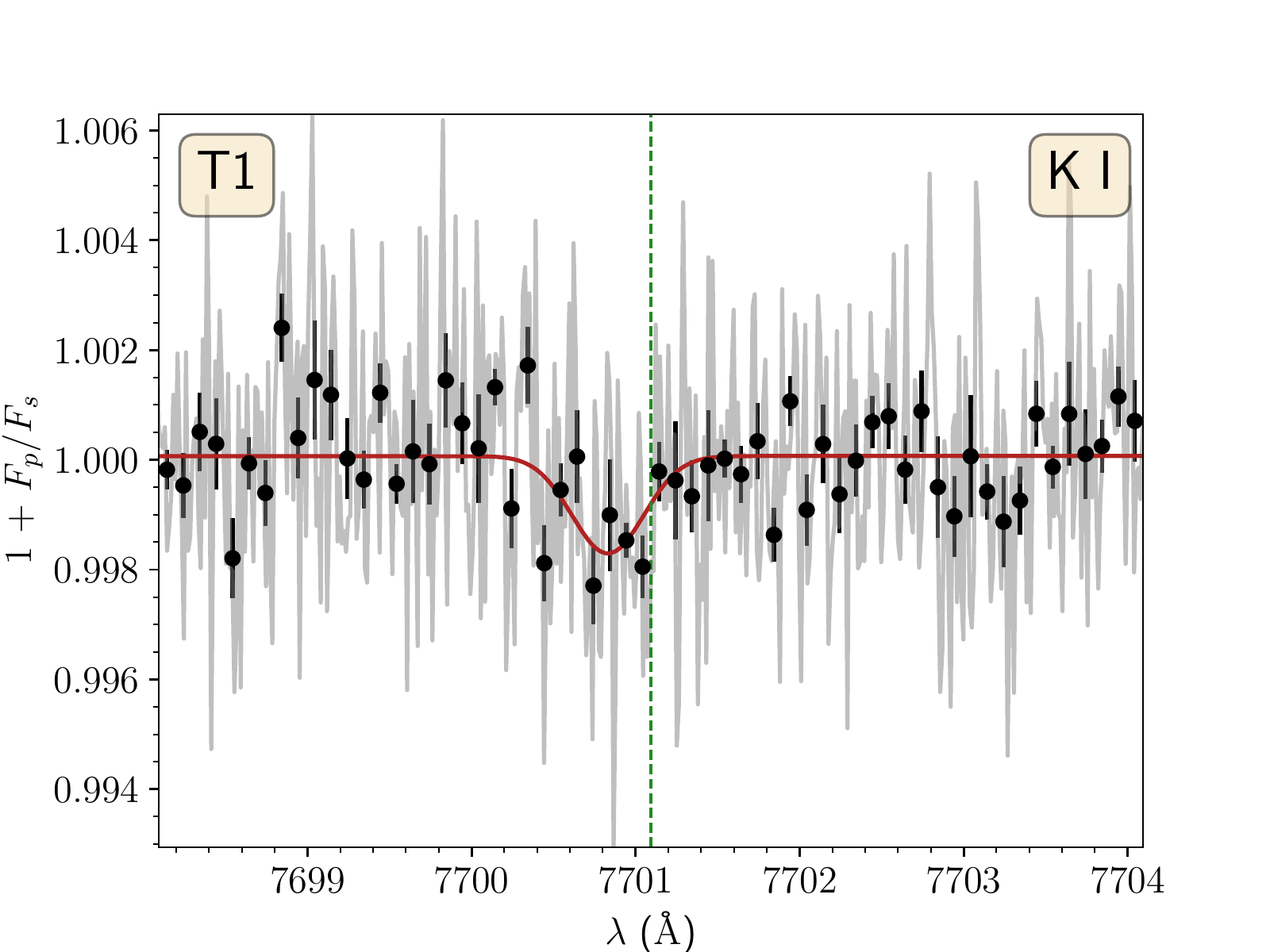}
    \includegraphics[width=0.48\textwidth]{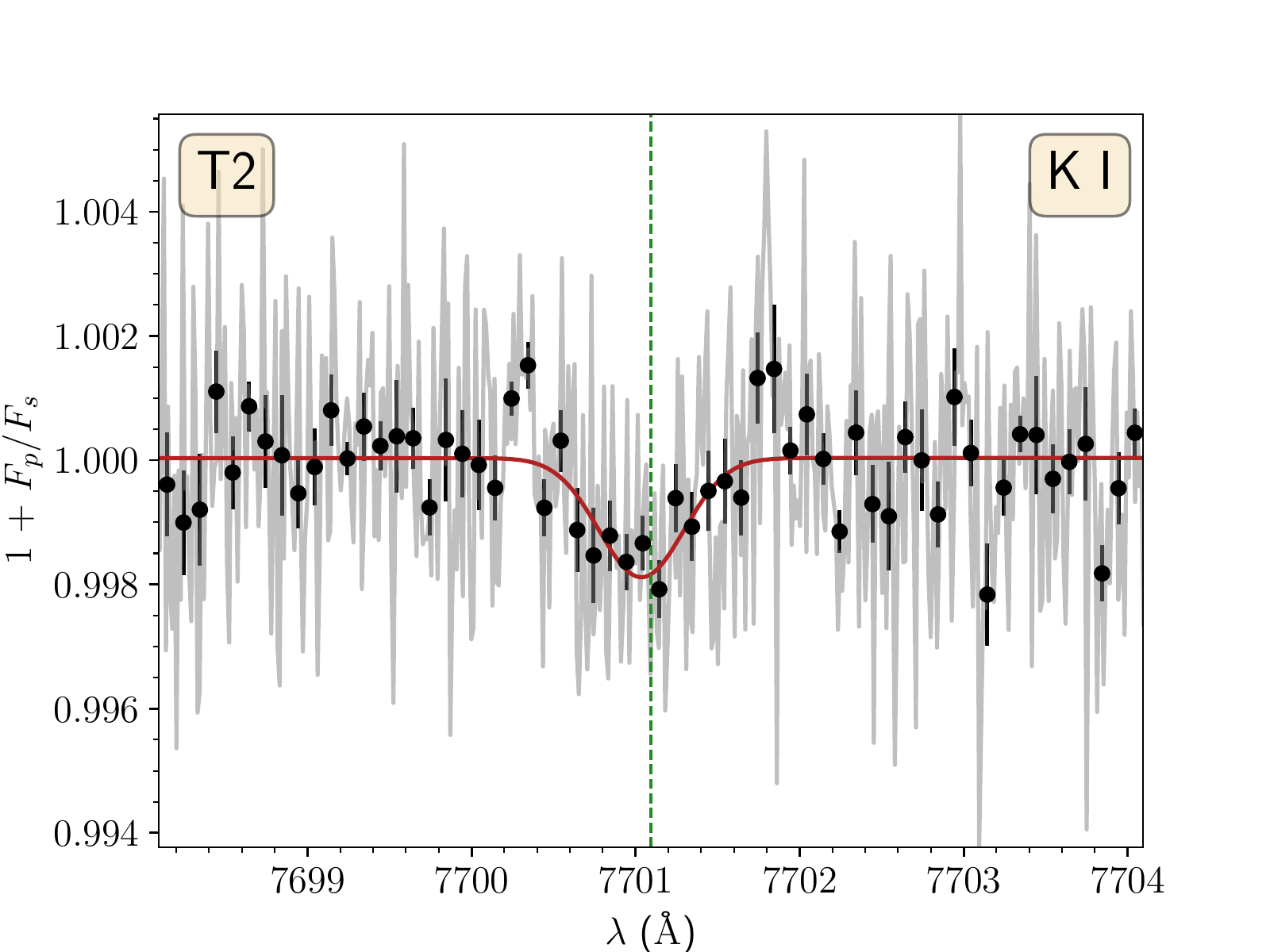} 
    \caption{ Same as Fig.~\ref{weak_lines_planet_1} but for \ion{Mg}{i}   and \ion{K}{i}.}
    \label{weak_lines_planet_2}
\end{figure*}
\begin{table*}
\caption{Properties of the atomic lines studied in this work: line centre in the rest frame ($\lambda$), line depth (h), Doppler shift of the line ($V_{wind}$), Full width half maximum (FWHM), and the significance of the detection. All wavelengths are in vacuum. Lines corresponding to the \ion{Mg}{i}~b~T, \ion{Mn}{i}, and \ion{Fe}{i} have been combined to strengthen the signal to noise.}             
\label{line_intent}      
\begin{tabular}{llccccccc}   
\hline
\hline                 
Line &  $\lambda$ & Transit & h & V$_{\rm wind}$ & FWHM  & R$_{\lambda}$ & significance\\
     &  [\AA]        &         &    [\%]          &  [km~s$^{-1}$]  &   [km~s$^{-1}$] &  [$R_p$]                                &  [$\sigma$]\\
\hline   
\hline
\noalign{\smallskip}
\ion{Ca}{ii}~K & 3934.78 & T1 & 1.75~$\pm$~0.25 &  4.1~$\pm$~5.1 & 77.8~$\pm$~11.2 & 1.57~$\pm$~0.26  & 7.1  \\
\ion{Ca}{ii}~K  & 3934.78 & T2 & 2.67~$\pm$~0.32 &  1.0~$\pm$~3.0 & 51.3~$\pm$~6.9 & 1.80~$\pm$~0.30  & 8.4\\
\hline
\ion{Ca}{ii}~H  & 3969.59  & T1 &  2.56~$\pm$~0.30 & -4.4~$\pm$~2.5 & 43.2~$\pm$~5.8 & 1.78~$\pm$~0.29  &  8.5\\
\ion{Ca}{ii}~H  & 3969.59  & T2 &  2.76~$\pm$~0.49  & -2.1~$\pm$~1.9 & 21.5~$\pm$~4.4 & 1.82$\pm$~0.45  & 5.6\\
 \hline
 \noalign{\smallskip}
  \ion{Mn}{i}  & $\sim$4033.91 & T1 & 0.305~$\pm$~0.065 &    -12.2~$\pm$~1.4 & 13.9~$\pm$~3.5 & 1.12~$\pm$0.10 & 4.7 \\
  \ion{Mn}{i} &  $\sim$4033.91 & T2 & 0.271~$\pm$~0.054 &   -7.4~$\pm$~2.2 & 23.1~$\pm$~5.3 & 1.108~$\pm$0.082 & 5.0 \\
   \hline
 \noalign{\smallskip}
  \ion{Fe}{i}  & $\sim$4402.58 & T1 & 0.144~$\pm$~0.031 & -4.3~$\pm$~2.2  & 20.8~$\pm$~5.3 & 1.059~$\pm$~0.049 & 4.7 \\
  \ion{Fe}{i} & $\sim$4402.58 & T2 & 0.203~$\pm$~0.030 & -8.3~$\pm$~1.6  & 22.4~$\pm$~3.9 & 1.082~$\pm$~0.047 &  6.7\\
 \hline
 \noalign{\smallskip}
 \ion{Mg}{i}  & 4572.38 & T1 & 0.201~$\pm$~0.072 &  -9.9~$\pm$~2.8 & 17.4~$\pm$~7.4  & 1.08~$\pm$~0.11 & 2.8\\
  \ion{Mg}{i}  & 4572.38 & T2 & 0.213~$\pm$~0.058 & -9.4~$\pm$~2.3 & 15.9~$\pm$~5.6 & 1.086~$\pm$~0.091 & 3.7 \\
\hline
\noalign{\smallskip} 
\ion{Mg}{i}~b  & $\sim$5175.97 & T1 &  0.142~$\pm$~0.028 & -4.1~$\pm$~2.1 &  21.4~$\pm$~5.2 & 1.058~$\pm$~0.045  & 5.0 \\
\ion{Mg}{i}~b  & $\sim$5175.97 & T2 &  0.288~$\pm$~0.038  & -7.31~$\pm$~0.71  & 11.1~$\pm$~1.7 & 1.115~$\pm$~0.058 & 7.5 \\
\hline
\noalign{\smallskip}
\ion{Na}{i}~D2  & 5891.58 & T1 &  0.449~$\pm$~0.049 & -5.8~$\pm$~1.0 & 24.6~$\pm$~3.1 & 1.174~$\pm$~0.070  & 9.2 \\
\ion{Na}{i}~D2  & 5891.58 & T2 &  0.246~$\pm$~0.037  & -5.3~$\pm$~1.4  & 31.7~$\pm$~5.7 & 1.099~$\pm$~0.057& 6.7 \\

\hline
\noalign{\smallskip}
\ion{Na}{i}~D1  & 5897.56 & T1 & 0.385~$\pm$~0.051 & -5.8~$\pm$~1.0 & 21.2~$\pm$~3.3 & 1.148~$\pm$~0.074    & 7.5\\
\ion{Na}{i}~D1  & 5897.56 & T2 &  0.294~$\pm$~0.042 & -5.3~$\pm$~1.4  & 23.6~$\pm$~4.1  & 1.117~$\pm$~0.063 & 7.0\\
\hline
H$\alpha$  & 6564.61 & T1 & 0.48~$\pm$~0.12 &  -7.48~$\pm$~0.73 & 5.9~$\pm$~1.8 & 1.18~$\pm$~0.16  &  4.0\\
H$\alpha$  & 6564.61 & T2 & $<$~0.2 &  NA & NA & NA & NA \\
\hline
\ion{Li}{i}  & 6709.61 & T1 & 0.173~$\pm$~0.039 &-1.8~$\pm$~3.0 &   26.6~$\pm$~7.6 & 1.070~$\pm$~0.061  &  4.4\\
\ion{Li}{i}  & 6709.61 & T2 & 0.235~$\pm$~0.042      &-5.9 ~$\pm$~1.6 &  18.7~$\pm$~4.0 &   1.094~$\pm$~0.064 &  5.7 \\
\hline
\ion{K}{i} &7701.09   & T1 &  0.185~$\pm$~0.048& -10.1~$\pm$~2.7 & 21.9~$\pm$~6.8 & 1.075~$\pm$~0.076 & 3.8 \\
\ion{K}{i}  & 7701.09 & T2 & 0.212~$\pm$~0.037 &  -2.2~$\pm$~2.1 & 25.5~$\pm$~5.4 & 1.086~$\pm$~0.057 & 5.7 \\
\hline
\end{tabular}
\end{table*}

\begin{table}
  \caption{Cross-correlation data for the analysed atomic and molecular species}             
  \label{ccf_intent}
  \begin{tabular}{llccc}
  \hline
  \hline
Element &   Transit & depth  & V$_{\rm wind}$ & FWHM  \\ 
        &           &  (ppm) &   (km~s$^{-1}$) & (km~s$^{-1}$)     \\
  \hline
   \noalign{\smallskip}  
\ion{Ti}{i}       & T1 & $<$~61      & NA  &  NA \\
  \ion{Ti}{i}       & T2 & $<$~60     & NA   & NA \\ 
\hline
 \noalign{\smallskip}  
\ion{Cr}{i}       & T1  & $<$~77   & NA   & NA \\
\ion{Cr}{i}       & T2  & $<$~78 &  NA  & NA \\
\hline
 \noalign{\smallskip}  
\ion{Fe}{i}       & T1 &   255~$\pm$~20 & -8.27$~\pm$0.25 &     6.62~$\pm$~0.59 \\
\ion{Fe}{i}       & T2 &  182~$\pm$~12  & -8.75$~\pm$0.56  & 18.0~$\pm$~1.3\\
\hline
 \noalign{\smallskip}  
\ion{Ni}{i}       & T1 &    $<$~130  & NA  & NA \\
  \ion{Ni}{i} & T2 &  $<$~122  & NA & NA\\ 
  \hline
 TiO & T1 &  $<$~6  & NA  & NA\\       
 TiO   & T2 &   $<$~6    &  NA & NA \\ 
\hline
  VO & T1 &  $<$~9  &  NA  & NA \\                   
  VO       & T2 &    $<$~8    &  NA & NA\\
  \hline
  ZrO   & T1 &  $<$~9    & NA & NA \\  
    ZrO   & T2 & $<$~8   &  NA  & NA\\  
\hline

\end{tabular}
\end{table}

\section{Summary and conclusions}
\label{sect_sum}

We have analysed two transits of the UHJ WASP-76b using ESPRESSO at the VLT. In this work, we generated two independent transmission spectra covering the available wavelength range. Using these spectra we have been able to detect features that were not reported in previous studies: \ion{Li}{i}, \ion{Mg}{i}, \ion{K}{i}, \ion{Ca}{ii}, \ion{Mn}{i}. In addition, our works strengthens the previous detections of  \ion{Na}{i} \citep{sei19,ess20} and \ion{Fe}{i} \citep{ehr20}.\\

\begin{figure*}
    \centering
     \includegraphics[width=0.48\textwidth]{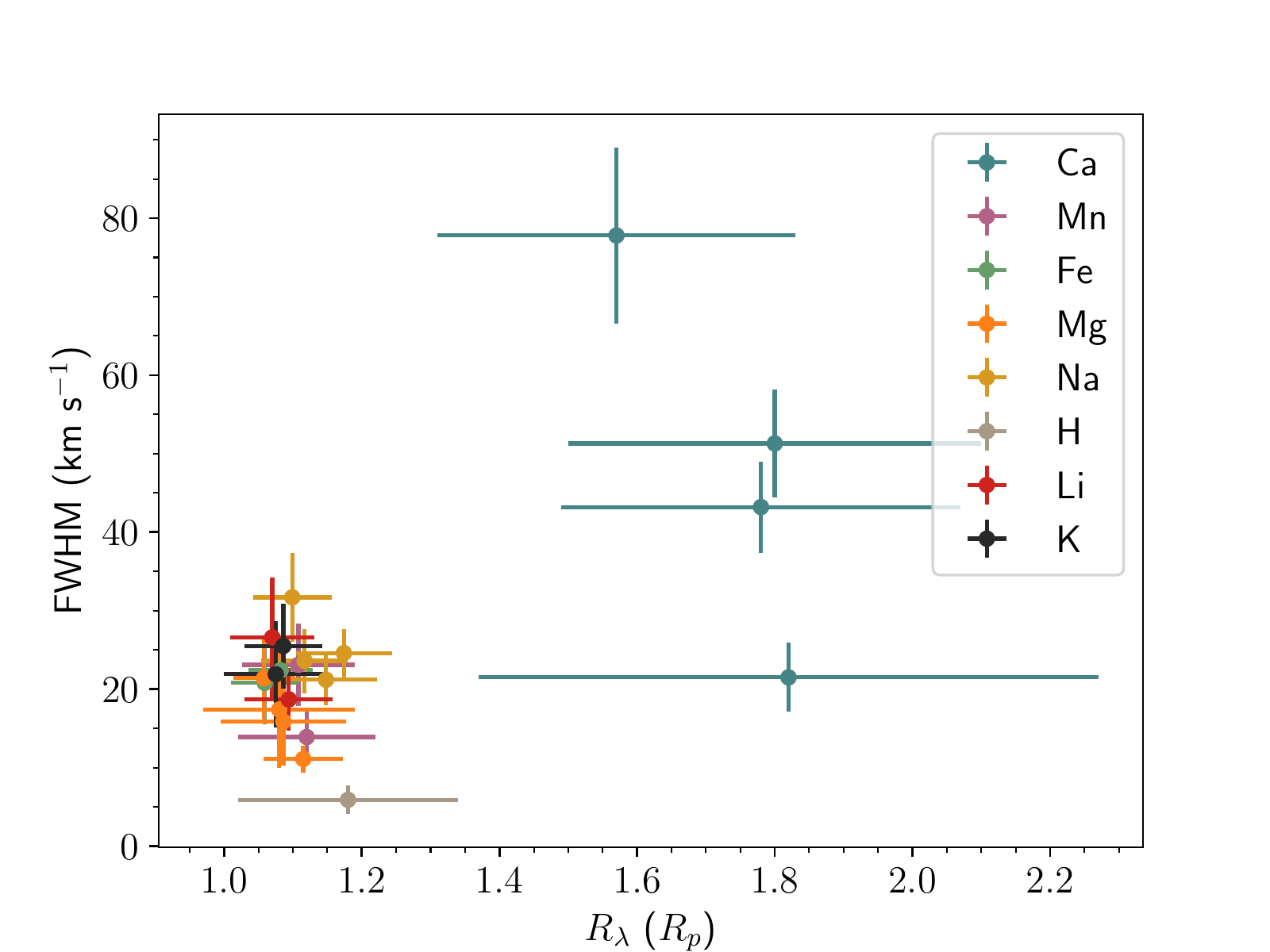}
      \includegraphics[width=0.48\textwidth]{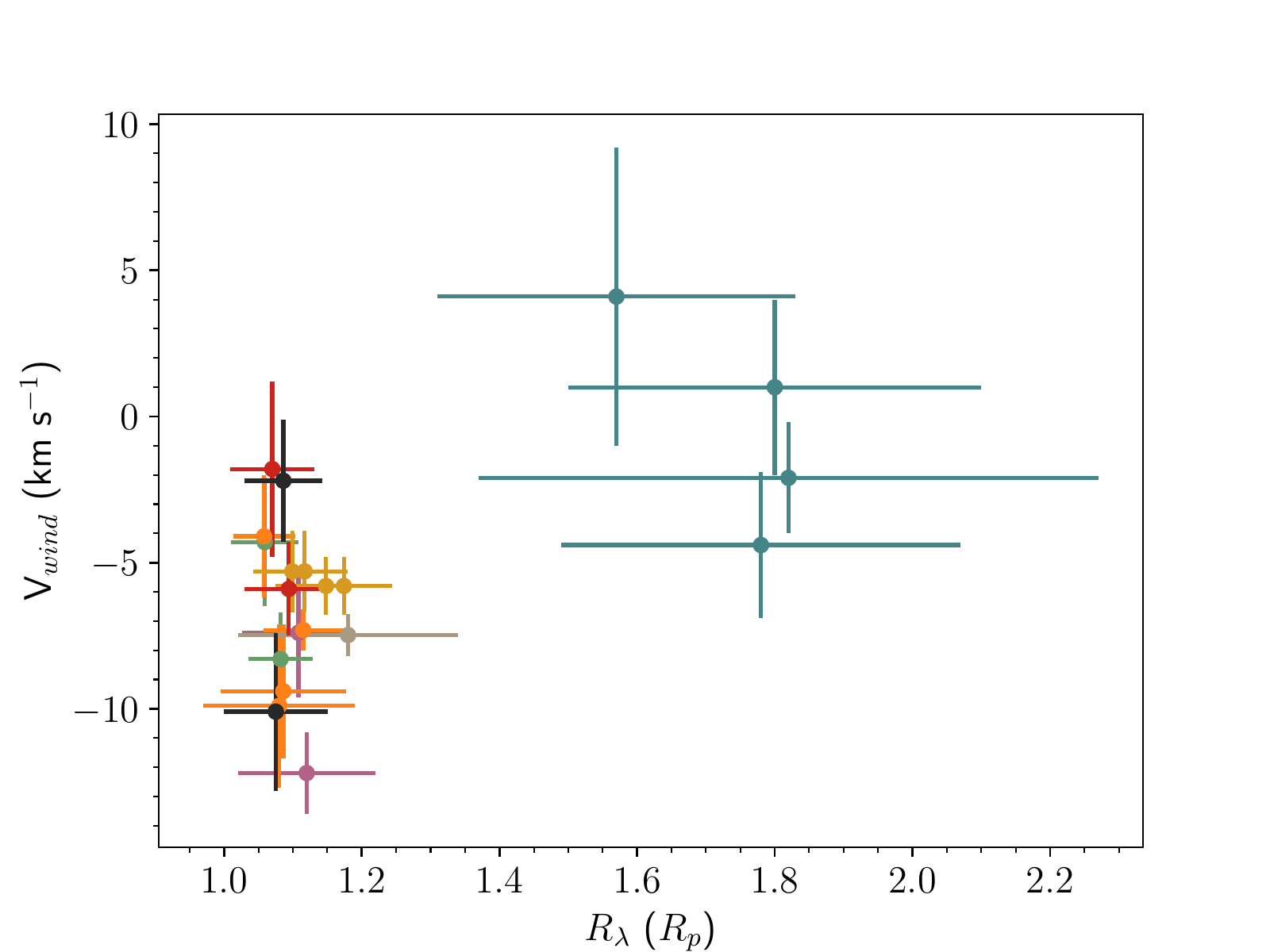}
    \caption{ FWHM and $V_{wind}$ for the lines studied in this work vs $R_{\lambda}$.}
    \label{prop_lines}
\end{figure*}

We have found that most lines are blueshifted with respect to their rest-frame wavelengths (see Fig.~\ref{prop_lines}). The median is -5.6~$\pm$~4.2~km~s$^{-1}$ for T1, and -5.2~$\pm$~3.1~km~s$^{-1}$ for T2. These shifts are probably due to planetary winds and are frequently reported in the literature \citep[see, e.g.][]{cas19,hoe19,gib20,ehr20}.\\

Interestingly, our calculations indicate that the \ion{Ca}{ii} lines are formed in higher layers than the other lines ($\sim$1.6$-$1.8~$R_{p}$, see Table~\ref{line_intent}) which points towards an extended exosphere. This has been already reported for other planets such as MASCARA-2b \citep{cas19}, WASP-33b \citep{yan19}, and WASP-121b  \citep{bor20}. In addition, their intrinsic width is in all instances higher than $\sim$~20~km~s$^{-1}$ (see Fig.~\ref{prop_lines}). In addition, the H$\alpha$ line seen in T1 is much narrow compared to the other atomic features. Its width is $\sim$~6~km~s$^{-1}$ that is greater than the resolving power of our data (R~$=$~140000, FWHM~$=$~2.1~km~s$^{-1}$). In all, the measured FWHM for this line is in principle physically possible and its absence in T2 might be indicative of atmospheric variability. However, the current data are not sufficient to fully explore this scenario.\\

In addition, our data show that the \ion{K}{i} absorption line is weaker than the \ion{Li}{i} feature, a fact that is unexpected for a near-solar composition where K is much more abundant than Li (both atoms have very similar electronic structure). The observations suggest that Li is at least as abundant as K in the investigated planetary atmospheric layers. This requires scenarios that have not been explored in exoplanetary atmospheres so far, for example, lithium production in situ or a strongly inhomogeneous distribution of the chemical abundances within the atmosphere.\\

Regarding the molecular species in WASP-76b, we could not find any signs of TiO, VO, or ZrO. They are either not present in the atmosphere of WASP-76b or their intensity is well below the minimum noise level in these observations.\\

Finally, we have demonstrated that we are able to reduce the noise of the exoplanetary data using ESPRESSO to a degree that we can detect many planetary atomic features by means of a single transmission spectrum.

\newpage

\begin{acknowledgements}

This work was supported by FCT - Fundação para a Ciência e a Tecnologia 
through national funds and by FEDER through COMPETE2020 - Programa 
Operacional Competitividade e Internacionalização by these grants: 
UID/FIS/04434/2019; UIDB/04434/2020; UIDP/04434/2020; 
PTDC/FIS-AST/32113/2017 \& POCI-01-0145-FEDER-032113; 
PTDC/FIS-AST/28953/2017 \& POCI-01-0145-FEDER-028953; 
PTDC/FIS-AST/28987/2017 \& POCI-01-0145-FEDER-028987. V.A., S.G.S., 
S.C.C.B. acknowledge support from FCT through Investigador FCT contracts 
nsº IF/00650/2015/CP1273/CT0001; IF/00028/2014/CP1215/CT0002; 
IF/01312/2014/CP1215/CT0004. S.G.S acknowledges the support from FCT through Investigador FCT contract nr. CEECIND/00826/2018 and POPH/FSE (EC).   J.P.F., O.D., and J.H.C.M. acknowledge support from FCT 
through national funds in the form of a work contract with the 
references DL 57/2016/CP1364/CT0005; DL 57/2016/CP1364/CT0004; DL 57/2016/CP1364/CT0007. HMT and MRZO acknowledge financial support from the Spanish Ministerio de Ciencia, Innovaci\'{o}n y Universidades (MICIU) through project AYA2016-79425-C3-2.
A.S.M., R.R., J.I.G.H., C.A.P. acknowledge financial support from the Spanish MICIU project AYA2017-86389-P. J.I.G.H. acknowledges financial support from the Spanish MICIU under the 2013 Ram\'on y Cajal program RYC-2013-14875. This work has been carried out within the framework of the National Centre of Competence in Research PlanetS supported by the Swiss National Science Foundation. The authors acknowledge the financial support of the SNSF. The INAF authors acknowledge financial support of the Italian Ministry of Education, University, and Research with PRIN 201278X4FL and the "Progetti Premiali" funding scheme. This project has received funding from the European Research Council (ERC) under the European Union’s Horizon 2020 research and innovation programme (project {\sc Four Aces}; grant agreement No 724427). N.J.N. acknowledges support from FCT through Investigador FCT contract and exploratory project IF/00852/2015, and project PTDC/FIS-OUT/29048/2017. J.V.S. acknowledges funding from the European Research Council (ERC) under the European Union’s Horizon 2020 research and innovation programme (project Four Aces; grant agreement No. 724427).

\end{acknowledgements}

\bibliographystyle{aa} 
\bibliography{WASP76}                                                           

\begin{thebibliography}{76}
\expandafter\ifx\csname natexlab\endcsname\relax\def\natexlab#1{#1}\fi

\bibitem[{{Alam} {et~al.}(2015){Alam}, {Albareti}, {Allende Prieto}, {Anders},
  {Anderson}, {Anderton}, {Andrews}, {Armengaud}, {Aubourg}, {Bailey}, {Basu},
  {Bautista}, {Beaton}, {Beers}, {Bender}, {Berlind}, {Beutler}, {Bhardwaj},
  {Bird}, {Bizyaev}, {Blake}, {Blanton}, {Blomqvist}, {Bochanski}, {Bolton},
  {Bovy}, {Shelden Bradley}, {Brandt}, {Brauer}, {Brinkmann}, {Brown},
  {Brownstein}, {Burden}, {Burtin}, {Busca}, {Cai}, {Capozzi}, {Carnero
  Rosell}, {Carr}, {Carrera}, {Chambers}, {Chaplin}, {Chen}, {Chiappini},
  {Chojnowski}, {Chuang}, {Clerc}, {Comparat}, {Covey}, {Croft}, {Cuesta},
  {Cunha}, {da Costa}, {Da Rio}, {Davenport}, {Dawson}, {De Lee}, {Delubac},
  {Deshpande}, {Dhital}, {Dutra-Ferreira}, {Dwelly}, {Ealet}, {Ebelke},
  {Edmondson}, {Eisenstein}, {Ellsworth}, {Elsworth}, {Epstein}, {Eracleous},
  {Escoffier}, {Esposito}, {Evans}, {Fan}, {Fern{\'a}ndez-Alvar}, {Feuillet},
  {Filiz Ak}, {Finley}, {Finoguenov}, {Flaherty}, {Fleming}, {Font-Ribera},
  {Foster}, {Frinchaboy}, {Galbraith-Frew}, {Garc{\'\i}a},
  {Garc{\'\i}a-Hern{\'a}ndez}, {Garc{\'\i}a P{\'e}rez}, {Gaulme}, {Ge},
  {G{\'e}nova-Santos}, {Georgakakis}, {Ghezzi}, {Gillespie}, {Girardi},
  {Goddard}, {Gontcho}, {Gonz{\'a}lez Hern{\'a}ndez}, {Grebel}, {Green},
  {Grieb}, {Grieves}, {Gunn}, {Guo}, {Harding}, {Hasselquist}, {Hawley},
  {Hayden}, {Hearty}, {Hekker}, {Ho}, {Hogg}, {Holley-Bockelmann}, {Holtzman},
  {Honscheid}, {Huber}, {Huehnerhoff}, {Ivans}, {Jiang}, {Johnson},
  {Kinemuchi}, {Kirkby}, {Kitaura}, {Klaene}, {Knapp}, {Kneib}, {Koenig},
  {Lam}, {Lan}, {Lang}, {Laurent}, {Le Goff}, {Leauthaud}, {Lee}, {Lee},
  {Licquia}, {Liu}, {Long}, {L{\'o}pez-Corredoira}, {Lorenzo-Oliveira},
  {Lucatello}, {Lundgren}, {Lupton}, {Mack}, {Mahadevan}, {Maia}, {Majewski},
  {Malanushenko}, {Malanushenko}, {Manchado}, {Manera}, {Mao}, {Maraston},
  {Marchwinski}, {Margala}, {Martell}, {Martig}, {Masters}, {Mathur},
  {McBride}, {McGehee}, {McGreer}, {McMahon}, {M{\'e}nard}, {Menzel},
  {Merloni}, {M{\'e}sz{\'a}ros}, {Miller}, {Miralda-Escud{\'e}}, {Miyatake},
  {Montero-Dorta}, {More}, {Morganson}, {Morice-Atkinson}, {Morrison},
  {Mosser}, {Muna}, {Myers}, {Nand ra}, {Newman}, {Neyrinck}, {Nguyen},
  {Nichol}, {Nidever}, {Noterdaeme}, {Nuza}, {O'Connell}, {O'Connell},
  {O'Connell}, {Ogando}, {Olmstead}, {Oravetz}, {Oravetz}, {Osumi}, {Owen},
  {Padgett}, {Padmanabhan}, {Paegert}, {Palanque-Delabrouille}, {Pan},
  {Parejko}, {P{\^a}ris}, {Park}, {Pattarakijwanich}, {Pellejero-Ibanez},
  {Pepper}, {Percival}, {P{\'e}rez-Fournon}, {Ṕrez-Ra`fols}, {Petitjean},
  {Pieri}, {Pinsonneault}, {Porto de Mello}, {Prada}, {Prakash},
  {Price-Whelan}, {Protopapas}, {Raddick}, {Rahman}, {Reid}, {Rich}, {Rix},
  {Robin}, {Rockosi}, {Rodrigues}, {Rodr{\'\i}guez-Torres}, {Roe}, {Ross},
  {Ross}, {Rossi}, {Ruan}, {Rubi{\~n}o-Mart{\'\i}n}, {Rykoff},
  {Salazar-Albornoz}, {Salvato}, {Samushia}, {S{\'a}nchez}, {Santiago},
  {Sayres}, {Schiavon}, {Schlegel}, {Schmidt}, {Schneider}, {Schultheis},
  {Schwope}, {Sc{\'o}ccola}, {Scott}, {Sellgren}, {Seo}, {Serenelli}, {Shane},
  {Shen}, {Shetrone}, {Shu}, {Silva Aguirre}, {Sivarani}, {Skrutskie},
  {Slosar}, {Smith}, {Sobreira}, {Souto}, {Stassun}, {Steinmetz}, {Stello},
  {Strauss}, {Streblyanska}, {Suzuki}, {Swanson}, {Tan}, {Tayar}, {Terrien},
  {Thakar}, {Thomas}, {Thomas}, {Thompson}, {Tinker}, {Tojeiro}, {Troup},
  {Vargas-Maga{\~n}a}, {Vazquez}, {Verde}, {Viel}, {Vogt}, {Wake}, {Wang},
  {Weaver}, {Weinberg}, {Weiner}, {White}, {Wilson}, {Wisniewski},
  {Wood-Vasey}, {Ye`che}, {York}, {Zakamska}, {Zamora}, {Zasowski}, {Zehavi},
  {Zhao}, {Zheng}, {Zhou}, {Zhou}, {Zou}, \& {Zhu}}]{ala15}
{Alam}, S., {Albareti}, F.~D., {Allende Prieto}, C., {et~al.} 2015, \apjs, 219,
  12

\bibitem[{{Allart} {et~al.}(2019){Allart}, {Bourrier}, {Lovis}, {Ehrenreich},
  {Aceituno}, {Guijarro}, {Pepe}, {Sing}, {Spake}, \& {Wyttenbach}}]{all19}
{Allart}, R., {Bourrier}, V., {Lovis}, C., {et~al.} 2019, \aap, 623, A58

\bibitem[{{Allart} {et~al.}(2018){Allart}, {Bourrier}, {Lovis}, {Ehrenreich},
  {Spake}, {Wyttenbach}, {Pino}, {Pepe}, {Sing}, \& {Lecavelier des
  Etangs}}]{all18}
{Allart}, R., {Bourrier}, V., {Lovis}, C., {et~al.} 2018, Science, 362, 1384

\bibitem[{{Allart} {et~al.}(2017){Allart}, {Lovis}, {Pino}, {Wyttenbach},
  {Ehrenreich}, \& {Pepe}}]{all17}
{Allart}, R., {Lovis}, C., {Pino}, L., {et~al.} 2017, \aap, 606, A144

\bibitem[{{Alonso-Floriano} {et~al.}(2019){Alonso-Floriano}, {Snellen},
  {Czesla}, {Bauer}, {Salz}, {Lamp{\'o}n}, {Lara}, {Nagel},
  {L{\'o}pez-Puertas}, {Nortmann}, {S{\'a}nchez-L{\'o}pez}, {Sanz-Forcada},
  {Caballero}, {Reiners}, {Ribas}, {Quirrenbach}, {Amado}, {Aceituno},
  {Anglada-Escud{\'e}}, {B{\'e}jar}, {Brinkm{\"o}ller}, {Hatzes}, {Henning},
  {Kaminski}, {K{\"u}rster}, {Labarga}, {Montes}, {Pall{\'e}}, {Schmitt}, \&
  {Zapatero Osorio}}]{alo19}
{Alonso-Floriano}, F.~J., {Snellen}, I.~A.~G., {Czesla}, S., {et~al.} 2019,
  \aap, 629, A110

\bibitem[{{Baraffe} {et~al.}(2015){Baraffe}, {Homeier}, {Allard}, \&
  {Chabrier}}]{bar15}
{Baraffe}, I., {Homeier}, D., {Allard}, F., \& {Chabrier}, G. 2015, \aap, 577,
  A42

\bibitem[{{Barklem} {et~al.}(1998){Barklem}, {Anstee}, \& {O'Mara}}]{abo98}
{Barklem}, P.~S., {Anstee}, S.~D., \& {O'Mara}, B.~J. 1998, \pasa, 15, 336

\bibitem[{{Barman} {et~al.}(2015){Barman}, {Konopacky}, {Macintosh}, \&
  {Marois}}]{barm15}
{Barman}, T.~S., {Konopacky}, Q.~M., {Macintosh}, B., \& {Marois}, C. 2015,
  \apj, 804, 61

\bibitem[{{Blanco-Cuaresma} {et~al.}(2014){Blanco-Cuaresma}, {Soubiran},
  {Heiter}, \& {Jofr{\'e}}}]{cua14}
{Blanco-Cuaresma}, S., {Soubiran}, C., {Heiter}, U., \& {Jofr{\'e}}, P. 2014,
  \aap, 569, A111

\bibitem[{{Borsa} {et~al.}(2020){Borsa}, {Allart}, {Casasayas-Barris},
  {Tabernero}, {Zapatero Osorio}, {Cristiani}, {Pepe}, {Rebolo}, {Santos},
  {Adibekyan}, {Bourrier}, {Demangeon}, {Ehrenreich}, {Pall{\'e}}, {Sousa},
  {Lillo-Box}, {Lovis}, {Micela}, {Oshagh}, {Poretti}, {Sozzetti}, {Allende
  Prieto}, {Alibert}, {Amate}, {Benz}, {Bouchy}, {Cabral}, {Dekker},
  {D'Odorico}, {Di Marcantonio}, {Figueira}, {Genova Santos}, {Gonz{\'a}lez
  Hern{\'a}ndez}, {Lo Curto}, {Manescau}, {Martins}, {M{\'e}gevand}, {Mehner},
  {Molaro}, {Nunes}, {Riva}, {Su{\'a}rez Mascare{\~n}o}, {Udry}, \&
  {Zerbi}}]{bor20}
{Borsa}, F., {Allart}, R., {Casasayas-Barris}, N., {et~al.} 2020, arXiv
  e-prints, arXiv:2011.01245

\bibitem[{{Borucki} {et~al.}(2010){Borucki}, {Koch}, {Basri}, {Batalha},
  {Brown}, {Caldwell}, {Caldwell}, {Christensen-Dalsgaard}, {Cochran},
  {DeVore}, {Dunham}, {Dupree}, {Gautier}, {Geary}, {Gilliland}, {Gould},
  {Howell}, {Jenkins}, {Kondo}, {Latham}, {Marcy}, {Meibom}, {Kjeldsen},
  {Lissauer}, {Monet}, {Morrison}, {Sasselov}, {Tarter}, {Boss}, {Brownlee},
  {Owen}, {Buzasi}, {Charbonneau}, {Doyle}, {Fortney}, {Ford}, {Holman},
  {Seager}, {Steffen}, {Welsh}, {Rowe}, {Anderson}, {Buchhave}, {Ciardi},
  {Walkowicz}, {Sherry}, {Horch}, {Isaacson}, {Everett}, {Fischer}, {Torres},
  {Johnson}, {Endl}, {MacQueen}, {Bryson}, {Dotson}, {Haas}, {Kolodziejczak},
  {Van Cleve}, {Chandrasekaran}, {Twicken}, {Quintana}, {Clarke}, {Allen},
  {Li}, {Wu}, {Tenenbaum}, {Verner}, {Bruhweiler}, {Barnes}, \&
  {Prsa}}]{kepler}
{Borucki}, W.~J., {Koch}, D., {Basri}, G., {et~al.} 2010, Science, 327, 977

\bibitem[{{Bressan} {et~al.}(2012){Bressan}, {Marigo}, {Girardi}, {Salasnich},
  {Dal Cero}, {Rubele}, \& {Nanni}}]{bre12}
{Bressan}, A., {Marigo}, P., {Girardi}, L., {et~al.} 2012, \mnras, 427, 127

\bibitem[{{Casasayas-Barris} {et~al.}(2019){Casasayas-Barris}, {Pall{\'e}},
  {Yan}, {Chen}, {Kohl}, {Stangret}, {Parviainen}, {Helling}, {Watanabe},
  {Czesla}, {Fukui}, {Monta{\~n}{\'e}s-Rodr{\'\i}guez}, {Nagel}, {Narita},
  {Nortmann}, {Nowak}, {Schmitt}, \& {Zapatero Osorio}}]{cas19}
{Casasayas-Barris}, N., {Pall{\'e}}, E., {Yan}, F., {et~al.} 2019, \aap, 628,
  A9

\bibitem[{{Casasayas-Barris} {et~al.}(2020){Casasayas-Barris}, {Pall{\'e}},
  {Yan}, {Chen}, {Luque}, {Stangret}, {Nagel}, {Zechmeister}, {Oshagh},
  {Sanz-Forcada}, {Nortmann}, {Alonso-Floriano}, {Amado}, {Caballero},
  {Czesla}, {Khalafinejad}, {L{\'o}pez-Puertas}, {L{\'o}pez-Santiago},
  {Molaverdikhani}, {Montes}, {Quirrenbach}, {Reiners}, {Ribas},
  {S{\'a}nchez-L{\'o}pez}, \& {Zapatero Osorio}}]{cas20}
{Casasayas-Barris}, N., {Pall{\'e}}, E., {Yan}, F., {et~al.} 2020, \aap, 635,
  A206

\bibitem[{{Chabrier} {et~al.}(2000){Chabrier}, {Baraffe}, {Allard}, \&
  {Hauschildt}}]{cha00}
{Chabrier}, G., {Baraffe}, I., {Allard}, F., \& {Hauschildt}, P. 2000, \apj,
  542, 464

\bibitem[{{Charbonneau} {et~al.}(2002){Charbonneau}, {Brown}, {Noyes}, \&
  {Gilliland}}]{char02}
{Charbonneau}, D., {Brown}, T.~M., {Noyes}, R.~W., \& {Gilliland}, R.~L. 2002,
  \apj, 568, 377

\bibitem[{{Charnay} {et~al.}(2015){Charnay}, {Meadows}, \& {Leconte}}]{char15}
{Charnay}, B., {Meadows}, V., \& {Leconte}, J. 2015, \apj, 813, 15

\bibitem[{{Chen} {et~al.}(2018){Chen}, {Pall{\'e}}, {Welbanks},
  {Prieto-Arranz}, {Madhusudhan}, {Gandhi}, {Casasayas-Barris}, {Murgas},
  {Nortmann}, {Crouzet}, {Parviainen}, \& {Gandolfi}}]{chen18}
{Chen}, G., {Pall{\'e}}, E., {Welbanks}, L., {et~al.} 2018, \aap, 616, A145

\bibitem[{{Cosentino} {et~al.}(2012){Cosentino}, {Lovis}, {Pepe}, {Collier
  Cameron}, {Latham}, {Molinari}, {Udry}, {Bezawada}, {Black}, {Born},
  {Buchschacher}, {Charbonneau}, {Figueira}, {Fleury}, {Galli}, {Gallie},
  {Gao}, {Ghedina}, {Gonzalez}, {Gonzalez}, {Guerra}, {Henry}, {Horne},
  {Hughes}, {Kelly}, {Lodi}, {Lunney}, {Maire}, {Mayor}, {Micela}, {Ordway},
  {Peacock}, {Phillips}, {Piotto}, {Pollacco}, {Queloz}, {Rice}, {Riverol},
  {Riverol}, {San Juan}, {Sasselov}, {Segransan}, {Sozzetti}, {Sosnowska},
  {Stobie}, {Szentgyorgyi}, {Vick}, \& {Weber}}]{cos12}
{Cosentino}, R., {Lovis}, C., {Pepe}, F., {et~al.} 2012, Society of
  Photo-Optical Instrumentation Engineers (SPIE) Conference Series, Vol. 8446,
  {Harps-N: the new planet hunter at TNG}, 84461V

\bibitem[{{da Silva} {et~al.}(2006){da Silva}, {Girardi}, {Pasquini},
  {Setiawan}, {von der L{\"u}he}, {de Medeiros}, {Hatzes}, {D{\"o}llinger}, \&
  {Weiss}}]{sil06}
{da Silva}, L., {Girardi}, L., {Pasquini}, L., {et~al.} 2006, \aap, 458, 609

\bibitem[{{Edwards} {et~al.}(2020){Edwards}, {Changeat}, {Baeyens}, {Tsiaras},
  {Al-Refaie}, {Taylor}, {Yip}, {Bieger}, {Blain}, {Gressier}, {Guilluy},
  {Jaziri}, {Kiefer}, {Modirrousta-Galian}, {Morvan}, {Mugnai}, {Pluriel},
  {Poveda}, {Skaf}, {Whiteford}, {Wright}, {Zingales}, {Charnay}, {Drossart},
  {Leconte}, {Venot}, {Waldmann}, \& {Beaulieu}}]{edw20}
{Edwards}, B., {Changeat}, Q., {Baeyens}, R., {et~al.} 2020, \aj, 160, 8

\bibitem[{{Ehrenreich} {et~al.}(2020){Ehrenreich}, {Lovis}, {Allart}, {Zapatero
  Osorio}, {Pepe}, {Cristiani}, {Rebolo}, {Santos}, {Borsa}, {Demangeon},
  {Dumusque}, {Gonz{\'a}lez Hern{\'a}ndez}, {Casasayas-Barris},
  {S{\'e}gransan}, {Sousa}, {Abreu}, {Adibekyan}, {Affolter}, {Allende Prieto},
  {Alibert}, {Aliverti}, {Alves}, {Amate}, {Avila}, {Baldini}, {Bandy}, {Benz},
  {Bianco}, {Bolmont}, {Bouchy}, {Bourrier}, {Broeg}, {Cabral}, {Calderone},
  {Pall{\'e}}, {Cegla}, {Cirami}, {Coelho}, {Conconi}, {Coretti}, {Cumani},
  {Cupani}, {Dekker}, {Delabre}, {Deiries}, {D'Odorico}, {Di Marcantonio},
  {Figueira}, {Fragoso}, {Genolet}, {Genoni}, {G{\'e}nova Santos}, {Hara},
  {Hughes}, {Iwert}, {Kerber}, {Knudstrup}, {Land oni}, {Lavie}, {Lizon},
  {Lendl}, {Lo Curto}, {Maire}, {Manescau}, {Martins}, {M{\'e}gevand },
  {Mehner}, {Micela}, {Modigliani}, {Molaro}, {Monteiro}, {Monteiro},
  {Moschetti}, {M{\"u}ller}, {Nunes}, {Oggioni}, {Oliveira}, {Pariani},
  {Pasquini}, {Poretti}, {Rasilla}, {Redaelli}, {Riva}, {Santana Tschudi},
  {Santin}, {Santos}, {Segovia Milla}, {Seidel}, {Sosnowska}, {Sozzetti},
  {Span{\`o}}, {Su{\'a}rez Mascare{\~n}o}, {Tabernero}, {Tenegi}, {Udry},
  {Zanutta}, \& {Zerbi}}]{ehr20}
{Ehrenreich}, D., {Lovis}, C., {Allart}, R., {et~al.} 2020, \nat, 580, 597

\bibitem[{{Gaia Collaboration} {et~al.}(2018){Gaia Collaboration}, {Brown},
  {Vallenari}, {Prusti}, {de Bruijne}, {Babusiaux}, {Bailer-Jones}, {Biermann},
  {Evans}, {Eyer}, {Jansen}, {Jordi}, {Klioner}, {Lammers}, {Lindegren},
  {Luri}, {Mignard}, {Panem}, {Pourbaix}, {Randich}, {Sartoretti}, {Siddiqui},
  {Soubiran}, {van Leeuwen}, {Walton}, {Arenou}, {Bastian}, {Cropper},
  {Drimmel}, {Katz}, {Lattanzi}, {Bakker}, {Cacciari}, {Casta{\~n}eda},
  {Chaoul}, {Cheek}, {De Angeli}, {Fabricius}, {Guerra}, {Holl}, {Masana},
  {Messineo}, {Mowlavi}, {Nienartowicz}, {Panuzzo}, {Portell}, {Riello},
  {Seabroke}, {Tanga}, {Th{\'e}venin}, {Gracia-Abril}, {Comoretto},
  {Garcia-Reinaldos}, {Teyssier}, {Altmann}, {Andrae}, {Audard},
  {Bellas-Velidis}, {Benson}, {Berthier}, {Blomme}, {Burgess}, {Busso},
  {Carry}, {Cellino}, {Clementini}, {Clotet}, {Creevey}, {Davidson}, {De
  Ridder}, {Delchambre}, {Dell'Oro}, {Ducourant},
  {Fern{\'a}ndez-Hern{\'a}ndez}, {Fouesneau}, {Fr{\'e}mat}, {Galluccio},
  {Garc{\'\i}a-Torres}, {Gonz{\'a}lez-N{\'u}{\~n}ez}, {Gonz{\'a}lez-Vidal},
  {Gosset}, {Guy}, {Halbwachs}, {Hambly}, {Harrison}, {Hern{\'a}ndez},
  {Hestroffer}, {Hodgkin}, {Hutton}, {Jasniewicz}, {Jean-Antoine-Piccolo},
  {Jordan}, {Korn}, {Krone-Martins}, {Lanzafame}, {Lebzelter}, {L{\"o}ffler},
  {Manteiga}, {Marrese}, {Mart{\'\i}n-Fleitas}, {Moitinho}, {Mora}, {Muinonen},
  {Osinde}, {Pancino}, {Pauwels}, {Petit}, {Recio-Blanco}, {Richards},
  {Rimoldini}, {Robin}, {Sarro}, {Siopis}, {Smith}, {Sozzetti}, {S{\"u}veges},
  {Torra}, {van Reeven}, {Abbas}, {Abreu Aramburu}, {Accart}, {Aerts},
  {Altavilla}, {{\'A}lvarez}, {Alvarez}, {Alves}, {Anderson}, {Andrei},
  {Anglada Varela}, {Antiche}, {Antoja}, {Arcay}, {Astraatmadja}, {Bach},
  {Baker}, {Balaguer-N{\'u}{\~n}ez}, {Balm}, {Barache}, {Barata}, {Barbato},
  {Barblan}, {Barklem}, {Barrado}, {Barros}, {Barstow}, {Bartholom{\'e}
  Mu{\~n}oz}, {Bassilana}, {Becciani}, {Bellazzini}, {Berihuete}, {Bertone},
  {Bianchi}, {Bienaym{\'e}}, {Blanco-Cuaresma}, {Boch}, {Boeche}, {Bombrun},
  {Borrachero}, {Bossini}, {Bouquillon}, {Bourda}, {Bragaglia}, {Bramante},
  {Breddels}, {Bressan}, {Brouillet}, {Br{\"u}semeister}, {Brugaletta},
  {Bucciarelli}, {Burlacu}, {Busonero}, {Butkevich}, {Buzzi}, {Caffau},
  {Cancelliere}, {Cannizzaro}, {Cantat-Gaudin}, {Carballo}, {Carlucci},
  {Carrasco}, {Casamiquela}, {Castellani}, {Castro-Ginard}, {Charlot},
  {Chemin}, {Chiavassa}, {Cocozza}, {Costigan}, {Cowell}, {Crifo}, {Crosta},
  {Crowley}, {Cuypers}, {Dafonte}, {Damerdji}, {Dapergolas}, {David}, {David},
  {de Laverny}, {De Luise}, {De March}, {de Martino}, {de Souza}, {de Torres},
  {Debosscher}, {del Pozo}, {Delbo}, {Delgado}, {Delgado}, {Di Matteo},
  {Diakite}, {Diener}, {Distefano}, {Dolding}, {Drazinos}, {Dur{\'a}n},
  {Edvardsson}, {Enke}, {Eriksson}, {Esquej}, {Eynard Bontemps}, {Fabre},
  {Fabrizio}, {Faigler}, {Falc{\~a}o}, {Farr{\`a}s Casas}, {Federici},
  {Fedorets}, {Fernique}, {Figueras}, {Filippi}, {Findeisen}, {Fonti},
  {Fraile}, {Fraser}, {Fr{\'e}zouls}, {Gai}, {Galleti}, {Garabato},
  {Garc{\'\i}a-Sedano}, {Garofalo}, {Garralda}, {Gavel}, {Gavras}, {Gerssen},
  {Geyer}, {Giacobbe}, {Gilmore}, {Girona}, {Giuffrida}, {Glass}, {Gomes},
  {Granvik}, {Gueguen}, {Guerrier}, {Guiraud}, {Guti{\'e}rrez-S{\'a}nchez},
  {Haigron}, {Hatzidimitriou}, {Hauser}, {Haywood}, {Heiter}, {Helmi}, {Heu},
  {Hilger}, {Hobbs}, {Hofmann}, {Holland}, {Huckle}, {Hypki}, {Icardi},
  {Jan{\ss}en}, {Jevardat de Fombelle}, {Jonker}, {Juh{\'a}sz}, {Julbe},
  {Karampelas}, {Kewley}, {Klar}, {Kochoska}, {Kohley}, {Kolenberg},
  {Kontizas}, {Kontizas}, {Koposov}, {Kordopatis}, {Kostrzewa-Rutkowska},
  {Koubsky}, {Lambert}, {Lanza}, {Lasne}, {Lavigne}, {Le Fustec}, {Le
  Poncin-Lafitte}, {Lebreton}, {Leccia}, {Leclerc}, {Lecoeur-Taibi},
  {Lenhardt}, {Leroux}, {Liao}, {Licata}, {Lindstr{\o}m}, {Lister}, {Livanou},
  {Lobel}, {L{\'o}pez}, {Managau}, {Mann}, {Mantelet}, {Marchal}, {Marchant},
  {Marconi}, {Marinoni}, {Marschalk{\'o}}, {Marshall}, {Martino}, {Marton},
  {Mary}, {Massari}, {Matijevi{\v{c}}}, {Mazeh}, {McMillan}, {Messina},
  {Michalik}, {Millar}, {Molina}, {Molinaro}, {Moln{\'a}r}, {Montegriffo},
  {Mor}, {Morbidelli}, {Morel}, {Morris}, {Mulone}, {Muraveva}, {Musella},
  {Nelemans}, {Nicastro}, {Noval}, {O'Mullane}, {Ord{\'e}novic},
  {Ord{\'o}{\~n}ez-Blanco}, {Osborne}, {Pagani}, {Pagano}, {Pailler},
  {Palacin}, {Palaversa}, {Panahi}, {Pawlak}, {Piersimoni}, {Pineau}, {Plachy},
  {Plum}, {Poggio}, {Poujoulet}, {Pr{\v{s}}a}, {Pulone}, {Racero}, {Ragaini},
  {Rambaux}, {Ramos-Lerate}, {Regibo}, {Reyl{\'e}}, {Riclet}, {Ripepi}, {Riva},
  {Rivard}, {Rixon}, {Roegiers}, {Roelens}, {Romero-G{\'o}mez}, {Rowell},
  {Royer}, {Ruiz-Dern}, {Sadowski}, {Sagrist{\`a} Sell{\'e}s}, {Sahlmann},
  {Salgado}, {Salguero}, {Sanna}, {Santana-Ros}, {Sarasso}, {Savietto},
  {Schultheis}, {Sciacca}, {Segol}, {Segovia}, {S{\'e}gransan}, {Shih},
  {Siltala}, {Silva}, {Smart}, {Smith}, {Solano}, {Solitro}, {Sordo}, {Soria
  Nieto}, {Souchay}, {Spagna}, {Spoto}, {Stampa}, {Steele},
  {Steidelm{\"u}ller}, {Stephenson}, {Stoev}, {Suess}, {Surdej}, {Szabados},
  {Szegedi-Elek}, {Tapiador}, {Taris}, {Tauran}, {Taylor}, {Teixeira},
  {Terrett}, {Teyssand ier}, {Thuillot}, {Titarenko}, {Torra Clotet}, {Turon},
  {Ulla}, {Utrilla}, {Uzzi}, {Vaillant}, {Valentini}, {Valette}, {van Elteren},
  {Van Hemelryck}, {van Leeuwen}, {Vaschetto}, {Vecchiato}, {Veljanoski},
  {Viala}, {Vicente}, {Vogt}, {von Essen}, {Voss}, {Votruba}, {Voutsinas},
  {Walmsley}, {Weiler}, {Wertz}, {Wevers}, {Wyrzykowski}, {Yoldas},
  {{\v{Z}}erjal}, {Ziaeepour}, {Zorec}, {Zschocke}, {Zucker}, {Zurbach}, \&
  {Zwitter}}]{GDR2}
{Gaia Collaboration}, {Brown}, A.~G.~A., {Vallenari}, A., {et~al.} 2018, \aap,
  616, A1

\bibitem[{{Gardner} {et~al.}(2006){Gardner}, {Mather}, {Clampin}, {Doyon},
  {Greenhouse}, {Hammel}, {Hutchings}, {Jakobsen}, {Lilly}, {Long}, {Lunine},
  {McCaughrean}, {Mountain}, {Nella}, {Rieke}, {Rieke}, {Rix}, {Smith},
  {Sonneborn}, {Stiavelli}, {Stockman}, {Windhorst}, \& {Wright}}]{gar06}
{Gardner}, J.~P., {Mather}, J.~C., {Clampin}, M., {et~al.} 2006, \ssr, 123, 485

\bibitem[{{Gibson} {et~al.}(2020){Gibson}, {Merritt}, {Nugroho}, {Cubillos},
  {de Mooij}, {Mikal-Evans}, {Fossati}, {Lothringer}, {Nikolov}, {Sing},
  {Spake}, {Watson}, \& {Wilson}}]{gib20}
{Gibson}, N.~P., {Merritt}, S., {Nugroho}, S.~K., {et~al.} 2020, arXiv
  e-prints, arXiv:2001.06430

\bibitem[{{Gustafsson} {et~al.}(2008){Gustafsson}, {Edvardsson}, {Eriksson},
  {J{\o}rgensen}, {Nordlund}, \& {Plez}}]{gus08}
{Gustafsson}, B., {Edvardsson}, B., {Eriksson}, K., {et~al.} 2008, \aap, 486,
  951

\bibitem[{{Heiter} {et~al.}(2015){Heiter}, {Lind}, {Asplund}, {Barklem},
  {Bergemann}, {Magrini}, {Masseron}, {Mikolaitis}, {Pickering}, \&
  {Ruffoni}}]{hei15b}
{Heiter}, U., {Lind}, K., {Asplund}, M., {et~al.} 2015, \physscr, 90, 054010

\bibitem[{{Hoeijmakers} {et~al.}(2018){Hoeijmakers}, {Ehrenreich}, {Heng},
  {Kitzmann}, {Grimm}, {Allart}, {Deitrick}, {Wyttenbach}, {Oreshenko}, {Pino},
  {Rimmer}, {Molinari}, \& {Di Fabrizio}}]{hoe18}
{Hoeijmakers}, H.~J., {Ehrenreich}, D., {Heng}, K., {et~al.} 2018, Nature, 560,
  453

\bibitem[{{Hoeijmakers} {et~al.}(2019){Hoeijmakers}, {Ehrenreich}, {Kitzmann},
  {Allart}, {Grimm}, {Seidel}, {Wyttenbach}, {Pino}, {Nielsen}, {Fisher},
  {Rimmer}, {Bourrier}, {Cegla}, {Lavie}, {Lovis}, {Patzer}, {Stock}, {Pepe},
  \& {Heng}}]{hoe19}
{Hoeijmakers}, H.~J., {Ehrenreich}, D., {Kitzmann}, D., {et~al.} 2019, \aap,
  627, A165

\bibitem[{{Hoeijmakers} {et~al.}(2020){Hoeijmakers}, {Seidel}, {Pino},
  {Kitzmann}, {Sindel}, {Ehrenreich}, {Oza}, {Bourrier}, {Allart}, {Gebek},
  {Lovis}, {Yurchenko}, {Astudillo-Defru}, {Bayliss}, {Cegla}, {Lavie},
  {Lendl}, {Melo}, {Murgas}, {Nascimbeni}, {Pepe}, {S{\'e}gransan}, {Udry},
  {Wyttenbach}, \& {Heng}}]{hoe20}
{Hoeijmakers}, H.~J., {Seidel}, J.~V., {Pino}, L., {et~al.} 2020, arXiv
  e-prints, arXiv:2006.11308

\bibitem[{{Kausch} {et~al.}(2015){Kausch}, {Noll}, {Smette}, {Kimeswenger},
  {Barden}, {Szyszka}, {Jones}, {Sana}, {Horst}, \& {Kerber}}]{kau15}
{Kausch}, W., {Noll}, S., {Smette}, A., {et~al.} 2015, \aap, 576, A78

\bibitem[{{Kurucz}(1993)}]{kur93}
{Kurucz}, R. 1993, ATLAS9 Stellar Atmosphere Programs and 2 km/s grid. Kurucz
  CD-ROM No. 13. Cambridge, 13

\bibitem[{Mahadevan {et~al.}(2014)Mahadevan, Ramsey, Terrien, Halverson, Roy,
  Hearty, Levi, Stefansson, Robertson, Bender, Schwab, \& Nelson}]{hpf}
Mahadevan, S., Ramsey, L.~W., Terrien, R., {et~al.} 2014, in Ground-based and
  Airborne Instrumentation for Astronomy V, ed. S.~K. Ramsay, I.~S. McLean, \&
  H.~Takami, Vol. 9147, International Society for Optics and Photonics (SPIE),
  543 -- 552

\bibitem[{{Malik} {et~al.}(2017){Malik}, {Grosheintz}, {Mendon{\c{c}}a},
  {Grimm}, {Lavie}, {Kitzmann}, {Tsai}, {Burrows}, {Kreidberg}, {Bedell},
  {Bean}, {Stevenson}, \& {Heng}}]{mal17}
{Malik}, M., {Grosheintz}, L., {Mendon{\c{c}}a}, J.~M., {et~al.} 2017, \aj,
  153, 56

\bibitem[{{Malik} {et~al.}(2019){Malik}, {Kitzmann}, {Mendon{\c{c}}a}, {Grimm},
  {Marleau}, {Linder}, {Tsai}, \& {Heng}}]{mal19}
{Malik}, M., {Kitzmann}, D., {Mendon{\c{c}}a}, J.~M., {et~al.} 2019, \aj, 157,
  170

\bibitem[{{Mayor} {et~al.}(2003){Mayor}, {Pepe}, {Queloz}, {Bouchy},
  {Rupprecht}, {Lo Curto}, {Avila}, {Benz}, {Bertaux}, {Bonfils}, {Dall},
  {Dekker}, {Delabre}, {Eckert}, {Fleury}, {Gilliotte}, {Gojak}, {Guzman},
  {Kohler}, {Lizon}, {Longinotti}, {Lovis}, {Megevand}, {Pasquini}, {Reyes},
  {Sivan}, {Sosnowska}, {Soto}, {Udry}, {van Kesteren}, {Weber}, \&
  {Weilenmann}}]{may03}
{Mayor}, M., {Pepe}, F., {Queloz}, D., {et~al.} 2003, The Messenger, 114, 20

\bibitem[{{McKemmish} {et~al.}(2019){McKemmish}, {Masseron}, {Hoeijmakers},
  {P{\'e}rez-Mesa}, {Grimm}, {Yurchenko}, \& {Tennyson}}]{tioexomol}
{McKemmish}, L.~K., {Masseron}, T., {Hoeijmakers}, H.~J., {et~al.} 2019,
  \mnras, 488, 2836

\bibitem[{{McKemmish} {et~al.}(2016){McKemmish}, {Yurchenko}, \&
  {Tennyson}}]{voexomol}
{McKemmish}, L.~K., {Yurchenko}, S.~N., \& {Tennyson}, J. 2016, \mnras, 463,
  771

\bibitem[{{Moutou} {et~al.}(2015){Moutou}, {Boisse}, {H{\'e}brard},
  {H{\'e}brard}, {Donati}, {Delfosse}, \& {Kouach}}]{mou15}
{Moutou}, C., {Boisse}, I., {H{\'e}brard}, G., {et~al.} 2015, in SF2A-2015:
  Proceedings of the Annual meeting of the French Society of Astronomy and
  Astrophysics, 205--212

\bibitem[{{Nortmann} {et~al.}(2018){Nortmann}, {Pall{\'e}}, {Salz},
  {Sanz-Forcada}, {Nagel}, {Alonso-Floriano}, {Czesla}, {Yan}, {Chen},
  {Snellen}, {Zechmeister}, {Schmitt}, {L{\'o}pez-Puertas}, {Casasayas-Barris},
  {Bauer}, {Amado}, {Caballero}, {Dreizler}, {Henning}, {Lamp{\'o}n}, {Montes},
  {Molaverdikhani}, {Quirrenbach}, {Reiners}, {Ribas}, {S{\'a}nchez-L{\'o}pez},
  {Schneider}, \& {Zapatero Osorio}}]{nor18}
{Nortmann}, L., {Pall{\'e}}, E., {Salz}, M., {et~al.} 2018, Science, 362, 1388

\bibitem[{{Nugroho} {et~al.}(2017){Nugroho}, {Kawahara}, {Masuda}, {Hirano},
  {Kotani}, \& {Tajitsu}}]{nug17}
{Nugroho}, S.~K., {Kawahara}, H., {Masuda}, K., {et~al.} 2017, \aj, 154, 221

\bibitem[{{Parmentier} {et~al.}(2018){Parmentier}, {Line}, {Bean}, {Mansfield},
  {Kreidberg}, {Lupu}, {Visscher}, {D{\'e}sert}, {Fortney}, {Deleuil},
  {Arcangeli}, {Showman}, \& {Marley}}]{par18}
{Parmentier}, V., {Line}, M.~R., {Bean}, J.~L., {et~al.} 2018, \aap, 617, A110

\bibitem[{{Pepe} {et~al.}(2020){Pepe}, {Cristiani}, {Rebolo}, {Santos},
  {Dekker}, {Cabral}, {Di Marcantonio}, {Figueira}, {Lo Curto}, {Lovis},
  {Mayor}, {M{\'e}gevand}, {Molaro}, {Riva}, {Zapatero Osorio}, {Amate},
  {Manescau}, {Pasquini}, {Zerbi}, {Adibekyan}, {Abreu}, {Affolter}, {Alibert},
  {Aliverti}, {Allart}, {Allende Prieto}, {{\'A}lvarez}, {Alves}, {Avila},
  {Baldini}, {Bandy}, {Barros}, {Benz}, {Bianco}, {Borsa}, {Bourrier},
  {Bouchy}, {Broeg}, {Calderone}, {Cirami}, {Coelho}, {Conconi}, {Coretti},
  {Cumani}, {Cupani}, {D'Odorico}, {Damasso}, {Deiries}, {Delabre},
  {Demangeon}, {Dumusque}, {Ehrenreich}, {Faria}, {Fragoso}, {Genolet},
  {Genoni}, {G{\'e}nova Santos}, {Gonz{\'a}lez Hern{\'a}ndez}, {Hughes},
  {Iwert}, {Kerber}, {Knudstrup}, {Landoni}, {Lavie}, {Lillo-Box}, {Lizon},
  {Maire}, {Martins}, {Mehner}, {Micela}, {Modigliani}, {Monteiro}, {Monteiro},
  {Moschetti}, {Murphy}, {Nunes}, {Oggioni}, {Oliveira}, {Oshagh}, {Pall{\'e}},
  {Pariani}, {Poretti}, {Rasilla}, {Rebord{\~a}o}, {Redaelli}, {Santana
  Tschudi}, {Santin}, {Santos}, {S{\'e}gransan}, {Schmidt}, {Segovia},
  {Sosnowska}, {Sozzetti}, {Sousa}, {Span{\`o}}, {Su{\'a}rez Mascare{\~n}o},
  {Tabernero}, {Tenegi}, {Udry}, \& {Zanutta}}]{pepe20}
{Pepe}, F., {Cristiani}, S., {Rebolo}, R., {et~al.} 2020, arXiv e-prints,
  arXiv:2010.00316

\bibitem[{{Pepe} {et~al.}(2002){Pepe}, {Mayor}, {Galland}, {Naef}, {Queloz},
  {Santos}, {Udry}, \& {Burnet}}]{pepe02}
{Pepe}, F., {Mayor}, M., {Galland}, F., {et~al.} 2002, \aap, 388, 632

\bibitem[{{Pepe} {et~al.}(2010){Pepe}, {Cristiani}, {Rebolo Lopez}, {Santos},
  {Amorim}, {Avila}, {Benz}, {Bonifacio}, {Cabral}, {Carvas}, {Cirami},
  {Coelho}, {Comari}, {Coretti}, {De Caprio}, {Dekker}, {Delabre}, {Di
  Marcantonio}, {D'Odorico}, {Fleury}, {Garc{\'{\i}}a}, {Herreros Linares},
  {Hughes}, {Iwert}, {Lima}, {Lizon}, {Lo Curto}, {Lovis}, {Manescau},
  {Martins}, {M{\'e}gevand}, {Moitinho}, {Molaro}, {Monteiro}, {Monteiro},
  {Pasquini}, {Mordasini}, {Queloz}, {Rasilla}, {Rebord{\~a}o}, {Santana
  Tschudi}, {Santin}, {Sosnowska}, {Span{\`o}}, {Tenegi}, {Udry}, {Vanzella},
  {Viel}, {Zapatero Osorio}, \& {Zerbi}}]{pepe10}
{Pepe}, F.~A., {Cristiani}, S., {Rebolo Lopez}, R., {et~al.} 2010, in
  \procspie, Vol. 7735, Ground-based and Airborne Instrumentation for Astronomy
  III, 77350F

\bibitem[{{Plez}(1998)}]{tioplez}
{Plez}, B. 1998, \aap, 337, 495

\bibitem[{{Plez}(2003)}]{zroplez}
{Plez}, B. 2003, Astronomical Society of the Pacific Conference Series, Vol.
  298, {Cool star atmospheres and spectra for GAIA: MARCS models}, ed.
  U.~{Munari}, 189

\bibitem[{{Plez}(2012)}]{ple12}
{Plez}, B. 2012, {Turbospectrum: Code for spectral synthesis}, Astrophysics
  Source Code Library

\bibitem[{{Press} {et~al.}(2002){Press}, {Teukolsky}, {Vetterling}, \&
  {Flannery}}]{pre02}
{Press}, W.~H., {Teukolsky}, S.~A., {Vetterling}, W.~T., \& {Flannery}, B.~P.
  2002, {Numerical recipes in C++ : the art of scientific computing}

\bibitem[{{Quirrenbach} {et~al.}(2016){Quirrenbach}, {Amado}, {Caballero},
  {Mundt}, {Reiners}, {Ribas}, {Seifert}, {Abril}, {Aceituno},
  {Alonso-Floriano}, {Anwand-Heerwart}, {Azzaro}, {Bauer}, {Barrado},
  {Becerril}, {Bejar}, {Benitez}, {Berdinas}, {Brinkm{\"o}ller}, {Cardenas},
  {Casal}, {Claret}, {Colom{\'e}}, {Cortes-Contreras}, {Czesla}, {Doellinger},
  {Dreizler}, {Feiz}, {Fernandez}, {Ferro}, {Fuhrmeister}, {Galadi},
  {Gallardo}, {G{\'a}lvez-Ortiz}, {Garcia-Piquer}, {Garrido}, {Gesa},
  {G{\'o}mez Galera}, {Gonz{\'a}lez Hern{\'a}ndez}, {Gonzalez Peinado},
  {Gr{\"o}zinger}, {Gu{\`a}rdia}, {Guenther}, {de Guindos}, {Hagen}, {Hatzes},
  {Hauschildt}, {Helmling}, {Henning}, {Hermann}, {Hern{\'a}ndez Arabi},
  {Hern{\'a}ndez Casta{\~n}o}, {Hern{\'a}ndez Hernando}, {Herrero}, {Huber},
  {Huber}, {Huke}, {Jeffers}, {de Juan}, {Kaminski}, {Kehr}, {Kim}, {Klein},
  {Kl{\"u}ter}, {K{\"u}rster}, {Lafarga}, {Lara}, {Lamert}, {Laun},
  {Launhardt}, {Lemke}, {Lenzen}, {Llamas}, {Lopez del Fresno},
  {L{\'o}pez-Puertas}, {L{\'o}pez-Santiago}, {Lopez Salas}, {Magan
  Madinabeitia}, {Mall}, {Mandel}, {Mancini}, {Marin Molina}, {Maroto
  Fern{\'a}ndez}, {Mart{\'{\i}}n}, {Mart{\'{\i}}n-Ruiz}, {Marvin}, {Mathar},
  {Mirabet}, {Montes}, {Morales}, {Morales Mu{\~n}oz}, {Nagel}, {Naranjo},
  {Nowak}, {Palle}, {Panduro}, {Passegger}, {Pavlov}, {Pedraz}, {Perez},
  {P{\'e}rez-Medialdea}, {Perger}, {Pluto}, {Ram{\'o}n}, {Rebolo}, {Redondo},
  {Reffert}, {Reinhart}, {Rhode}, {Rix}, {Rodler}, {Rodr{\'{\i}}guez},
  {Rodr{\'{\i}}guez L{\'o}pez}, {Rohloff}, {Rosich}, {Sanchez Carrasco},
  {Sanz-Forcada}, {Sarkis}, {Sarmiento}, {Sch{\"a}fer}, {Schiller}, {Schmidt},
  {Schmitt}, {Sch{\"o}fer}, {Schweitzer}, {Shulyak}, {Solano}, {Stahl},
  {Storz}, {Tabernero}, {Tala}, {Tal-Or}, {Ulbrich}, {Veredas}, {Vico Linares},
  {Vilardell}, {Wagner}, {Winkler}, {Zapatero Osorio}, {Zechmeister},
  {Ammler-von Eiff}, {Anglada-Escud{\'e}}, {del Burgo}, {Garcia-Vargas},
  {Klutsch}, {Lizon}, {Lopez-Morales}, {Ofir}, {P{\'e}rez-Calpena}, {Perryman},
  {S{\'a}nchez-Blanco}, {Strachan}, {St{\"u}rmer}, {Su{\'a}rez}, {Trifonov},
  {Tulloch}, \& {Xu}}]{quir16}
{Quirrenbach}, A., {Amado}, P.~J., {Caballero}, J.~A., {et~al.} 2016, in
  \procspie, Vol. 9908, Ground-based and Airborne Instrumentation for Astronomy
  VI, 990812

\bibitem[{{Rando} {et~al.}(2018){Rando}, {Asquier}, {Corral Van Damme},
  {Isaak}, {Ratti}, {Verhoeve}, {Safa}, {Southworth}, {Broeg}, \&
  {Benz}}]{ran18}
{Rando}, N., {Asquier}, J., {Corral Van Damme}, C., {et~al.} 2018, in Society
  of Photo-Optical Instrumentation Engineers (SPIE) Conference Series, Vol.
  10698, \procspie, 106980K

\bibitem[{{Rauer} {et~al.}(2014){Rauer}, {Catala}, {Aerts}, {Appourchaux},
  {Benz}, {Brandeker}, {Christensen-Dalsgaard}, {Deleuil}, {Gizon}, {Goupil},
  {G{\"u}del}, {Janot-Pacheco}, {Mas-Hesse}, {Pagano}, {Piotto}, {Pollacco},
  {Santos}, {Smith}, {Su{\'a}rez}, {Szab{\'o}}, {Udry}, {Adibekyan}, {Alibert},
  {Almenara}, {Amaro-Seoane}, {Eiff}, {Asplund}, {Antonello}, {Barnes},
  {Baudin}, {Belkacem}, {Bergemann}, {Bihain}, {Birch}, {Bonfils}, {Boisse},
  {Bonomo}, {Borsa}, {Brand {\~a}o}, {Brocato}, {Brun}, {Burleigh}, {Burston},
  {Cabrera}, {Cassisi}, {Chaplin}, {Charpinet}, {Chiappini}, {Church},
  {Csizmadia}, {Cunha}, {Damasso}, {Davies}, {Deeg}, {D{\'\i}az}, {Dreizler},
  {Dreyer}, {Eggenberger}, {Ehrenreich}, {Eigm{\"u}ller}, {Erikson}, {Farmer},
  {Feltzing}, {de Oliveira Fialho}, {Figueira}, {Forveille}, {Fridlund},
  {Garc{\'\i}a}, {Giommi}, {Giuffrida}, {Godolt}, {Gomes da Silva}, {Granzer},
  {Grenfell}, {Grotsch-Noels}, {G{\"u}nther}, {Haswell}, {Hatzes},
  {H{\'e}brard}, {Hekker}, {Helled}, {Heng}, {Jenkins}, {Johansen},
  {Khodachenko}, {Kislyakova}, {Kley}, {Kolb}, {Krivova}, {Kupka}, {Lammer},
  {Lanza}, {Lebreton}, {Magrin}, {Marcos-Arenal}, {Marrese}, {Marques},
  {Martins}, {Mathis}, {Mathur}, {Messina}, {Miglio}, {Montalban}, {Montalto},
  {Monteiro}, {Moradi}, {Moravveji}, {Mordasini}, {Morel}, {Mortier},
  {Nascimbeni}, {Nelson}, {Nielsen}, {Noack}, {Norton}, {Ofir}, {Oshagh},
  {Ouazzani}, {P{\'a}pics}, {Parro}, {Petit}, {Plez}, {Poretti}, {Quirrenbach},
  {Ragazzoni}, {Raimondo}, {Rainer}, {Reese}, {Redmer}, {Reffert},
  {Rojas-Ayala}, {Roxburgh}, {Salmon}, {Santerne}, {Schneider}, {Schou},
  {Schuh}, {Schunker}, {Silva-Valio}, {Silvotti}, {Skillen}, {Snellen}, {Sohl},
  {Sousa}, {Sozzetti}, {Stello}, {Strassmeier}, {{\v{S}}vanda}, {Szab{\'o}},
  {Tkachenko}, {Valencia}, {Van Grootel}, {Vauclair}, {Ventura}, {Wagner},
  {Walton}, {Weingrill}, {Werner}, {Wheatley}, \& {Zwintz}}]{rau14}
{Rauer}, H., {Catala}, C., {Aerts}, C., {et~al.} 2014, Experimental Astronomy,
  38, 249

\bibitem[{{Redfield} {et~al.}(2008){Redfield}, {Endl}, {Cochran}, \&
  {Koesterke}}]{red08}
{Redfield}, S., {Endl}, M., {Cochran}, W.~D., \& {Koesterke}, L. 2008, The
  Astrophysical Journal, 673, L87

\bibitem[{{Ricker}(2014)}]{ric14}
{Ricker}, G.~R. 2014, Journal of the American Association of Variable Star
  Observers (JAAVSO), 42, 234

\bibitem[{{Ryabchikova} {et~al.}(2015){Ryabchikova}, {Piskunov}, {Kurucz},
  {Stempels}, {Heiter}, {Pakhomov}, \& {Barklem}}]{rad15}
{Ryabchikova}, T., {Piskunov}, N., {Kurucz}, R.~L., {et~al.} 2015, \physscr,
  90, 054005

\bibitem[{{Santos} {et~al.}(2013){Santos}, {Sousa}, {Mortier}, {Neves},
  {Adibekyan}, {Tsantaki}, {Delgado Mena}, {Bonfils}, {Israelian}, {Mayor}, \&
  {Udry}}]{san13}
{Santos}, N.~C., {Sousa}, S.~G., {Mortier}, A., {et~al.} 2013, \aap, 556, A150

\bibitem[{{Seidel} {et~al.}(2019){Seidel}, {Ehrenreich}, {Wyttenbach},
  {Allart}, {Lendl}, {Pino}, {Bourrier}, {Cegla}, {Lovis}, {Barrado},
  {Bayliss}, {Astudillo-Defru}, {Deline}, {Fisher}, {Heng}, {Joseph}, {Lavie},
  {Melo}, {Pepe}, {S{\'e}gransan}, \& {Udry}}]{sei19}
{Seidel}, J.~V., {Ehrenreich}, D., {Wyttenbach}, A., {et~al.} 2019, \aap, 623,
  A166

\bibitem[{{Sheppard} {et~al.}(2017){Sheppard}, {Mandell}, {Tamburo}, {Gand hi},
  {Pinhas}, {Madhusudhan}, \& {Deming}}]{she17}
{Sheppard}, K.~B., {Mandell}, A.~M., {Tamburo}, P., {et~al.} 2017, \apjl, 850,
  L32

\bibitem[{{Sitnova} {et~al.}(2013){Sitnova}, {Mashonkina}, \&
  {Ryabchikova}}]{sit13}
{Sitnova}, T.~M., {Mashonkina}, L.~I., \& {Ryabchikova}, T.~A. 2013, Astronomy
  Letters, 39, 126

\bibitem[{{Smette} {et~al.}(2015){Smette}, {Sana}, {Noll}, {Horst}, {Kausch},
  {Kimeswenger}, {Barden}, {Szyszka}, {Jones}, {Gallenne}, {Vinther},
  {Ballester}, \& {Taylor}}]{sme15}
{Smette}, A., {Sana}, H., {Noll}, S., {et~al.} 2015, \aap, 576, A77

\bibitem[{{Sneden}(1973)}]{sne73}
{Sneden}, C. 1973, \apj, 184, 839

\bibitem[{{Snellen} {et~al.}(2008){Snellen}, {Albrecht}, {de Mooij}, \& {Le
  Poole}}]{sne08}
{Snellen}, I.~A.~G., {Albrecht}, S., {de Mooij}, E.~J.~W., \& {Le Poole}, R.~S.
  2008, \aap, 487, 357

\bibitem[{{Sousa} {et~al.}(2018){Sousa}, {Adibekyan}, {Delgado-Mena}, {Santos},
  {Andreasen}, {Ferreira}, {Tsantaki}, {Barros}, {Demangeon}, {Israelian},
  {Faria}, {Figueira}, {Mortier}, {Brand{\~a}o}, {Montalto}, {Rojas-Ayala}, \&
  {Santerne}}]{sou18}
{Sousa}, S.~G., {Adibekyan}, V., {Delgado-Mena}, E., {et~al.} 2018, \aap, 620,
  A58

\bibitem[{{Sousa} {et~al.}(2015){Sousa}, {Santos}, {Adibekyan}, {Delgado-Mena},
  \& {Israelian}}]{sou15}
{Sousa}, S.~G., {Santos}, N.~C., {Adibekyan}, V., {Delgado-Mena}, E., \&
  {Israelian}, G. 2015, \aap, 577, A67

\bibitem[{{Sousa} {et~al.}(2008){Sousa}, {Santos}, {Mayor}, {Udry},
  {Casagrande}, {Israelian}, {Pepe}, {Queloz}, \& {Monteiro}}]{sou08}
{Sousa}, S.~G., {Santos}, N.~C., {Mayor}, M., {et~al.} 2008, \aap, 487, 373

\bibitem[{{Tabernero} {et~al.}(2020){Tabernero}, {Allende Prieto}, {Zapatero
  Osorio}, {Gonz{\'a}lez Hern{\'a}ndez}, {del Burgo}, {Garc{\'\i}a L{\'o}pez},
  {Rebolo}, {Abril-Abril}, {Barreto}, {Calvo Tovar}, {D{\'\i}az Torres},
  {Fern{\'a}ndez Izquierdo}, {G{\'o}mez-Re{\~n}asco}, {Gracia-T{\'e}mich},
  {Joven}, {Pe{\~n}ate Castro}, {Santana-Tschudi}, {Tenegi}, \& {Viera
  Mart{\'\i}n}}]{tab20}
{Tabernero}, H.~M., {Allende Prieto}, C., {Zapatero Osorio}, M.~R., {et~al.}
  2020, \mnras

\bibitem[{{Tabernero} {et~al.}(2019){Tabernero}, {Marfil}, {Montes}, \&
  {Gonz{\'a}lez Hern{\'a}ndez}}]{tab19}
{Tabernero}, H.~M., {Marfil}, E., {Montes}, D., \& {Gonz{\'a}lez
  Hern{\'a}ndez}, J.~I. 2019, \aap, 628, A131

\bibitem[{Tinetti {et~al.}(2016)Tinetti, Drossart, Eccleston, Hartogh, Heske,
  Leconte, Micela, Ollivier, Pilbratt, Puig, Turrini, Vandenbussche,
  Wolkenberg, Pascale, Beaulieu, Güdel, Min, Rataj, Ray, Ribas, Barstow,
  Bowles, Coustenis, du~Foresto, Decin, Encrenaz, Forget, Friswell, Griffin,
  Lagage, Malaguti, Moneti, Morales, Pace, Rocchetto, Sarkar, Selsis, Taylor,
  Tennyson, Venot, Waldmann, Wright, Zingales, \& Zapatero-Osorio}]{ariel}
Tinetti, G., Drossart, P., Eccleston, P., {et~al.} 2016, in Space Telescopes
  and Instrumentation 2016: Optical, Infrared, and Millimeter Wave, ed. H.~A.
  MacEwen, G.~G. Fazio, M.~Lystrup, N.~Batalha, N.~Siegler, \& E.~C. Tong, Vol.
  9904, International Society for Optics and Photonics (SPIE), 658 -- 667

\bibitem[{{Van Eck} {et~al.}(2017){Van Eck}, {Neyskens}, {Jorissen}, {Plez},
  {Edvardsson}, {Eriksson}, {Gustafsson}, {J{\o}rgensen}, \&
  {Nordlund}}]{vaneckgrid}
{Van Eck}, S., {Neyskens}, P., {Jorissen}, A., {et~al.} 2017, \aap, 601, A10

\bibitem[{{Virtanen} {et~al.}(2020){Virtanen}, {Gommers}, {Oliphant},
  {Haberland}, {Reddy}, {Cournapeau}, {Burovski}, {Peterson}, {Weckesser},
  {Bright}, {van der Walt}, {Brett}, {Wilson}, {Jarrod Millman}, {Mayorov},
  {Nelson}, {Jones}, {Kern}, {Larson}, {Carey}, {Polat}, {Feng}, {Moore}, {Vand
  erPlas}, {Laxalde}, {Perktold}, {Cimrman}, {Henriksen}, {Quintero}, {Harris},
  {Archibald}, {Ribeiro}, {Pedregosa}, {van Mulbregt}, \&
  {Contributors}}]{scipy}
{Virtanen}, P., {Gommers}, R., {Oliphant}, T.~E., {et~al.} 2020, Nature
  Methods, 17, 261

\bibitem[{{Vogt} {et~al.}(1994){Vogt}, {Allen}, {Bigelow}, {Bresee}, {Brown},
  {Cantrall}, {Conrad}, {Couture}, {Delaney}, {Epps}, {Hilyard}, {Hilyard},
  {Horn}, {Jern}, {Kanto}, {Keane}, {Kibrick}, {Lewis}, {Osborne},
  {Pardeilhan}, {Pfister}, {Ricketts}, {Robinson}, {Stover}, {Tucker}, {Ward},
  \& {Wei}}]{vog94}
{Vogt}, S.~S., {Allen}, S.~L., {Bigelow}, B.~C., {et~al.} 1994, Society of
  Photo-Optical Instrumentation Engineers (SPIE) Conference Series, Vol. 2198,
  {HIRES: the high-resolution echelle spectrometer on the Keck 10-m Telescope},
  ed. D.~L. {Crawford} \& E.~R. {Craine}, 362

\bibitem[{{von Essen} {et~al.}(2020){von Essen}, {Mallonn}, {Hermansen},
  {Nixon}, {Madhusudhan}, {Kjeldsen}, \& {Tautvai{\v{s}}ien{\.{e}}}}]{ess20}
{von Essen}, C., {Mallonn}, M., {Hermansen}, S., {et~al.} 2020, \aap, 637, A76

\bibitem[{{West} {et~al.}(2016){West}, {Hellier}, {Almenara}, {Anderson},
  {Barros}, {Bouchy}, {Brown}, {Collier Cameron}, {Deleuil}, {Delrez}, {Doyle},
  {Faedi}, {Fumel}, {Gillon}, {G{\'o}mez Maqueo Chew}, {H{\'e}brard}, {Jehin},
  {Lendl}, {Maxted}, {Pepe}, {Pollacco}, {Queloz}, {S{\'e}gransan}, {Smalley},
  {Smith}, {Southworth}, {Triaud}, \& {Udry}}]{wes16}
{West}, R.~G., {Hellier}, C., {Almenara}, J.-M., {et~al.} 2016, \aap, 585, A126

\bibitem[{Wildi {et~al.}(2017)Wildi, Blind, Reshetov, Hernandez, Genolet,
  Conod, Sordet, Segovilla, Rasilla, Brousseau, Thibault, Delabre, Bandy,
  Sarajlic, Cabral, Bovay, Vallée, Bouchy, Doyon, Artigau, Pepe, Hagelberg,
  Melo, Delfosse, Figueira, Santos, Hernández, de~Medeiros, Rebolo, Broeg,
  Benz, Boisse, Malo, Käufl, \& Saddlemyer}]{wil17}
Wildi, F., Blind, N., Reshetov, V., {et~al.} 2017, in Techniques and
  Instrumentation for Detection of Exoplanets VIII, ed. S.~Shaklan, Vol. 10400,
  International Society for Optics and Photonics (SPIE), 321 -- 335

\bibitem[{{Wyttenbach} {et~al.}(2015){Wyttenbach}, {Ehrenreich}, {Lovis},
  {Udry}, \& {Pepe}}]{wyt15}
{Wyttenbach}, A., {Ehrenreich}, D., {Lovis}, C., {Udry}, S., \& {Pepe}, F.
  2015, \aap, 577, A62

\bibitem[{{Yan} {et~al.}(2019){Yan}, {Casasayas-Barris}, {Molaverdikhani},
  {Alonso-Floriano}, {Reiners}, {Pall{\'e}}, {Henning}, {Molli{\`e}re}, {Chen},
  {Nortmann}, {Snellen}, {Ribas}, {Quirrenbach}, {Caballero}, {Amado},
  {Azzaro}, {Bauer}, {Cort{\'e}s Contreras}, {Czesla}, {Khalafinejad}, {Lara},
  {L{\'o}pez-Puertas}, {Montes}, {Nagel}, {Oshagh}, {S{\'a}nchez-L{\'o}pez},
  {Stangret}, \& {Zechmeister}}]{yan19}
{Yan}, F., {Casasayas-Barris}, N., {Molaverdikhani}, K., {et~al.} 2019, \aap,
  632, A69

\end{thebibliography}
                         
\newpage
\begin{appendix}
\newpage
\section{Appendix}

\begin{figure*}

\centering                                                                                                                                                                                            
\includegraphics[width=0.38\textwidth]{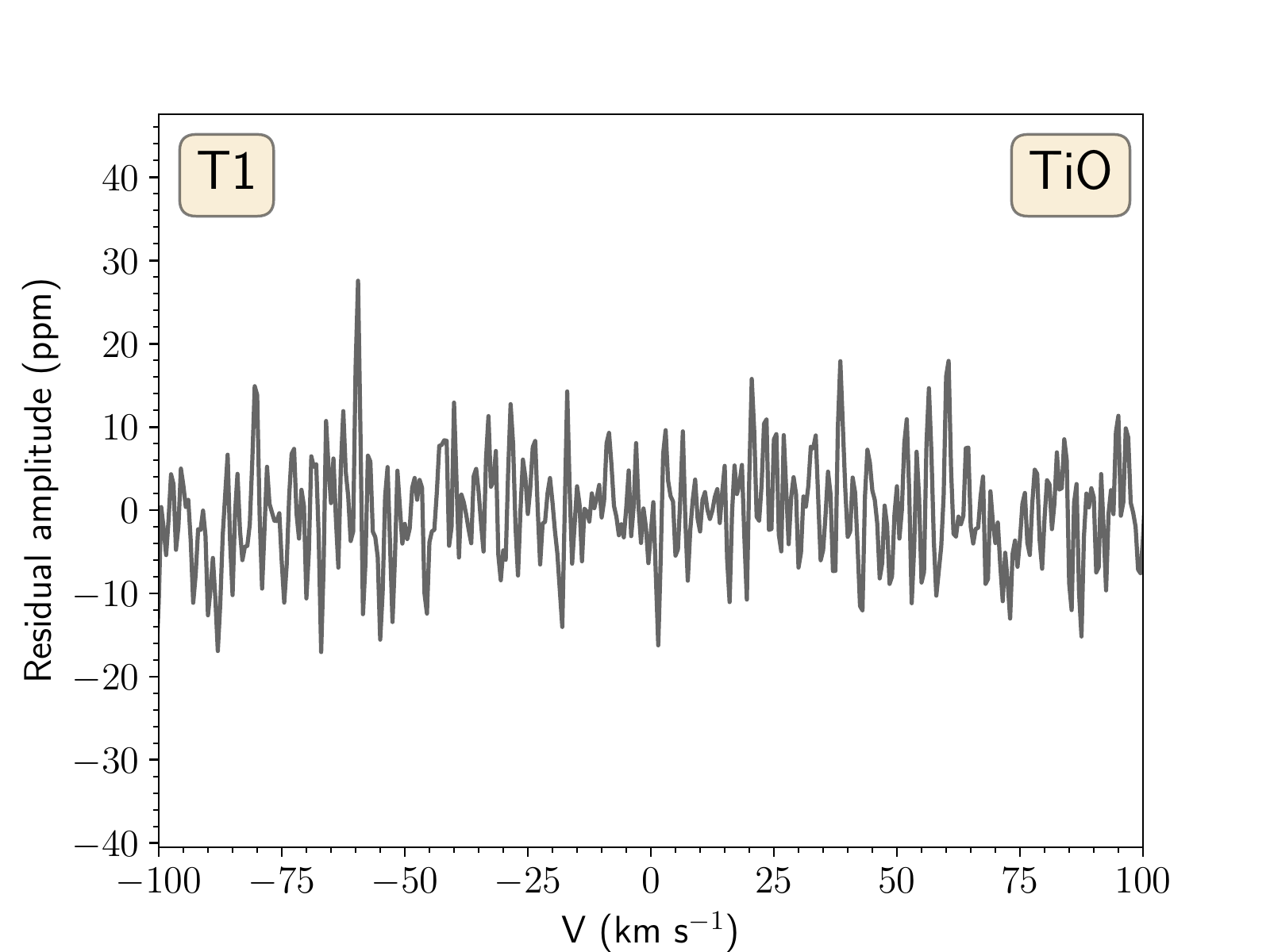} 
\includegraphics[width=0.38\textwidth]{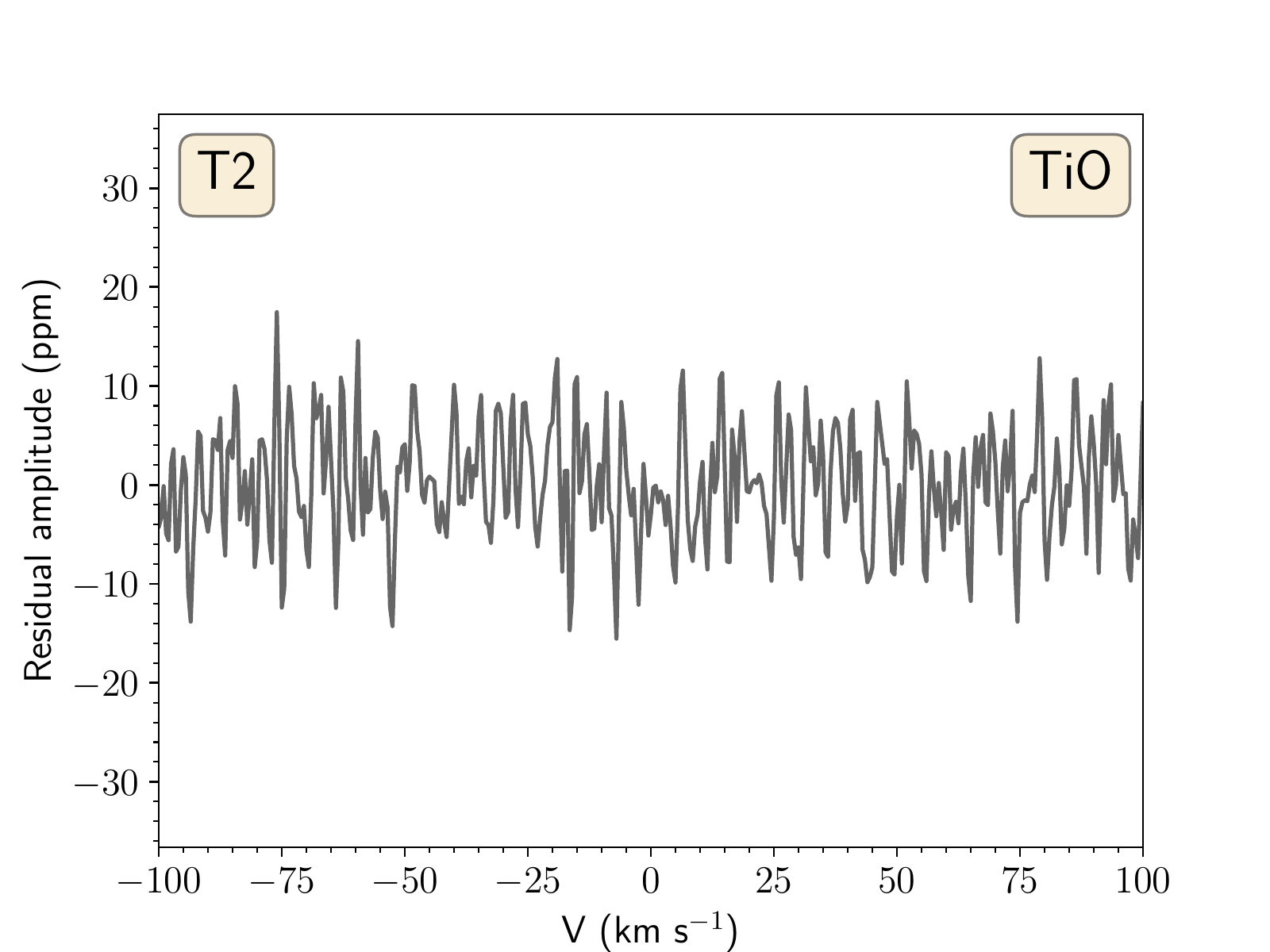} 
\includegraphics[width=0.38\textwidth]{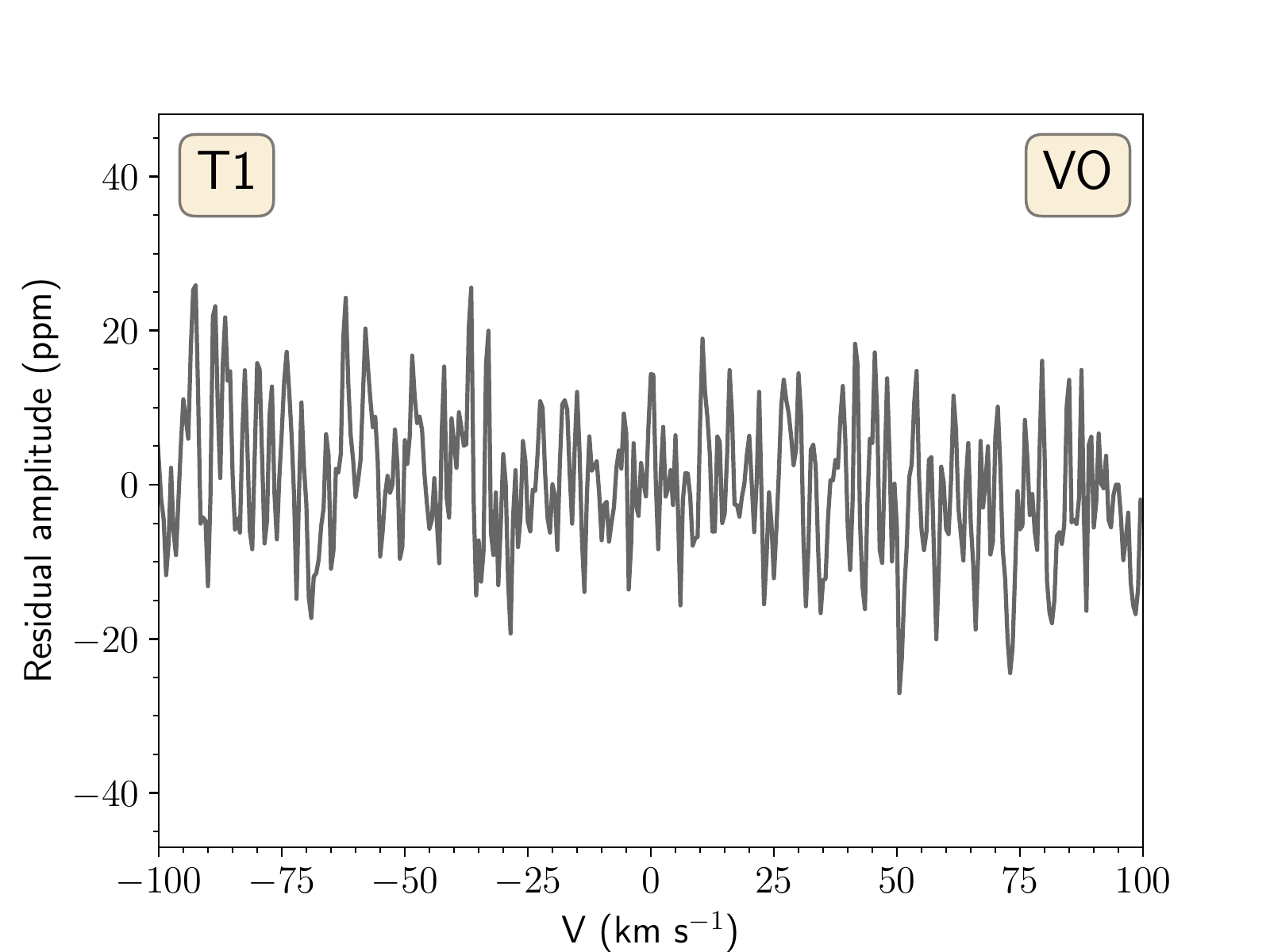}                                                                                                                                                   
\includegraphics[width=0.38\textwidth]{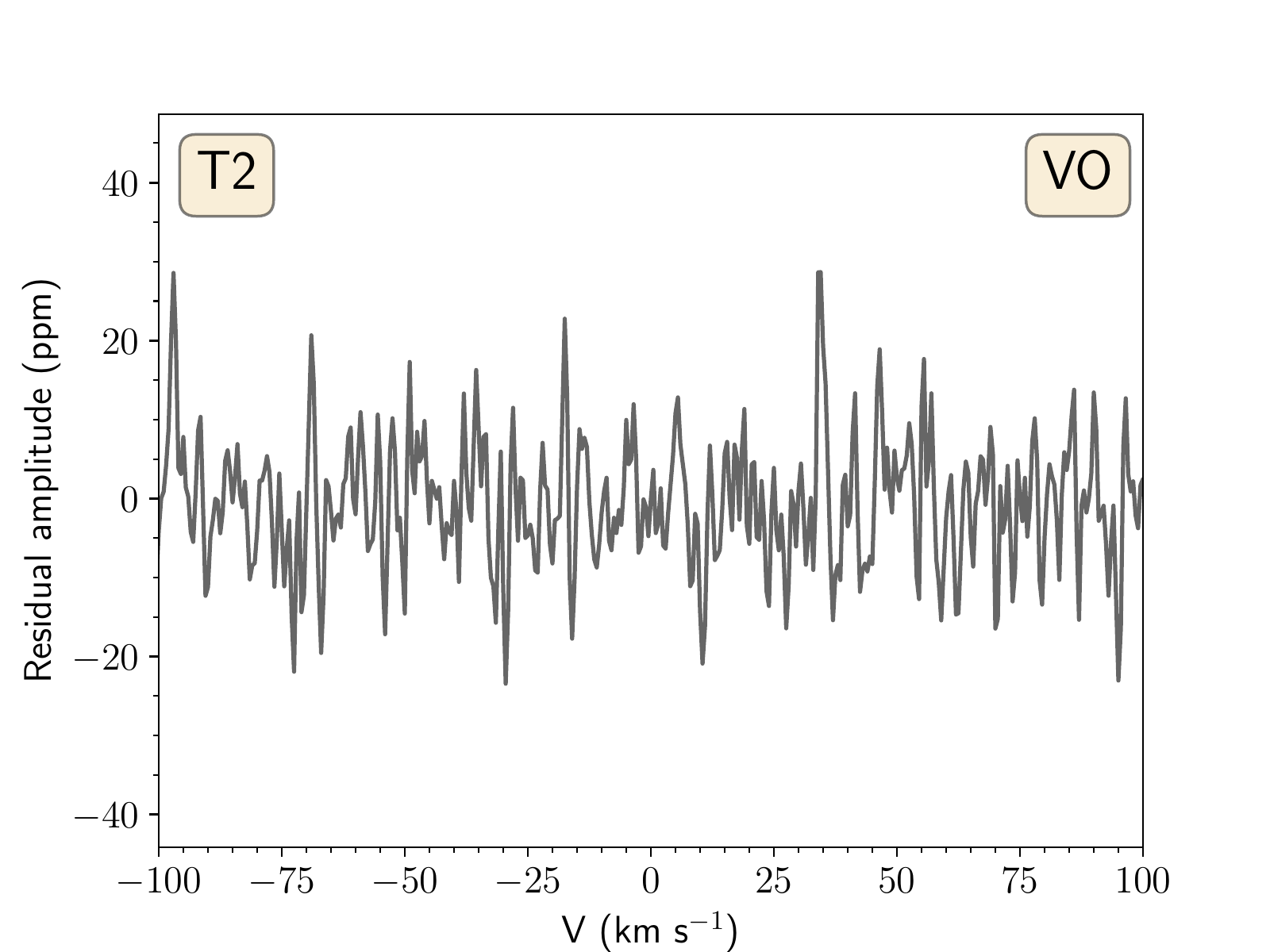}
\includegraphics[width=0.38\textwidth]{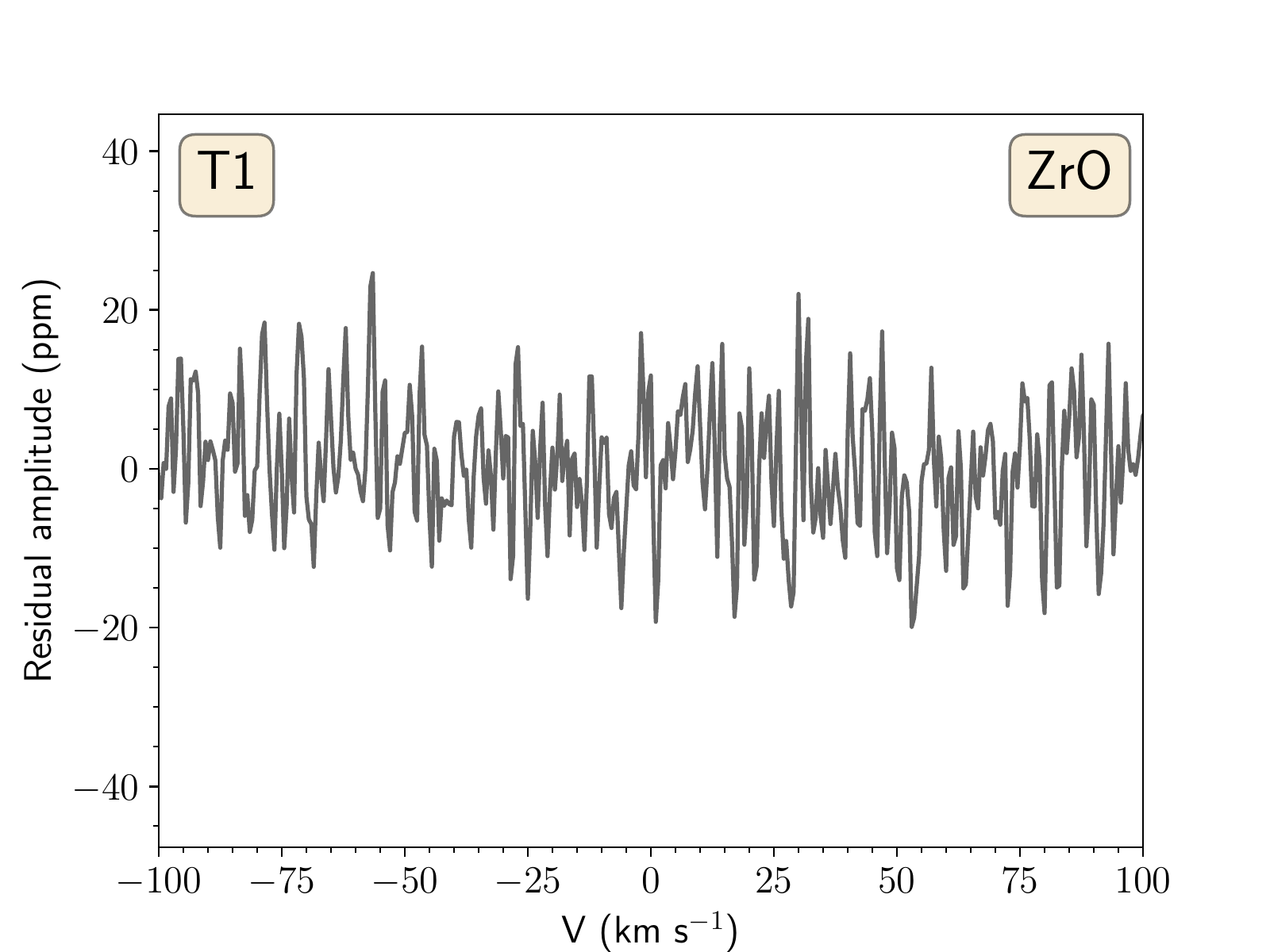} 
\includegraphics[width=0.38\textwidth]{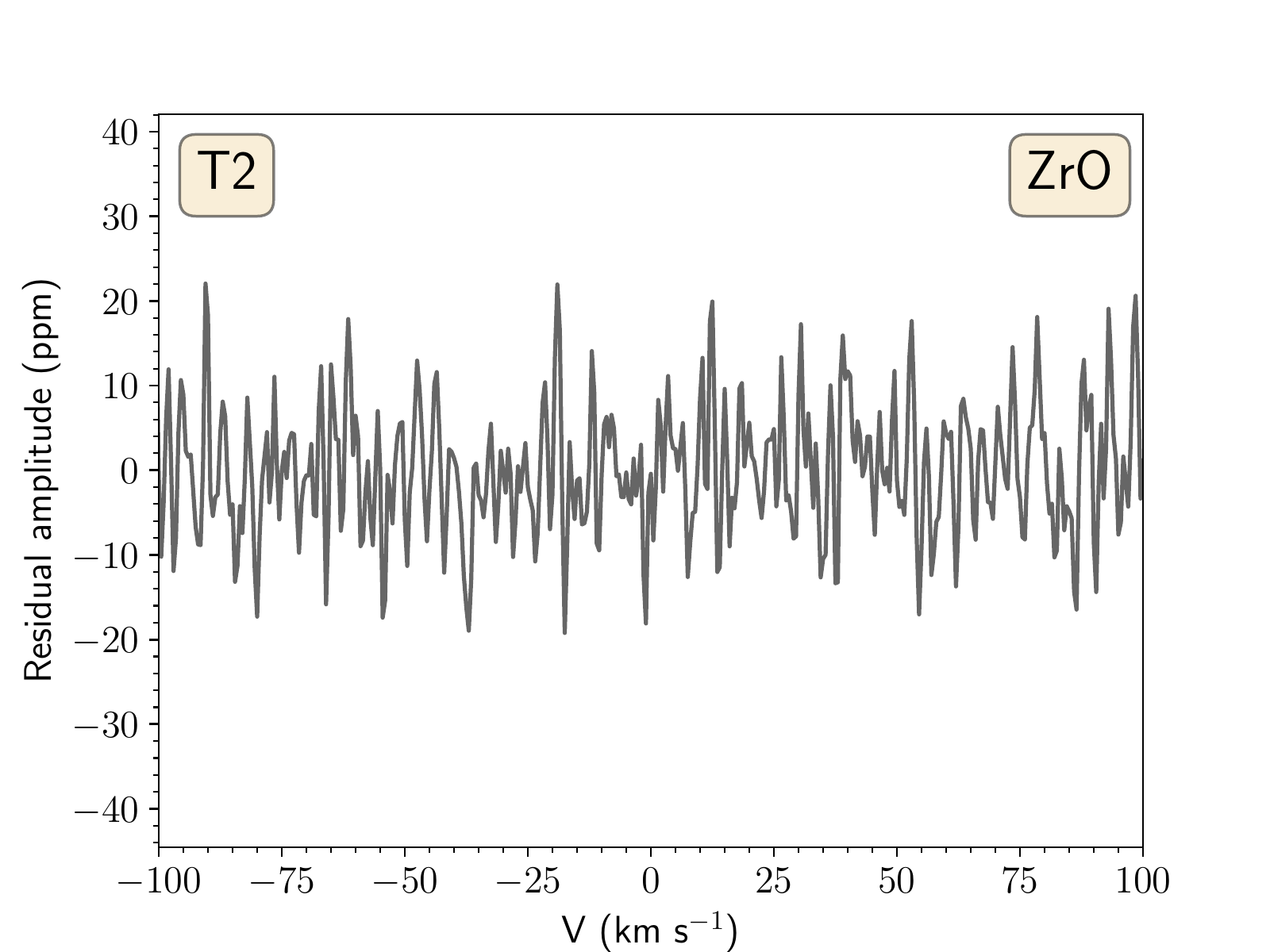}
\caption{ Same as Fig.~\ref{ccf_lines_planet} for the diatomic molecules. From top to bottom: TiO, VO, and ZrO.}
\label{ccf_lines_planet_mol}                                                                                                                                                                              
\end{figure*}

\begin{figure*}
\centering
    \includegraphics[width=0.48\textwidth]{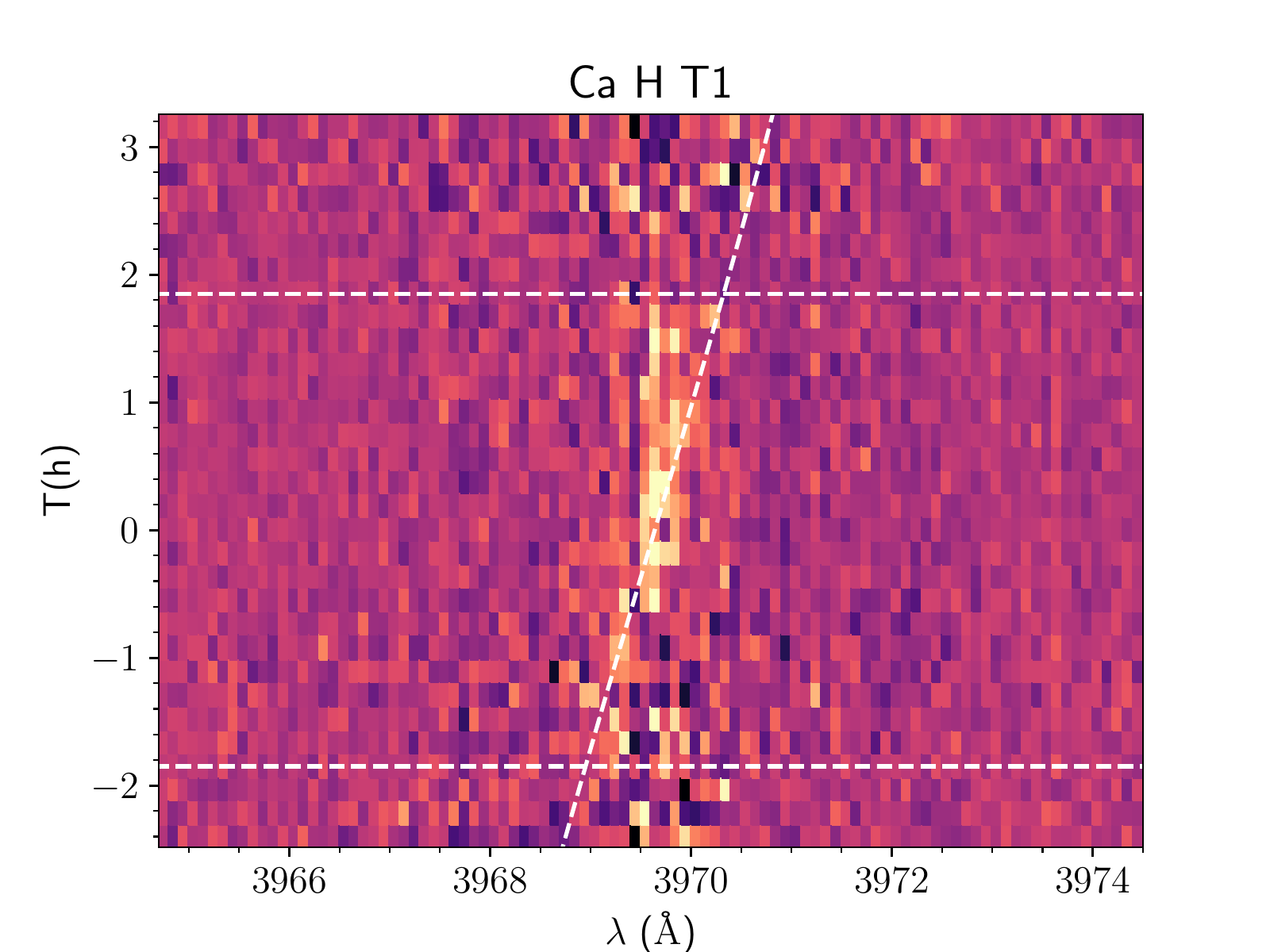}
    \includegraphics[width=0.48\textwidth]{map_CaH_T1.pdf}
    \includegraphics[width=0.48\textwidth]{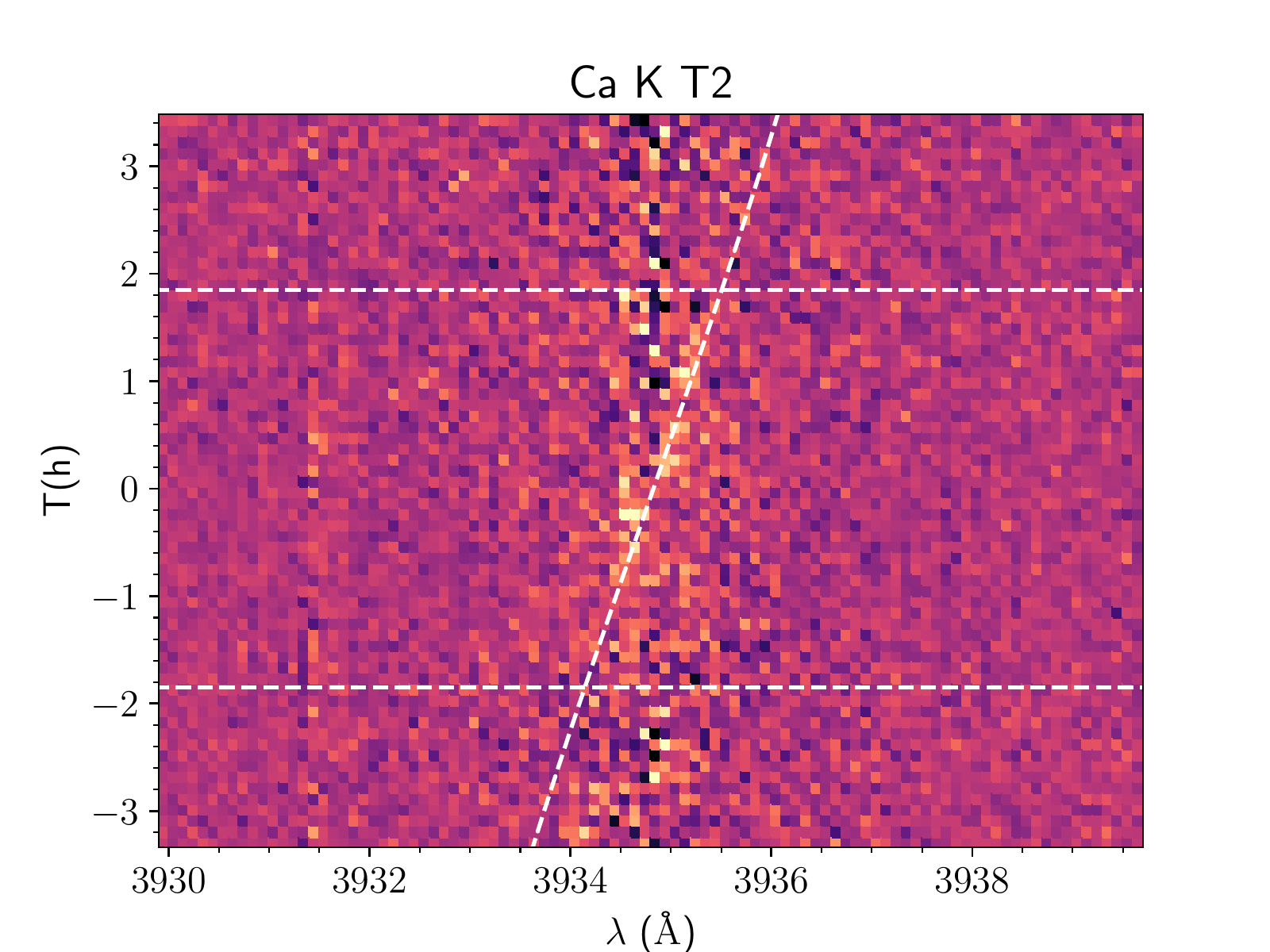}
    \includegraphics[width=0.48\textwidth]{map_CaK_T2.pdf}
    \caption{Same as Fig.~\ref{map_na_planet} but for  the \ion{Ca}{ii} HK lines}
    \label{map_HK}
\end{figure*}
\begin{figure*}
\includegraphics[width=0.48\textwidth]{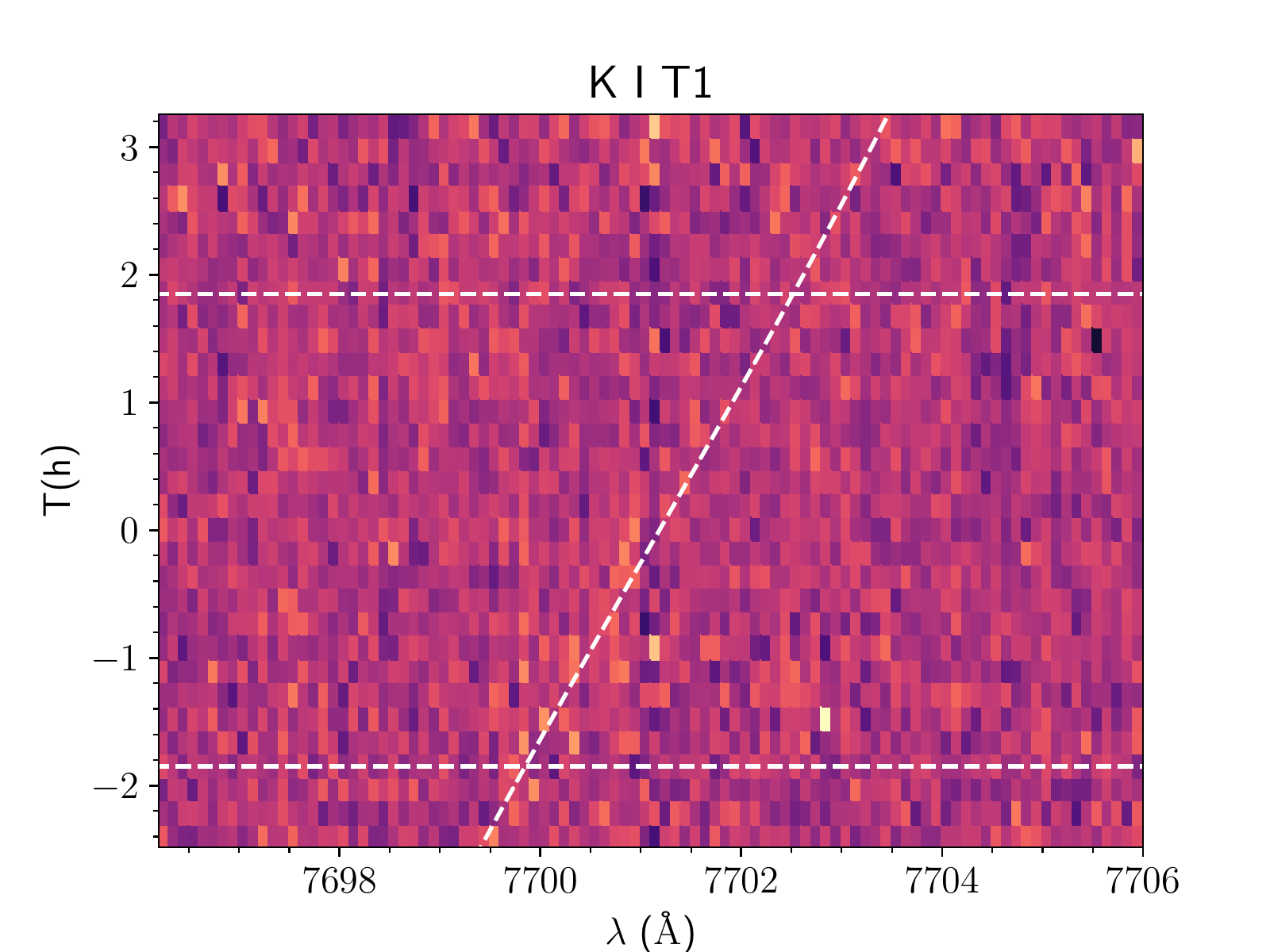}
\includegraphics[width=0.48\textwidth]{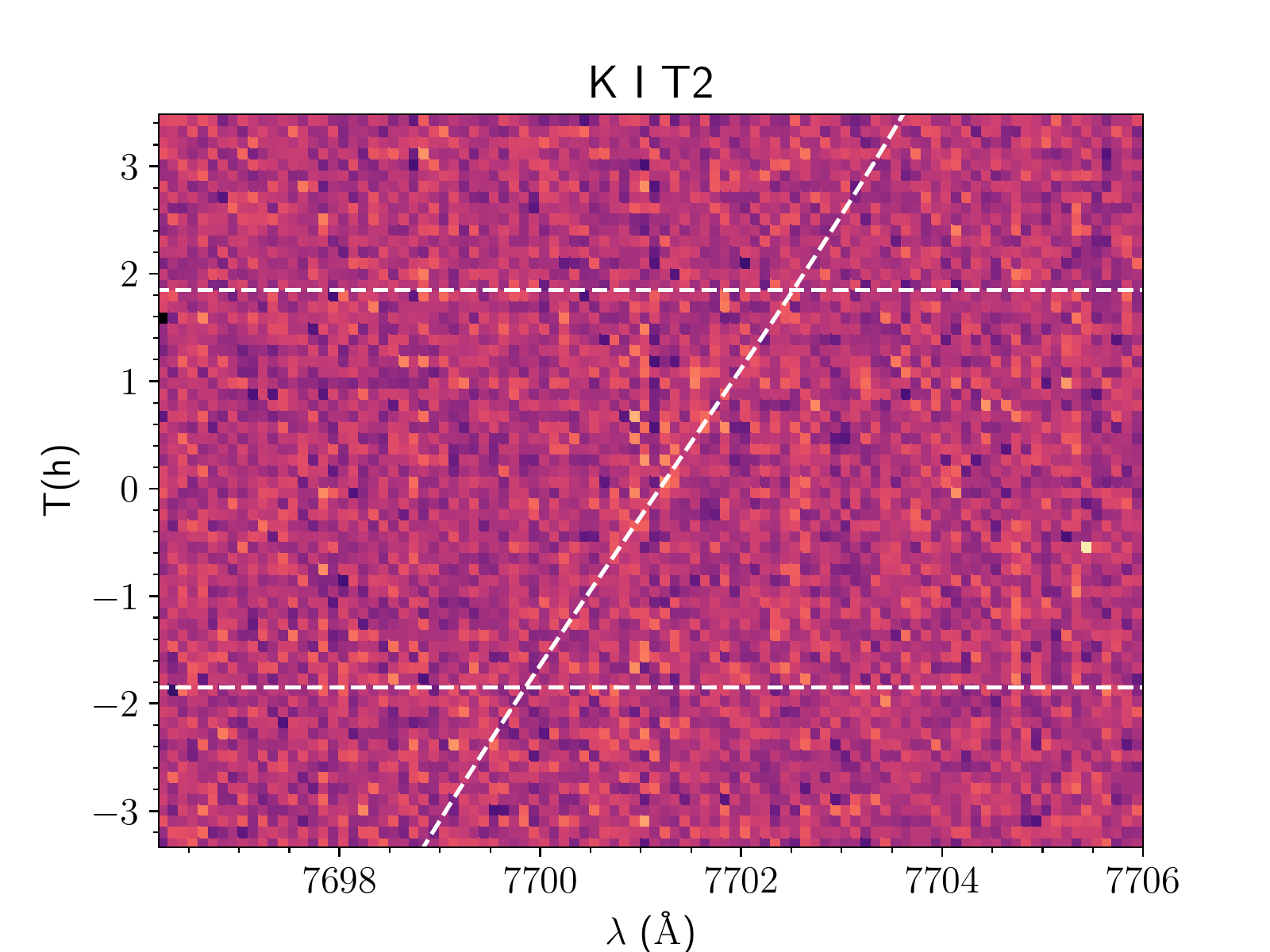}
\caption{Same as Fig.~\ref{map_na_planet} but for \ion{K}{i} }
\label{map_K}
\end{figure*}

\begin{figure*}
\includegraphics[width=0.48\textwidth]{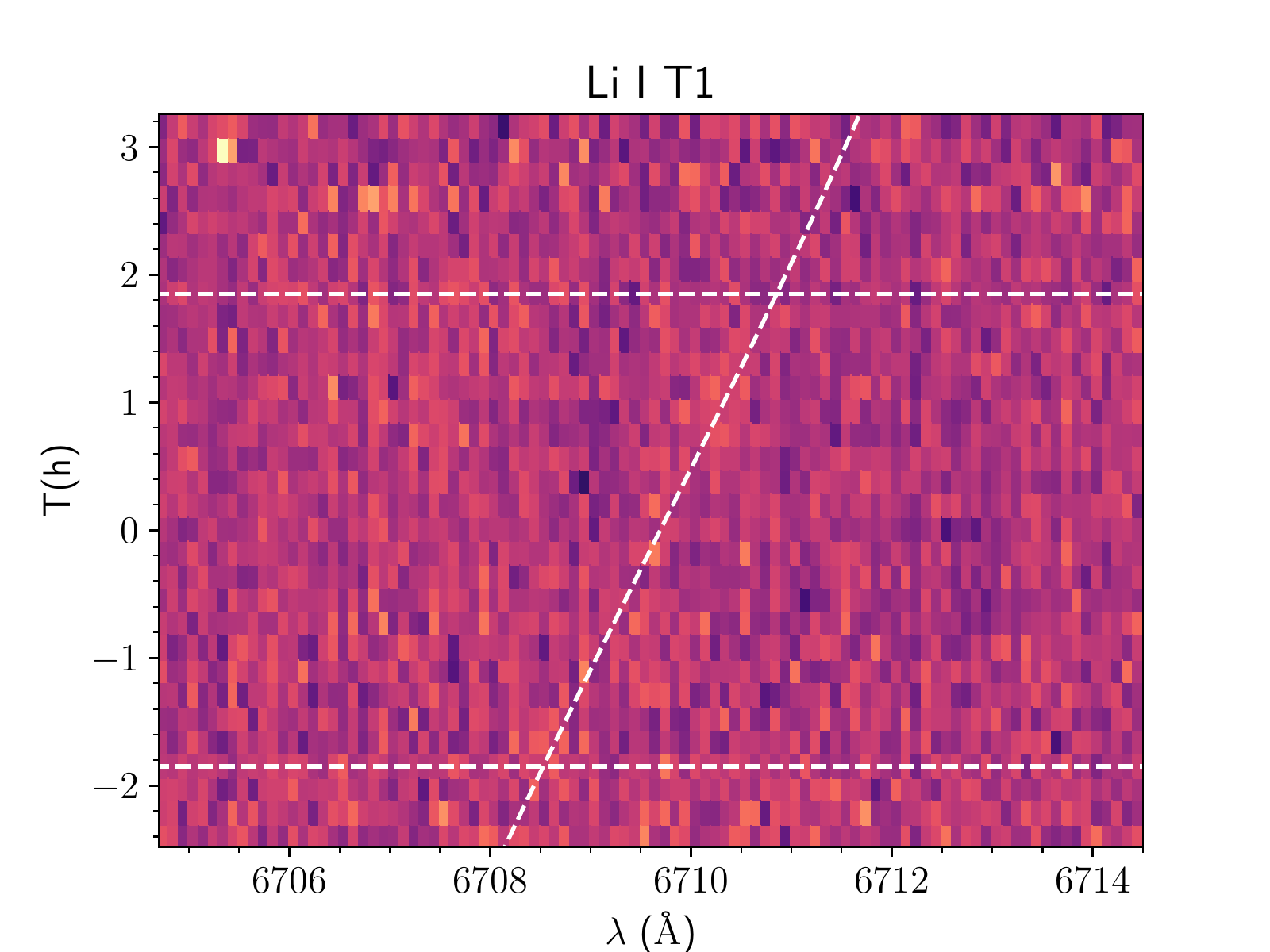}
\includegraphics[width=0.48\textwidth]{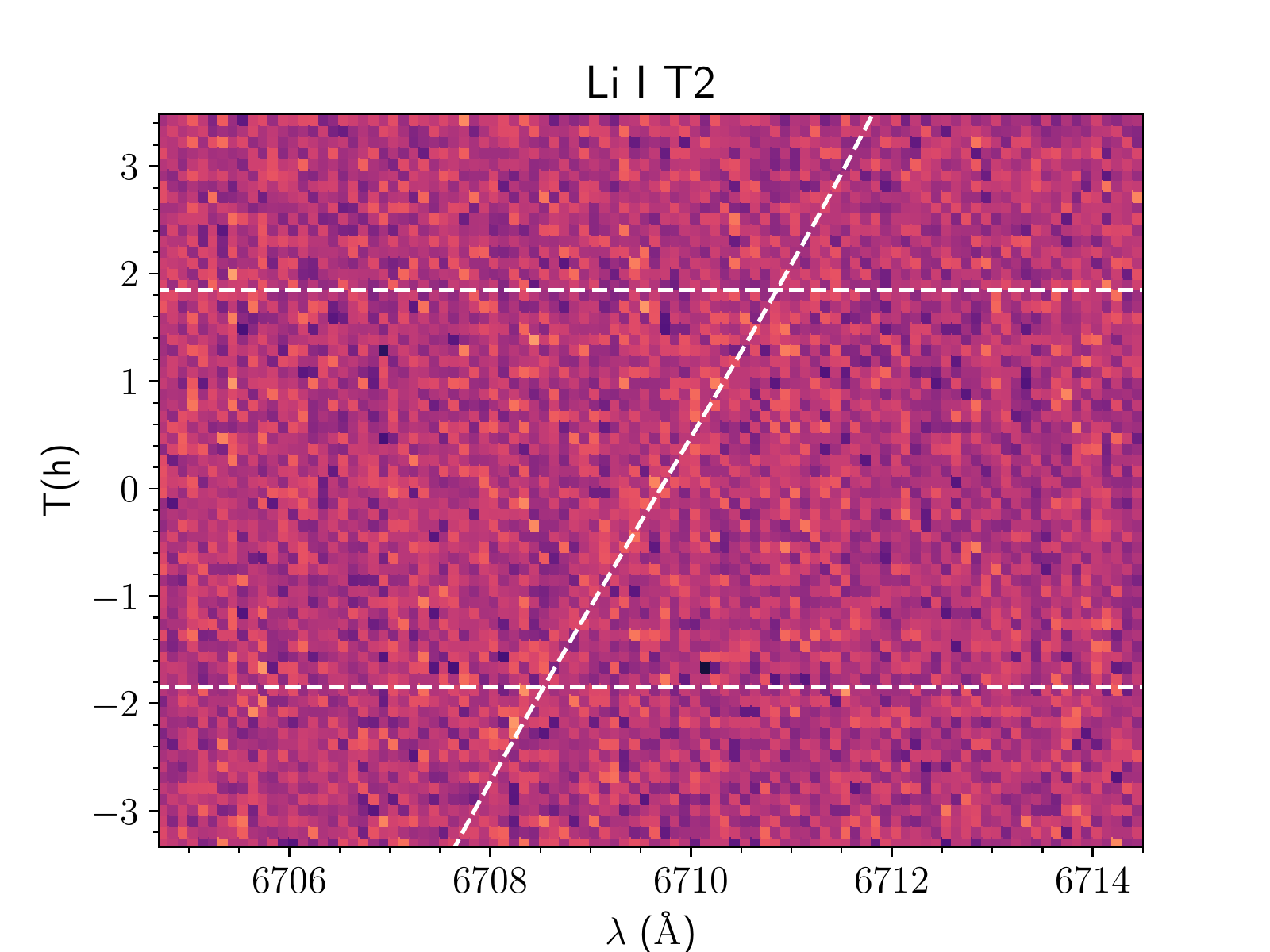}
    \caption{Same as Fig.~\ref{map_na_planet} but for \ion{Li}{i}}
	\label{map_Li}
\end{figure*}

 \begin{figure*}
 \includegraphics[width=0.48\textwidth]{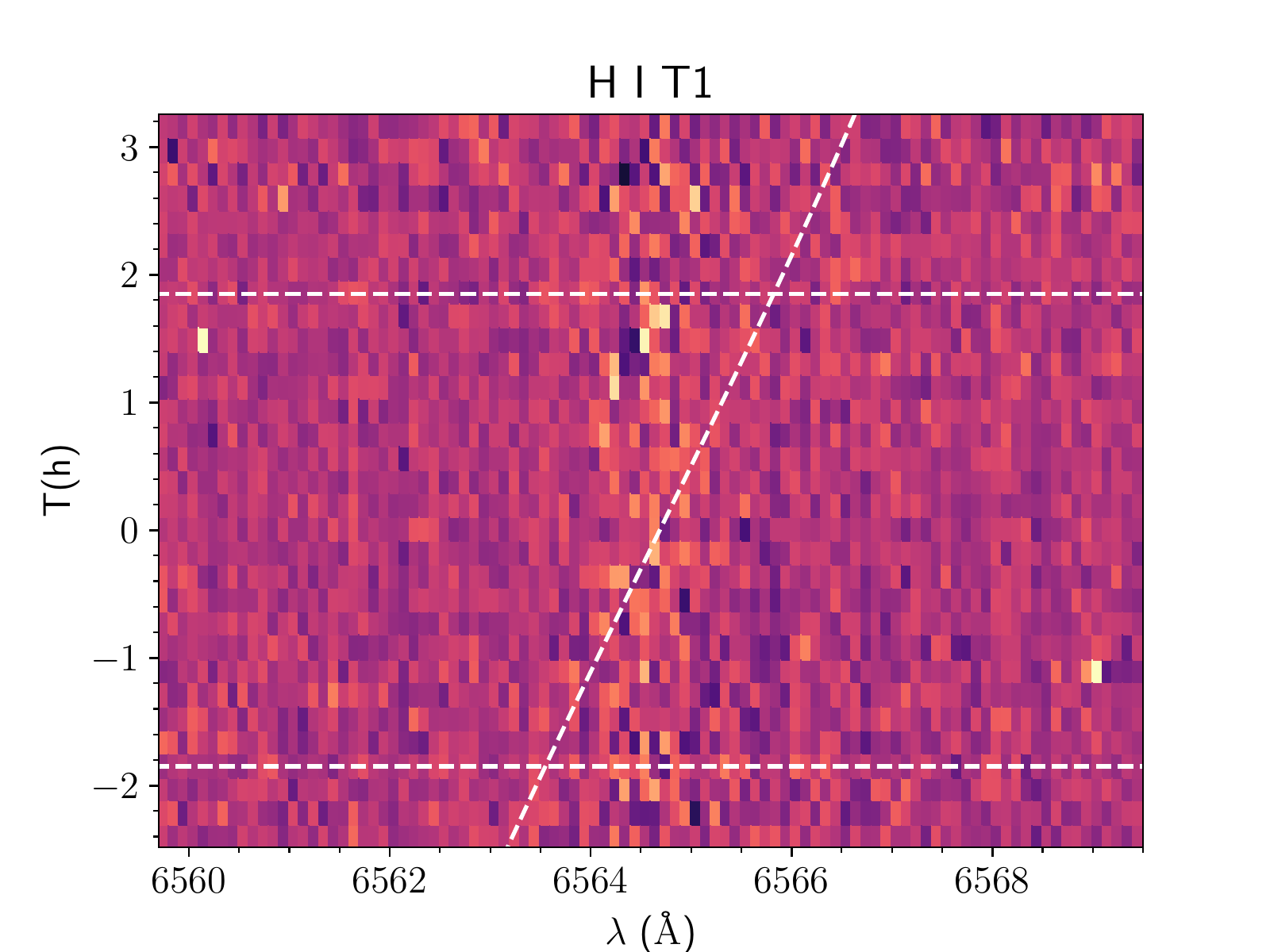}
\includegraphics[width=0.48\textwidth]{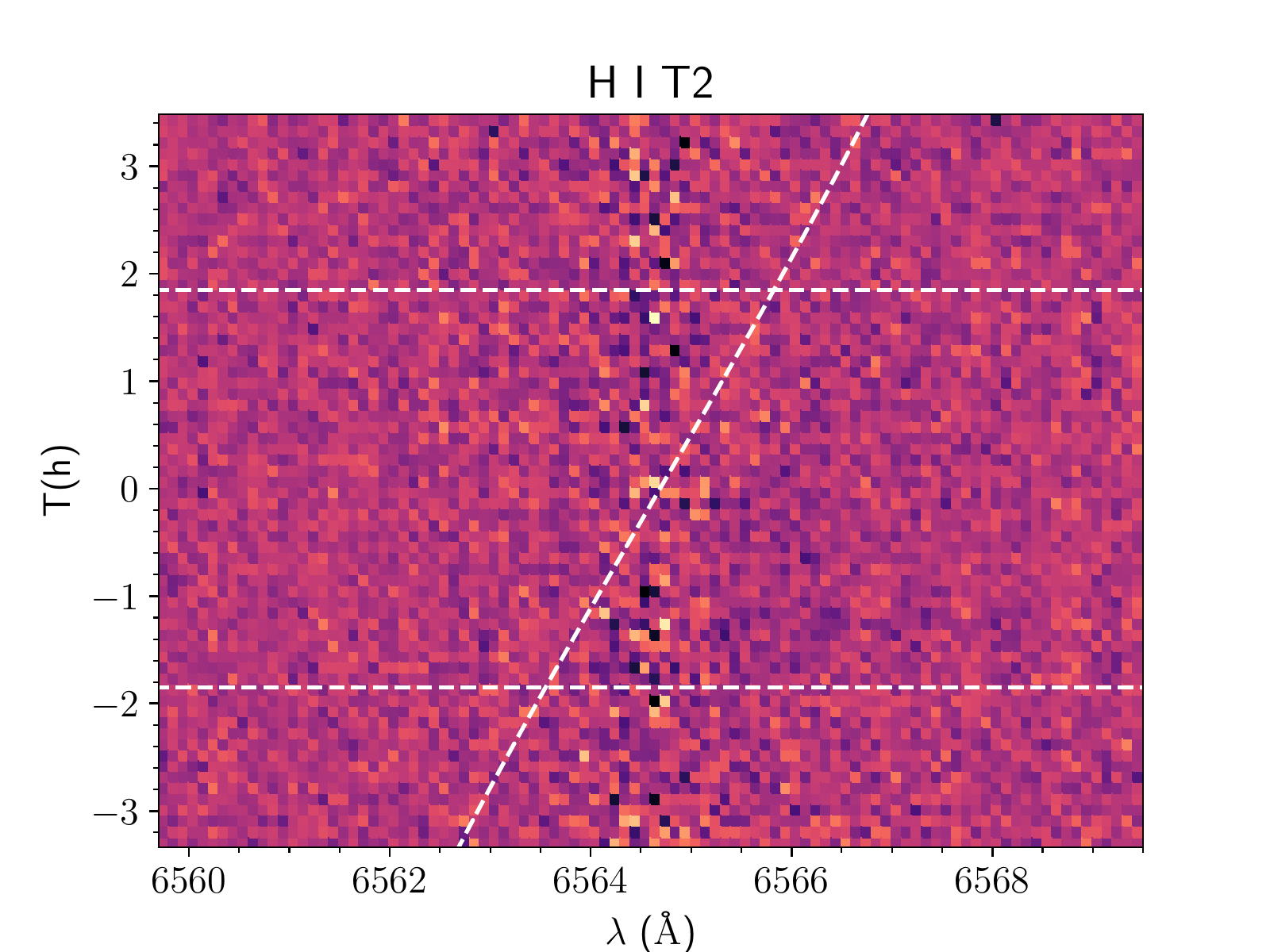}
      \caption{Same as Fig.~\ref{map_na_planet} but for H$\alpha$}                                                                                                                                        
			             \label{map_Ha}                                                                                                                                                       \end{figure*}  
\begin{table}
\small
\centering
\caption{Atomic data employed to calculate the chemical abundances of WASP-76.}  
\label{linelists}
\begin{tabular}{lccc}
\hline\hline                                                                                                                                                                                                
$\lambda$ & Species & $\chi_l$  & $\log{gf}$ \\ 
(\AA{})   &         &   (eV)    &  (dex) \\
\hline  \hline
\noalign{\smallskip}
6709.61  & \ion{Li}{i}  &  0.00  & 0.174 \\
5053.55  &  \ion{C}{i}  &  7.68  &  -1.304  \\
5381.82  &  \ion{C}{i} &  7.68  &  -1.615  \\
6589.43  &  \ion{C}{i} &  8.54  &  -1.021  \\
7774.08  &  \ion{O}{i} &  9.15  &  0.369  \\
7776.31  &  \ion{O}{i}  &  9.15  &  0.223  \\
7777.53  &  \ion{O}{i} &  9.15  &  0.002  \\
6155.93  &  \ion{Na}{i}  &  2.1  &  -1.547  \\
6162.45  &  \ion{Na}{i}  &  2.1  &  -1.246  \\
5712.67  &  \ion{Mg}{i}  &  4.35  &  -1.724  \\
6320.46  &  \ion{Mg}{i}  &  5.11  &  -2.103  \\
5519.07  &  \ion{Si}{i}  &  5.08  &  -2.611  \\
5647.18  &  \ion{Si}{i}  &  4.93  &  -2.043  \\
5667.13  &  \ion{Si}{i} &  4.92  &  -1.94  \\
5686.06  &  \ion{Si}{i}  &  4.95  &  -1.553  \\
5692.00 &  \ion{Si}{i} &  4.93  &  -1.773  \\
5702.69  &  \ion{Si}{i}  &  4.93  &  -1.953  \\
5709.98  &  \ion{Si}{i}  &  4.95  &  -1.37  \\
5749.26  &  \ion{Si}{i} &  5.61  &  -1.543  \\
5950.19  &  \ion{Si}{i} &  5.08  &  -1.13  \\
6126.72  &  \ion{Si}{i} &  5.61  &  -1.464  \\
6133.27  &  \ion{Si}{i}  &  5.62  &  -1.556  \\
6133.55  &  \ion{Si}{i} &  5.62  &  -1.615  \\
6144.18  &  \ion{Si}{i} &  5.62  &  -1.295  \\
6146.72  & \ion{Si}{i} &  5.62  &  -1.31  \\
6156.84  & \ion{Si}{i} &  5.62  &  -0.754  \\
6197.15  & \ion{Si}{i} &  5.87  &  -1.49  \\
6239.04  & \ion{Si}{i} &  5.61  &  -0.975  \\
6246.19  & \ion{Si}{i} &  5.62  &  -1.093  \\
6301.34  & \ion{Si}{i} &  5.98  &  -1.116  \\
6416.75  & \ion{Si}{i} &  5.87  &  -1.035  \\
6723.70  & \ion{Si}{i} &  5.86  &  -1.062  \\
6743.49  & \ion{Si}{i} &  5.98  &  -1.653  \\
7701.09  & \ion{K}{i} &  0.0  &  -0.154  \\
5263.17  & \ion{Ca}{i} &  2.52  &  -0.579  \\
5350.95  & \ion{Ca}{i} &  2.71  &  -0.31  \\
5514.51  & \ion{Ca}{i} &  2.93  &  -0.464  \\
5583.51  & \ion{Ca}{i} &  2.52  &  -0.555  \\
5591.67  & \ion{Ca}{i} &  2.52  &  -0.571  \\
5596.02  & \ion{Ca}{i} &  2.52  &  0.097  \\
5859.07  & \ion{Ca}{i} &  2.93  &  0.24  \\
5869.19  & \ion{Ca}{i} &  2.93  &  -1.57  \\
6168.15  & \ion{Ca}{i} &  2.52  &  -1.142  \\
6170.75  & \ion{Ca}{i} &  2.52  &  -0.797  \\
6171.27  & \ion{Ca}{i} &  2.53  &  -0.478  \\
6440.85  & \ion{Ca}{i} &  2.53  &  0.39  \\
6473.45  & \ion{Ca}{i} &  2.53  &  -0.686  \\
6495.58  & \ion{Ca}{i} &  2.52  &  -0.109  \\
6501.45  & \ion{Ca}{i} &  2.52  &  -0.818  \\
6510.65  & \ion{Ca}{i} &  2.53  &  -2.408  \\
6574.59  & \ion{Ca}{i} &  0.0  &  -4.24  \\
6719.54  & \ion{Ca}{i} &  2.71  &  -0.524  \\
\hline
\end{tabular}
\end{table}
\begin{table}
\small
\centering
\caption{Table~\ref{linelists} continued}  
\label{linelists2}
\begin{tabular}{lccc}
\hline\hline    
$\lambda$ & Species & $\chi_l$  & $\log{gf}$ \\ 
(\AA{})   &         &   (eV)    &  (dex) \\
\hline  \hline
\noalign{\smallskip}
4821.76  & \ion{Ti}{i} &  1.5  &  -0.380  \\
4914.98  & \ion{Ti}{i} &  1.87  &  0.220  \\
4979.58  & \ion{Ti}{i} &  1.97  &  -0.303  \\
5000.9  & \ion{Ti}{i} &  0.83  &  0.320  \\
5017.56  & \ion{Ti}{i} &  0.85  &  -0.480  \\
5024.27  & \ion{Ti}{i} &  0.83  &  -0.330  \\
5026.25  & \ion{Ti}{i} &  0.82  &  -0.530  \\
5211.83  & \ion{Ti}{i} &  0.05  &  -0.820  \\
5221.16  & \ion{Ti}{i} &  0.02  &  -2.220  \\
5691.04  & \ion{Ti}{i} &  2.30  &  -0.360  \\
5868.08  & \ion{Ti}{i} &  1.07  &  -0.790  \\
5920.18  & \ion{Ti}{i} &  1.07  &  -1.640  \\
5923.75  & \ion{Ti}{i} &  1.05  &  -1.380  \\
5980.2  & \ion{Ti}{i} &  1.87  &  -0.440  \\
6092.86  & \ion{Ti}{i} &  2.27  &  -0.320  \\
6127.91  & \ion{Ti}{i} &  1.07  &  -1.368  \\
6259.83  & \ion{Ti}{i} &  1.44  &  -0.390  \\
6262.83  & \ion{Ti}{i} &  1.43  &  -0.530  \\
5201.62  & \ion{Cr}{i} &  3.38  &  -0.580  \\
5240.42  & \ion{Cr}{i} &  2.71  &  -1.270  \\
5298.16  & \ion{Cr}{i} &  0.98  &  -1.360  \\
5299.49  & \ion{Cr}{i} &  2.90  &  0.099  \\
5299.75  & \ion{Cr}{i} &  0.98  &  -1.140  \\
5302.22  & \ion{Cr}{i} &  0.98  &  -2.000  \\
5305.66  & \ion{Cr}{i} &  3.46  &  -0.670  \\
5349.8  & \ion{Cr}{i} &  1.00  &  -1.210  \\
5703.89  & \ion{Cr}{i} &  3.45  &  -0.670  \\
6331.84  & \ion{Cr}{i} &  0.94  &  -2.787  \\
6015.16  & \ion{Mn}{i} &  3.07  &  -0.354  \\
6018.31  & \ion{Mn}{i} &  3.07  &  -0.181  \\
6023.46  & \ion{Mn}{i} &  3.08  &  -0.054  \\
4813.33  & \ion{Ni}{i} &  3.66  &  -1.450  \\
4905.78  & \ion{Ni}{i} &  3.54  &  -0.016  \\
4915.34  & \ion{Ni}{i} &  3.74  &  -0.500  \\
4937.21  & \ion{Ni}{i} &  3.94  &  -0.213  \\
4947.41  & \ion{Ni}{i} &  3.80  &  -1.151  \\
4954.58  & \ion{Ni}{i} &  3.74  &  -0.580  \\
5034.13  & \ion{Ni}{i} &  3.90  &  -1.398  \\
5083.76  & \ion{Ni}{i} &  3.66  &  -0.439  \\
5198.61  & \ion{Ni}{i} &  3.90  &  -1.291  \\
5437.37  & \ion{Ni}{i} &  1.99  &  -2.580  \\
5580.27  & \ion{Ni}{i} &  1.68  &  -2.830  \\
5595.29  & \ion{Ni}{i} &  3.90  &  -0.682  \\
5643.45  & \ion{Ni}{i} &  4.11  &  -1.046  \\
5696.56  & \ion{Ni}{i} &  4.09  &  -0.467  \\
5749.95  & \ion{Ni}{i} &  1.68  &  -3.240  \\
5848.61  & \ion{Ni}{i} &  1.68  &  -3.460  \\
6087.97  & \ion{Ni}{i} &  4.27  &  -0.410  \\
6109.81  & \ion{Ni}{i} &  1.68  &  -2.600  \\
6112.76  & \ion{Ni}{i} &  4.09  &  -0.865  \\
6177.08  & \ion{Ni}{i} &  4.09  &  -0.389  \\
6178.52  & \ion{Ni}{i} &  4.09  &  -0.260  \\
6188.42  & \ion{Ni}{i} &  4.11  &  -0.880  \\
6206.32  & \ion{Ni}{i} &  4.09  &  -1.080  \\
6225.7  & \ion{Ni}{i} &  4.11  &  -0.910  \\
6323.91  & \ion{Ni}{i} &  4.15  &  -1.115  \\
6329.35  & \ion{Ni}{i} &  1.68  &  -3.170  \\
6484.59  & \ion{Ni}{i} &  1.94  &  -2.630  \\
6588.13  & \ion{Ni}{i} &  1.95  &  -2.780  \\
6600.42  & \ion{Ni}{i} &  4.24  &  -0.821  \\
6636.96  & \ion{Ni}{i} &  4.42  &  -0.765  \\
6645.46  & \ion{Ni}{i} &  1.68  &  -2.220  \\
6769.64  & \ion{Ni}{i} &  1.83  &  -2.140  \\
6774.18  & \ion{Ni}{i} &  3.66  &  -0.797  \\

\hline
\end{tabular}
\end{table}
\end{appendix}

%
%

\end{document}